\RequirePackage{fix-cm}
\documentclass[twocolumn]{svjour2}          
\smartqed  
\usepackage{graphicx}
\usepackage{graphicx}
\usepackage{dcolumn}
\usepackage{bm}
\usepackage{amssymb}   
\usepackage{subfigure}
\usepackage{tikz}
\usepackage{epstopdf}
\usepackage{epsfig}
\usepackage{upgreek}

\usepackage{textcomp}
\usepackage{amsmath}
\usepackage{hyperref}

\newcommand{\rmax} { r_\mathrm{max} }  
\newcommand{\rmin} { r_\mathrm{min} } 

\newcommand{\stresstensor} {\boldsymbol\sigma}
\newcommand{\Qtensor} {{\mathrm{\boldsymbol{\mathrm Q}}}} 

\newcommand{\Cstar} {C^{*}}

\newcommand{\Cstariso} {C^*_0}
\newcommand{\Pres} {P}
\newcommand{\norPres} {p}
\newcommand{\shstress} {\Gamma}
\newcommand{\nstress} {\uptau}
\newcommand{\stressrat} {\nstress/\norPres}

\newcommand{\shstrain} {{\varepsilon}_{d}} 
\newcommand{\anisoF} {F_\mathrm{d}}

\newcommand{\strnrate} {{\mathrm{\dot{\boldsymbol{\mathrm E}}}}}
\newcommand{\Fv} {F_\mathrm{v}}
\newcommand{\volfrac} {\phi}
\newcommand{\volfracmax} {\phi^{\mathrm {max}}}
\newcommand{\volfracJ} {\phi_J}
\newcommand{\volfracJmax} {\phi_J^{\mathrm{max}}}  

\newcommand{\volfracJi} {\phi_{J,i}}
\newcommand{\MvolfracJi} {{}^M\phi_{J,i}}
\newcommand{\ovolfracJi} {{}^1\phi_{J,i}}
\newcommand{\volfracscaled} {\phi_{sc}}
\newcommand{\volfracSJ} {\phi_{c}}  
\newcommand{\volfract} {\phi_t}

\newcommand{\epsddot} { {\dot\epsilon}_\mathrm{d} }

\newcommand{\fNR} {f_\mathrm{NR}}
\newcommand{\favg} {f_\mathrm{avg}}

\newcommand{\Qd} { Q_\mathrm{d} }
\newcommand{\volfraccr} {\phi_\mathrm{cr}}

\usepackage[sort&compress, numbers]{natbib}

\begin{document}\sloppy

\title{Memory of jamming -- multiscale models for soft and granular matter
}


\author{Nishant Kumar \and Stefan Luding
}


\institute{Multi Scale Mechanics (MSM), \\
Faculty of Engineering Technology, MESA+, \\
P.O.Box 217, 7500 AE Enschede, The Netherlands\\
\vspace{3mm} 
\email{n.kumar@utwente.nl} \and   
\email{s.luding@utwente.nl}            \\
The authors declare no conflict of interest.
}

\date{\today}

\maketitle

\begin{abstract}
Soft, disordered, micro-structured materials are ubiquitous in nature and industry, 
and are different from ordinary fluids or solids, with unusual, interesting 
static and flow properties. 
The transition from fluid to solid -- at the so-called jamming density -- features a multitude 
of complex mechanisms, but there is no unified theoretical framework that explains them all.
In this study, a simple yet quantitative and predictive model is presented, 
which allows for a {\em variable, changing jamming density}, 
encompassing the memory of the deformation history and explaining a multitude 
of phenomena at and around jamming. 
The jamming density, now introduced as a {\em new state-variable}, changes due to the 
deformation history and relates the system's macroscopic response to its micro-structure.
The packing efficiency can increase {\em logarithmically slow} under 
gentle repeated (isotropic) compression, leading to an increase of the jamming
density. In contrast, shear deformations cause anisotropy, 
changing the packing efficiency {\em exponentially fast}
with either dilatancy or compactancy.
The memory of the system near jamming can be explained 
by a micro-statistical model that involves a multiscale, fractal energy 
landscape and links the microscopic particle picture to the macroscopic
continuum description, providing a unified explanation for the qualitatively 
different flow-behavior for different deformation modes. 
To complement our work, a recipe to extract the history-dependent jamming density 
from experimentally accessible data is proposed, and alternative state-variables are 
compared.
The proposed simple macroscopic constitutive model is calibrated with the memory of microstructure. 
Such approach can help understanding
predicting and mitigating failure of structures or geophysical hazards, and
will bring forward industrial process design and optimization, 
and help solving scientific challenges in fundamental research. 

\end{abstract}

\section{Introduction}\label{sec:introduction}
Granular materials are a special case of soft-matter with micro-structure, as also foams, colloidal systems, 
glasses, or emulsions \citep{denisvoc2013resolving, trappe2001jamming, walker2015self}.  
Particles can flow through a hopper or an hour-glass when shaken, but jam 
(solidify) when the shaking stops \citep{wambaugh2007response}. 
These materials jam above a ``certain''
volume fraction, or jamming density, referred to as the ``jamming point'' or ``jamming density''
\citep{liu1998nonlinear, song2008phase, bi2011jamming, 
zhang2005jamming, ohern2003jamming, silbert2010jamming, 
silbert2006structural, otsuki2011critical,  wang2012edwards, 
silbert2009normal, pica2009jamming, van2010jamming, banigan2013chaotic, liu2010jamming, cates1998jamming, majmudar2007jamming,coulais2014ideal, walker2015self,reichhardt2014aspects, torquato2010jammed}, 
and become mechanically stable with finite bulk- and shear-moduli 
\citep{pica2009jamming, ohern2003jamming, zhang2005jamming, otsuki2011critical, bohy2012soft, parisi2010hard,inagaki2011protocol, metayer2011shearing}. 
Notably, in the jammed state, these systems can ``flow'' by reorganizations of their micro-structure 
\citep{saitoh2015master, farhadi2014dynamics}.
Around the jamming transition, these systems display considerable inhomogeneity, 
such as reflected by 
over-population of weak/soft/slow mechanical 
oscillation modes \citep{silbert2006structural}, force-networks \citep{silbert2010jamming, radjai1998bimodal, snoeijer2006sheared}, 
diverging correlation lengths and relaxation time-scales 
\citep{ohern2003jamming, wang2012edwards, lerner2012toward, brown2009dynamic, reichhardt2014aspects, suzuki2015divergence, wyart2014discontinuous}, 
and some universal scaling behaviors \citep{otsuki2012critical, otsuki2014avalanche}. 
Related to jamming, but at all densities, other phenomena occur, like shear-strain localization 
\citep{otsuki2011critical, peyneau2008frictionless, van2010jamming, schall2009shear, singh2014effect}, anisotropic evolution of structure and stress 
\citep{silbert2006structural, ohern2003jamming, imole2013hydrostatic, 
wang2012edwards, bi2011jamming, ciamarra2011jamming, peyneau2008frictionless, radjai1998bimodal, snoeijer2006sheared, schall2009shear, singh2014effect, imole2014micro, kumar2014effects, zhao2013unique, guo2013signature}, 
and force chain inhomogeneity \citep{bi2011jamming, cates1998jamming, saitoh2015master}.
To gain a better understanding of the jamming transition concept, one needs to consider 
both the structure (positions and contacts) and contact forces. Both of them illustrate and reflect the transition, e.g.,\ with a strong force chain network percolating the full system and 
thus making unstable packings permanent, stable and rigid \citep{zhang2010jamming, bi2011jamming, wang2013regime, cates1998jamming, walker2014uncovering}.

For many years, scientists and researchers have considered the jamming transition 
in granular materials to occur at \textit{a particular} volume fraction, $\volfracJ$ \citep{brown1945packing}. 
In contrast, over the last decade, numerous experiments and computer simulations have suggested the existence of a broad range of $\volfracJ$, even for a given material. 
It was shown that the critical density for the jamming transition depends on 
the preparation protocol  
\citep{charbonneau2012universal, torquato2010jammed, liu2010jamming, olsson2011critical, olsson2013athermal, otsuki2011critical, otsuki2012critical, ozawa2012jamming, ohern2001force, torquato2000random, mari2009jamming, bandi2013fragility, reichhardt2014aspects}, 
and that this state-variable can be used to describe and scale macroscopic properties of the system \citep{inagaki2011protocol}. 
For example, rheological studies have shown that $\volfracJ$ decreases with increasing compression rate \citep{ashwin2013inherent, zhang2005jamming, mari2009jamming, vagberg2011glassiness} 
(or with increasing growth rate of the particles), 
with the critical scaling by the distance from the jamming point ($\volfrac - \volfracJ$) being universal and independent of $\volfracJ$ 
\citep{charbonneau2012universal, otsuki2012critical,chaudhuri2010jamming, zhao2011new, majmudar2007jamming}
Recently, the notion of an a-thermal isotropic jamming ``point'' was 
challenged due to its protocol dependence, suggesting the extension of the jamming point, to become a \textit{J-segment} \citep{ciamarra2010recent, ciamarra2011jamming, vagberg2011glassiness}. 
Furthermore, it was shown experimentally, that for a tapped, unjammed frictional 2D systems, 
shear can jam the system (known as ``shear jamming''), with force chain networks percolating throughout the system, making 
the assemblies jammed, rigid and stable \citep{zhang2010jamming, bi2011jamming, ren2013reynolds, wang2013regime, farhadi2014dynamics, grob2014jamming}, all highlighting a memory that makes the structure dependent on history $H$.
But to the best of our knowledge, quantitative characterization of the varying/moving/changing transition points, based on $H$, remains a major open challenge.

\subsection{Application examples}

In the fields of material science, civil engineering and geophysics, 
the materials behave highly hysteretic, non-linear and involve irreversibility (plasticity), possibly already 
at very small deformations, due to particle rearrangements, more visible near the jamming transition 
\cite{bardet1994observations,goddard1990nonlinear,kumar2014macroscopic,sibille2009analysis}. 
Many industrial and geotechnical applications that are crucial for our society involve 
structures that are designed to be far from failure (e.g.\ shallow foundations or underlying infrastructure),
since the understanding when failure and flow happens is not sufficient, but is
essential for the realistic prediction of ground movements \cite{clayton2011stiffness}.
Finite-element analyses of, for example tunnels, depend on the model adopted for the pre-failure 
soil behavior; when surface settlement is considered, the models predicting non-linear elasticity 
and history dependence become of utmost importance \cite{addenbrooke1998influence}.
Design and licensing of infrastructure such as nuclear plants and long span 
bridges are dependent on a robust knowledge of elastic properties in order to 
predict their response to seismic ground motion such as the risk of liquefaction 
and the effect of the presence of anisotropic strata. 
(Sediments are one example of anisotropic granular materials of particles of organic or inorganic origin 
that accumulate in a loose, unconsolidated form before they are compacted and solidified. 
Knowing their mechanical behavior is important in industrial, geotechnical and geophysical applications.
For instance, the elastic properties of high-porosity ocean-bottom sediments have a massive impact 
on unconventional resource exploration and exploitation by ocean drilling programs.)

When looking at natural flows, a complete description of the granular rheology 
should include an elastic regime \cite{campbell2006granular}, and the onset of failure 
(flow or unjamming) deserves particular attention in this context. 
The material parameters have a profound influence on the computed deformations prior to failure 
\cite{griffiths1999slope, einav2004pressure}, as the information on the material state is usually 
embedded in the parameters. Likewise, also for the onset of flow, the state of the material
is characterized by the value of the macroscopic friction angle, 
as obtained, e.g., from shear box experiments or tri-axial tests. 
Since any predictive model must describe the pre-failure deformation 
\cite{jamiolkowski1999pre} as well as the onset of flow (unjamming) of the material,
many studies have been devoted to the characteristics of geomaterials 
(e.g., tangent moduli, secant moduli, peak strength) and
to the post-failure regime \cite{tutluouglu2015relationship}
or the steady (critical) state flow rheology, see Refs.\ \cite{singh2014effect,singh2015rheology}
and references therein.

\subsection{Approach of this study}


Here, we consider frictionless sphere assemblies in a periodic system, 
which can help to elegantly probe the behavior of disordered bulk granular matter,
allowing to focus on the structure \citep{walker2015self}, without being disturbed by other 
non-linearities \citep{bi2011jamming, hartley2003logarithmic,farhadi2014dynamics} 
(as e.g.\ friction, cohesion, walls, environmental fluids or non-linear interaction laws). 
For frictionless assemblies, it is often assumed that the influence of memory 
is of little importance, maybe even negligible. If one really looks close enough,
however, its relevance becomes evident. We quantitatively explore its structural 
origin in systems where the re-arrangements of the micro-structure (contact network)
are the only possible mechanisms leading to the range of jamming densities (points), i.e.\ a 
variable state-variable jamming density. 

In this study, we probe the jamming transition concept by two pure deformation modes: 
isotropic compression  or ``tapping'' and deviatoric pure shear (volume conserving), 
which allow us to combine the \textit{J-segment} concept with a history 
dependent jamming density 
\footnote{Tapping or compression may not be 
technically equivalent to the protocol isotropic compression. 
In soil mechanics, the process of tapping may involve anisotropic compression or shear. 
The process of compression may be either isotropic or anisotropic or even involving shear. 
For example, a typical soil tests may include biaxial compression, conventional triaxial compression and true triaxial compression. 
In this work, in the context of compression, we always mean true isotropic in strain. 
In the context of tapping, we assume that the granular temperature, which is often assumed isotropic, 
does the work, even though the tapping process is normally not isotropic. 
So this is an oversimplification, and subject to future study since it was not detailed here.
}. 
Assuming that all other deformations can be superimposed by these two pure modes,
we coalesce the two concepts of isotropic and shear induced jamming, and provide
the unified model picture, involving a multiscale, fractal-type energy landscape  
\citep{krzakala2007landscape, xu2011direct, liu2010jamming,mobius2014ir}; in general,
deformation (or the preparation procedure) modify the landscape and its population; considering 
only changes of the population already allows 
to establish new configurations and to predict their evolution. %
The observations of different $\volfracJ$ of a single material require an alternative
interpretation of the classical ``jamming diagram'' \citep{liu1998nonlinear}.

Our results will provide a unified picture, including some
answers to the open questions from literature:  
$(i)$ What lies in between the jammed and flowing (unjammed) regime? -- as 
posed by \citet{ciamarra2010recent}.
$(ii)$ Is there an absolute minimum jamming density? -- as 
posed by \citet{ciamarra2010recent}.
$(iii)$ What protocols can generate jammed states?-- as 
posed by \citet{torquato2000random}.
$(iv)$ What happens to the jamming and shear jamming regime in 3D and is friction important to observe it? -- as posed by \citet{bi2011jamming}.
Eventually, accepting the fact that the jamming density is changing with 
deformation history, significant improvement of
continuum models is expected, not only for classical elasto-plastic or
rheology models, but also, e.g.,\ for anisotropic constitutive models 
\citep{rognon2008dense,sun2011constitutive,imole2013hydrostatic,kumar2014macroscopic}, 
GSH rate type models \citep{jiang2008incremental, jiang2015applying}, 
Cosserat micro-polar or hypoplastic models \citep{mohan2002frictional, goncu2012mechanics, tejchman2008shear}
or continuum models with a length scale and non-locality 
\citep{KamrinKoval2012,henann2013predictive}. 
For this purpose we provide a simple (usable) analytical macro/continuum model 
as generalization of continuum models by adding one isotropic state-variable. 
Only allowing $\volfracJ(H)$ to be dependent on history $H$, as key modification,
explains a multitude of reported observations and can be significant step forward to solve 
real-world problems in e.g. electronic industry related novel 
materials, geophysics or mechanical engineering.

Recent works showed already that,
along with the classical macroscopic properties (stress and volume fraction),
the structural anisotropy is an important 
\cite{luding2005anisotropy,imole2013hydrostatic,ezaoui2009experimental,
magnanimo2008characterizing,ragione2012contact, guo2013signature, zhao2013unique} state-variable for granular materials,
as quantified by the fabric tensor \cite{imole2014micro,kumar2014macroscopic}
that characterizes, on average, the geometric arrangement of the particles, the contacts 
and their network, i.e.\ the microstructure of the particle packing. Note that the anisotropy
alone is not enough to characterize the structure, but also an isotropic state-variable
is needed, as is the main message of this study.

\subsection{Overview}
The paper continues with the simulation method in section \ref{sec:Methodology}, before the micromechanical 
particle- and contact-scale observations are presented in section \ref{sec:results}, providing analytical
(quantitative) constitutive expressions for the change of the jamming density with different 
modes of deformation.
Section \ref{sec:slowmodel} is dedicated to a (qualitative) meso-scale stochastic model that explains
the different (slow versus fast) change of $\volfracJ(H)$ for different deformation modes
(isotropic versus deviatoric/shear). 
A quantitative predictive macroscale model is presented in section \ref{sec:model} and verified by
comparison with the microscale simulations, before an experimental validation
procedure is discussed in section \ref{appA} and the paper is summarized and conclusions
are given in section \ref{sec:interpretation}.

\section{Simulation method}
\label{sec:Methodology}

Discrete Element Method (DEM) simulations are 
used to model the deformation behavior of systems with $N = 9261$ soft frictionless spherical particles with average 
radius $\langle r \rangle= 1$ [mm], density $\rho= 2000$ [kg/m\textsuperscript{3}], and a uniform polydispersity width $w = \rmax/\rmin = 3$, 
using the linear visco-elastic contact model in a 3D box with periodic boundaries \citep{kumar2014macroscopic, kumar2014effects}. 
The particle stiffness is $k = 10^8$ [kg/s\textsuperscript{2}], contact viscosity is $\gamma= 1$ [kg/s]. 
A background dissipation force proportional to the moving  velocity is added 
with $\gamma_b= 0.1$ [kg/s]. The particle density is 
$\rho= 2000$ [kg/m\textsuperscript{3}]. 
The smallest time of contact is $t_c = 0.2279$ [$\mu$s] 
for a collision between two smallest sized particles \citep{imole2013hydrostatic}.

\subsection{Preparation procedure and main experiments}
\label{sec:prep}
For the preparation, the particles 
are generated with random velocities at volume (solid) fraction $\volfrac=0.3$ and are isotropically 
compressed to $\volfract=0.64$, and later relaxed. 
From such a relaxed, unjammed, stress free initial state with volume fraction, $\volfract = 0.64 < \volfracJ$, 
we compress isotropically further to a maximum volume fraction, $\volfracmax_i$, and decompress back 
to $\volfract$, during the latter unloading $\volfracJ$ is identified.
This process is repeated over $M$ ($100$) cycles, 
which provides different isotropic jamming densities (points) $\volfracJ=:\MvolfracJi$, 
related with $\volfracmax_i$ and $M$ (see section\ \ref{subsec:iso1}). 

Several isotropic configurations $\volfrac$, 
such that $\volfract < \volfrac < \ovolfracJi$ from the 
decompression branch are chosen as the initial configurations for shear experiments. 
We relax them and apply pure (volume conserving) shear (plane-strain) with the diagonal 
strain-rate tensor $\strnrate = \pm\epsddot \left(-1,1,0\right)$, for four cycles
\footnote{
This deformation mode represents the only fundamental 
deviatoric deformation motion (complementary to isotropic deformation), since axial strain can be 
superposed by two plane-strain modes, and because the plane-strain mode allows to study
the non-Newtonian out-of-shear-plane response of the system (pressure dilatancy), whereas
the axial mode does not. If superposition is allowed, as it seems to be the case for frictionless particles,
studying only these two modes is minimal effort, however, we cannot directly extrapolate to more realistic
materials.}. 
The $x$ and $y$ walls move, while the $z$ wall is stationary.
The strain rate of the (quasi-static) deformation is small, $\epsddot t_c < 3.10^{-6} $, to 
minimize transient behavior and dynamic effects
\footnote{
For the isotropic deformation tests, we move the (virtual) walls and 
for the shear tests, we move all the grains according to an affine motion compatible with the (virtual) wall motion. 
When only the (virtual) walls moves, some arching near the corners can be seen when there is a huge particle size dispersity or if there is a considerable particle friction (data not shown). 
For the small polydispersity and the frictionless spheres considered in this work, the system is and remains homogeneous and the macroscopic quantities are indistinguishable between the two methods, however,
this must not be taken for granted in the presence of friction or cohesion, where wall motions other than
by imposed homogeneous strain, can lead to undesired inhomogeneities in the periodic representative
volume element.}.

\subsection{Macroscopic (tensorial) quantities} 
\label{subsec:tensorial}
 Here, we focus on defining averaged tensorial macroscopic quantities 
 -- including strain-, stress- and fabric (structure) tensors -- that provide information about the state of the packing
and reveal interesting bulk features.

From DEM simulations, one can measure the `static' stress in the system \citep{christoffersen1981micromechanical} as 
\begin{equation}
\label{stresseqn}
\stresstensor=\left({1}/{V}\right)\sum_{c\in V}\mathbf{l}^{c}\otimes\mathbf{f}^{c}, 
\end{equation}
average over all the contacts in the volume $V$ of the dyadic products between the contact force $\mathbf{f}^{c}$
and the branch vector $\mathbf{l}^{c}$, where the contribution of the kinetic fluctuation energy has been neglected 
\citep{luding2005anisotropy, imole2013hydrostatic}. The dynamic component of the stress tensor is four orders of magnitude smaller than the former
and hence its contribution is neglected. 
The isotropic component of the stress is the pressure $\Pres = \mathrm{tr}(\stresstensor)/3$. 

In order to characterize the geometry/structure of the static aggregate at microscopic level, 
we will measure the fabric tensor, defined as 
\begin{equation}
\mathbf{F}=\frac{1}{V}\sum_{{\mathcal{P}}\in V}V^{{\mathcal{P}}}\sum_{c\in {\mathcal{P}}}\mathbf{n}^{c}\otimes\mathbf{n}^{c},
\label{eq:fabriceq}
\end{equation}
where $V^{\mathcal{P}}$ is the volume relative to particle ${\mathcal{P}}$, which lies inside the
averaging volume $V$, and $\mathbf{n}^{c}$ is the normal unit branch-vector
pointing from center of particle ${\mathcal{P}}$ to contact $c$ 
\citep{luding2005anisotropy, kumar2013evolution, zhang2010statistical}. 
Isotropic part of fabric is $\Fv = \mathrm{tr}(\mathbf{F})$. 
The corrected coordination number \citep{imole2013hydrostatic, bi2011jamming}
is $\Cstar = {M_4}/{N_4},$ where, $M_4$ is total contacts of 
the $N_4$ particles having at least 4 contacts, and the non-rattler fraction 
is $\fNR = N_4/N$. 
$C$ is the ratio of total non-rattler contacts $M_4$ and total number of particles $N$, i.e., $C=M_4/N = \left(M_4/N_4\right)\left(N_4/N\right) = \Cstar\fNR$, 
with corrected coordination number $\Cstar$ and fraction of non-rattlers $\fNR$. 
The isotropic fabric $\Fv$ is given by the relation 
$\Fv = g_3 \volfrac C$, as taken from \citet{imole2013hydrostatic}, with $g_3\cong1.22$ for the polydispersity used in the present work. %
For any tensor $\Qtensor$, its deviatoric part can be defined as
$\Qd  = \mathrm{sgn}\left(q_{yy} - q_{xx}\right)  \sqrt{ 3 q_{ij}q_{ij}/2}$, 
where $q_{ij}$ are the components of the deviator of $\Qtensor$, 
and the sign function accounts for the shear direction, in the system considered here,
where a more general formulation is given by \cite{kumar2014macroscopic}. 
Both pressure $\Pres$ and shear stress $\shstress$ are non-dimensionalized by 
${2\langle r \rangle}/{k}$ to give dimensionless pressure $p$ and shear stress $\nstress$. 

\section{Micromechanical results}
\label{sec:results}

\subsection{Isotropic deformation} 
\label{subsec:iso1}

In this section, we present a procedure to identify 
the jamming densities and their range. We also show the effect of cyclic 
over-compression to different target volume fractions  
and present a model that captures this phenomena.

\subsubsection{Identification of the jamming density} 
\label{subsubsec:iso2}

\begin{figure}
\centering
\subfigure[]{\includegraphics[width=0.35\textwidth , angle=270]{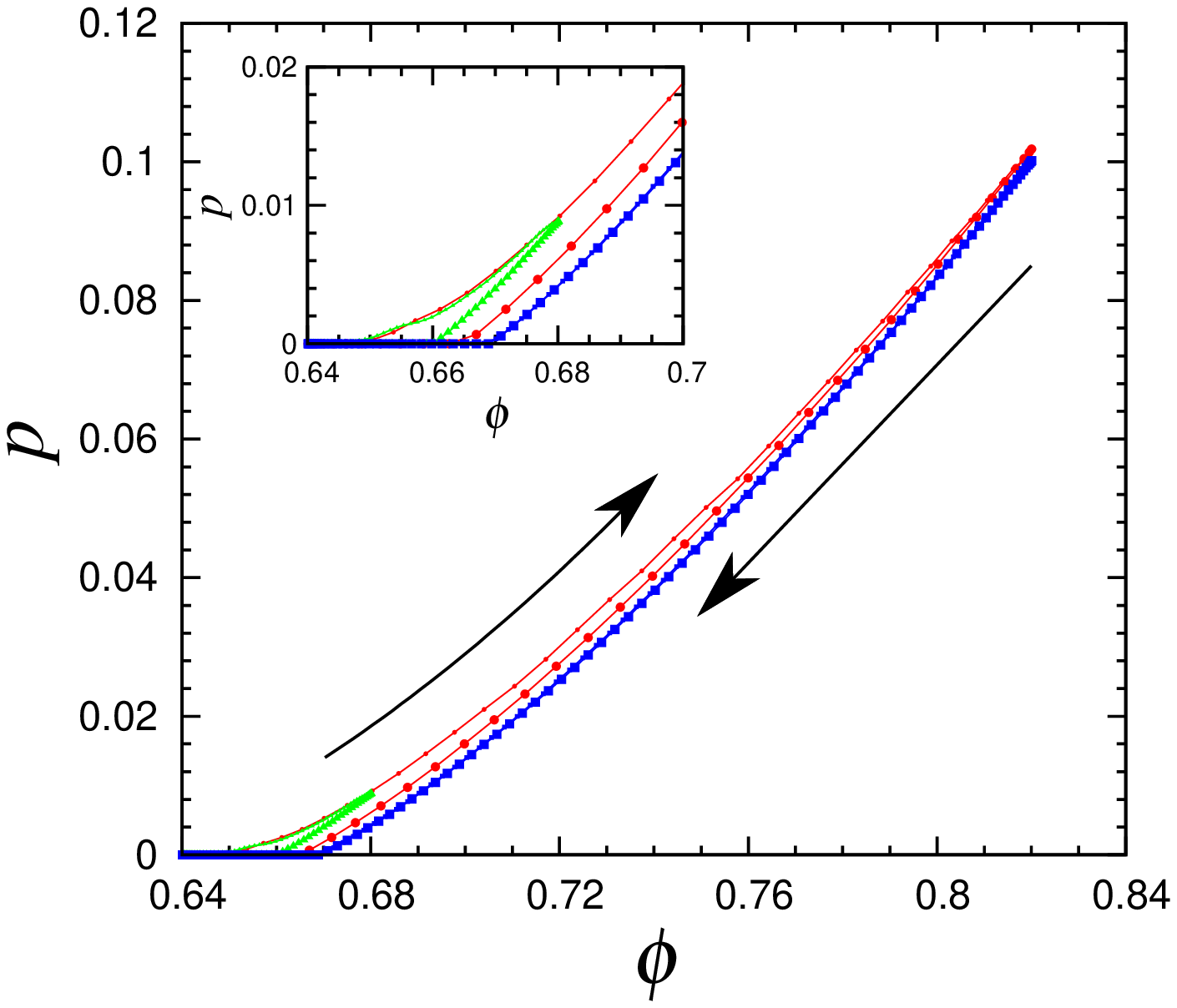}\label{pvsnu}}
\subfigure[]{\includegraphics[width=0.35\textwidth , angle=270]{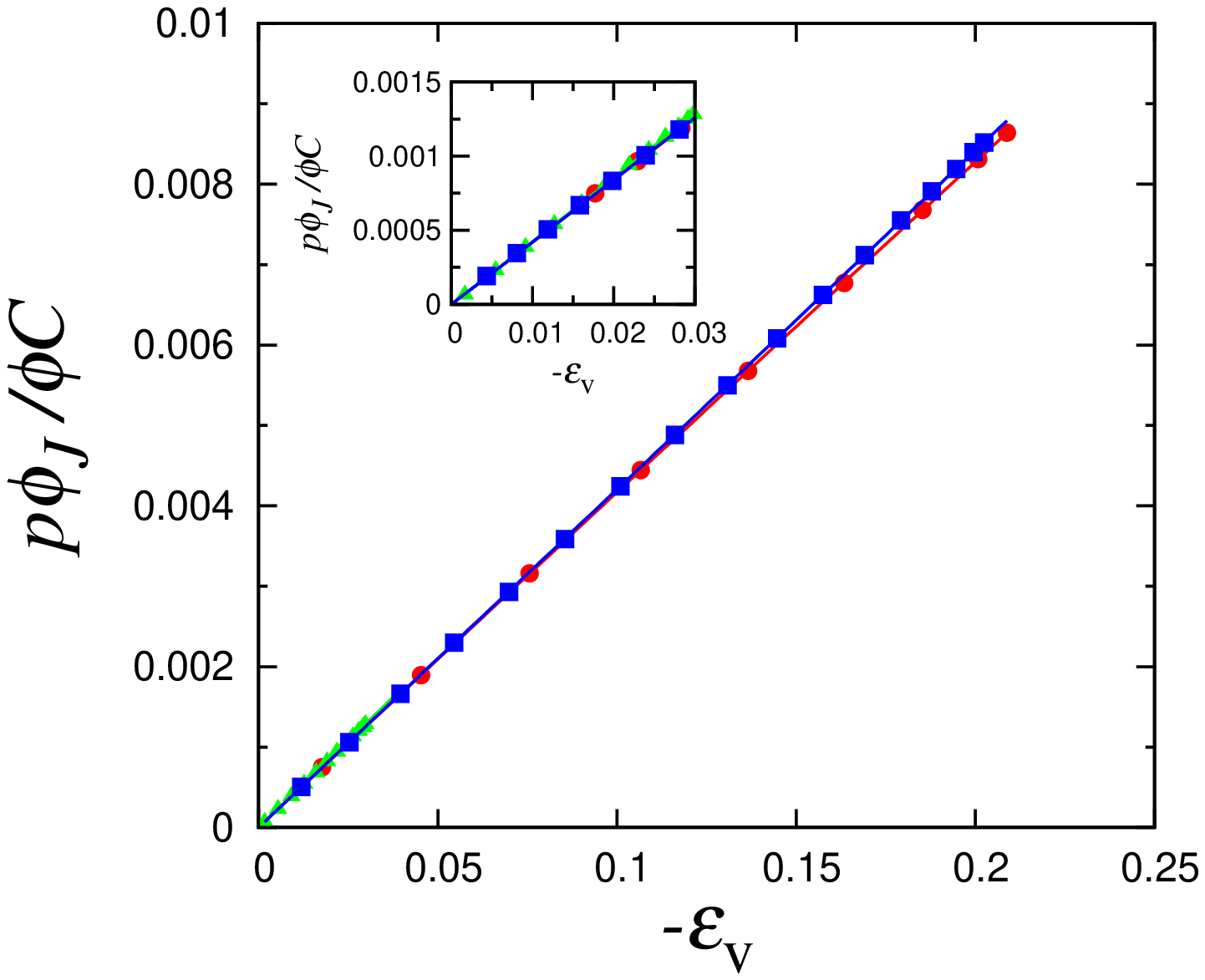}\label{pstarvsnu}}
\caption{(a) Dimensionless pressure $\norPres$ plotted against volume fraction $\volfrac$ and
for an isotropic compression starting 
from $\volfract = 0.64$ to $\volfracmax_i = 0.68$ (green `$\blacktriangledown$') and $\volfracmax_i = 0.82$ (red `$\bullet$') 
and decompression back to $\volfract$ for $M=1$, leading to ${}^1\volfracJ(\volfracmax_i = 0.68) = 0.66$ 
and ${}^1\volfracJ(\volfracmax_i = 0.82) = 0.6652$. The blue `$\blacksquare$' 
data points represent cyclic over-compression to $\volfracmax_i = 0.82$ for $M=100$, 
leading to ${}^{100}\volfracJ(\volfracmax_i = 0.82) = 0.6692$. 
The $\MvolfracJi$ are extracted using a fit to Eq.\ (\ref{eq:pstar}). 
The upward arrow indicates the loading path (small symbols) while the downward arrow indicates the 
unloading path (big symbols). The inset is the zoomed in regime near the jamming density,
and lines are just connecting the datasets.
(b) Scaled pressure $\norPres\volfracJ/\volfrac C$ plotted against volumetric strain 
$-\varepsilon_\mathrm{v}=\log(\volfrac/\volfracJ)$ for the same simulations as (a). 
Lines represents the scaled pressure, when Eq.\ (\ref{eq:pstar}) is 
used, with different $\gamma_p$ = -0.1, 0.07 and -0.01 for green, red and blue lines respectively. 
The inset is the zoomed in regime for small $-\varepsilon_\mathrm{v}$. }
\end{figure}

When a sample is over-compressed isotropically, the loading and unloading paths are different in pressure $\norPres$. 
This difference is most pronounced near the jamming density $\volfracJ$, and for the first cycle. 
It brings up the first question of how to identify a jamming density, $\volfracJ$. 
The unloading branch of a cyclic isotropic over-compression along volume fraction $\volfrac$ 
is well described by a linear relation in volumetric strain,
with a tiny quadratic correction 
\cite{goncu2010constitutive, kumar2014effects, kumar2015tuning}:
\begin{equation}
p=\frac{\volfrac C}{\volfracJ}p_0 (-\varepsilon_\mathrm{v})  \left [ 1-\gamma_p (-\varepsilon_\mathrm{v}) \right ],
\label{eq:pstar}
\end{equation}
where $p_0$, $\gamma_p$, as presented in Table~\ref{cpfnrparamter}, and the jamming density $\volfracJ$ 
are the fit parameters, and $-\varepsilon_\mathrm{v}=\log(\volfrac/\volfracJ)$ is the true or logarithmic 
volumetric strain of the system, defined relative to the reference where $p\to0$, i.e. jamming volume fraction.

\begin{table*}[t]
  \begin{center}
\def~{\hphantom{0}}
  \begin{tabular}{l@{\hskip .7in}c@{\hskip .7in}c}
      Quantity & Isotropic & Shear \\[3pt]
      $p$ & $p_0=0.042$; $\gamma_p=0\pm0.1$* & $p_0=0.042$; $\gamma_p=0\pm0.1$* \\[3mm]
    $\Cstar$ & $C_1=8.5 \pm 0.3$*; $\theta=0.58$ & $C_1=8.5 \pm 0.3$*; $\theta=0.58$ \\[3mm]
    $\fNR$ & $\varphi_c=0.13$; $\varphi_v=15$ & $\varphi_c=0.16$; $\varphi_v=15$ \\[3mm]
    \end{tabular}
  \caption{Parameters used in Eqs.\ (\ref{eq:cstareqn}), (\ref{eq:fNReqn}) and (\ref{eq:pstareqn}), 
  where `*' represents slightly different values than from \citet{imole2013hydrostatic}, modified slightly
  to have more simple numbers, without big deviation, and without loss of generality.}
  \label{cpfnrparamter}
  \end{center}
\end{table*}

Eq.\ (\ref{eq:pstar}), quantifies the scaled stress and is proportional to the
dimensionless deformation (overlap per particle size), as derived analytically \cite{goncu2010constitutive}
from the definition of stress and converges to $\norPres \to 0$ when $\volfrac \to \volfracJ$. 

We apply the same procedure for different over-compressions, 
$\volfracmax_i$, and many subsequent cycles $M$ to obtain $\MvolfracJi$, for which the results are discussed below. 
The material parameter $p_0$ is finite, almost constant, whereas $\gamma_p$ is small, sensitive to history and contributes mainly 
for large $-\varepsilon_\mathrm{v}$, with values ranging around $0\pm 0.1$; in particular, it is dependent on the over-compression $\volfracmax_i$ (data not shown). 
Unless strictly mentioned, we shall be using the values of $p_0$ and $\gamma_p$ given in Table~\ref{cpfnrparamter}.

Fig.\ \ref{pvsnu} shows the behavior of $\norPres$ with $\volfrac$ during one full over-compression cycle 
to display the dependence of the jamming density on the maximum over-compression volume fraction and the number of cycles. 
With increasing over-compression amplitude, e.g.\ comparing $\volfracmax_i = 0.68$ and $\volfracmax_i = 0.82$, the jamming density, as realized after unloading, is increasing. 
Also, with each cycle, from $M=1$ to $M=100$, the jamming density moves to larger values. 
Note that the difference between the loading and the unloading curves becomes smaller for subsequent over-compressions.  
Fig.\ \ref{pstarvsnu} shows the scaled pressure, i.e.,\ $\norPres$ normalized by $\volfrac C/\volfracJ$, 
which removes its non-linear behavior. $\norPres$ represents
the average deformation (overlap) of the particles at a given volume fraction, proportional to the distance from the 
jamming density $\volfracJ$ 
\footnote{
The grains are soft and overlap $\delta$ increases with increasing compression ($\volfrac$). 
For a linear contact model, it has been shown in Refs. \cite{goncu2010constitutive, kumar2015tuning} that 
$\langle \delta \rangle/\langle r \rangle \propto \mathrm{ln}\left( \volfrac/\volfracJ \right) = -\varepsilon_\mathrm{v}$ (volumetric strain).}.
In the small strain region, for all over-compression amplitude and cycles, the datasets collapse on a line with slope $p_0\sim0.042$.
Only for very strong over-compression, $-\varepsilon_\mathrm{v}>0.1$, a small deviation (from linear) 
of the simulation data is observed due to the tiny quadratic correction in Eq.\ (\ref{eq:pstar}).

\subsubsection{Isotropic cyclic over-compression} 
\label{subsubsec:iso3}

Many different isotropic jamming densities can be found 
in real systems and -- as shown here -- also for the
simplest model material in 3D. 
%
%
Fig.\ \ref{diffmaxvol} shows the evolution of these extracted isotropic jamming densities 
$\MvolfracJi$, which increase with increasing $M$ and with over-compression $\volfracmax_i$, 
for subsequent cycles $M$ of over-compressions, the jamming density $\MvolfracJi$
grows slower and slower and 
is best captured by a Kohlrausch-Williams-Watts (KWW) stretched exponential relation:
\begin{equation}
\begin{split}
\MvolfracJi :&= \volfracJ(\volfracmax_i, M) \\
&= {}^{\infty}\volfracJi  - \left({}^{\infty}\volfracJi - \volfracSJ\right)\exp\left[{-\left( {M}/{\mu_i}\right)^{\beta_i}}\right],
\label{eq:strexp}
\end{split}
\end{equation}
with the three universal ``material''-constants $\volfracSJ = 0.6567$ (section\ \ref{subsubsec:shear3}), 
$\mu_i =1$, and $\beta_i=0.3$, the lower limit of possible $\volfracJ$'s, the relaxation (cycle) scale and the stretched exponent parameters, respectively. 
Only ${}^{\infty}\volfracJi$, the equilibrium (steady-state or shakedown \citep{garcia2005characterization}) jamming density limit (extrapolated for $M \to \infty$), 
depends on the over-compressions $\volfracmax_i$. 
{$\volfracSJ$ is the critical density in the zero pressure 
limit without previous history, or after very long shear without temperature (which all are impossible to realize with experiments or simulation--only maybe with energy minimization).

Very little over-compression, $\volfracmax_i \gtrsim \volfracSJ$, does not lead to a 
significant increase in $\volfracJi$, giving us information about the lower limits 
of the isotropic jamming densities achievable by shear, which is the critical jamming density $\volfracSJ= 0.6567$. 
With each over-compression cycle, $\MvolfracJi$ increases, but for large $M$ it increases less and less. 
This is analogous to compaction by tapping, where the tapped density increases logarithmically slow with the 
number of taps. 
The limit value ${}^{\infty}\volfracJi$ with $\volfracmax_i$ can be fitted with a simple power law relation:
\begin{equation}
{}^{\infty}\volfracJi  =  \volfracSJ +  \alpha_{\rm max}  \left(\volfracmax_i / \volfracSJ - 1 \right)^{\beta} ,
\label{eq:asymptoticrelation}
\end{equation}
where the fit works perfect for $\volfracSJ < \volfracmax_i \le 0.9$, with parameters
$\volfracSJ=0.6567$, $\alpha_{\rm max}=0.02\pm2\%$, and $\beta=0.3$, while the few points for $\volfracmax_i \sim \volfracSJ$ are not well captured.  
The relation between the limit-value ${}^{\infty}\volfracJi$ and $\ovolfracJi$ is derived using Eq.\ (\ref{eq:strexp}):
\begin{equation}
{}^{\infty}\volfracJi  -\volfracSJ=  \frac{\ovolfracJi -\volfracSJ} {1-e^{-1}} \cong 1.58 \left({\ovolfracJi -\volfracSJ} \right),
\label{eq:asymptoticrelation_2}
\end{equation}
only by setting $M=1$, as shown in Fig.\ \ref{fullstretchedphiJlimits_prime}, with perfect match. With other words, using a single over-compression, 
Eq.\ (\ref{eq:asymptoticrelation_2}) allows to predict the limit value after first over-compression $\ovolfracJi$ (or subsequent over-compression cycles, using appropriate $M$). 

Thus, the isotropic jamming density $\volfracJ$  {\it{is not a unique point}}, 
not even for frictionless particle systems, and is dependent on the 
previous deformation history of the system 
\citep{ciamarra2010recent, ohern2002random, mobius2014ir}, e.g.\ over-compression 
or tapping/driving (data not shown).
Both (isotropic) modes of deformation lead to more compact, better packed 
configurations \citep{bi2011jamming, zhang2010jamming, rosato2010microstructure}. 
Considering different system sizes, and different preparation procedures, 
we confirmed that the jamming regime is the same (within fluctuations) 
for all the cases considered (not shown). 
All our data so far, for the material used, 
are consistent with a unique limit density $\volfracSJ$
that is reached after large strain, very slow shear, 
in the limit of vanishing confining pressure.
Unfortunately this limit is vaguely defined, since it is not directly
accessible, but rather corresponds to a virtual stress-free state.
The limit density is hard to determine experimentally 
and numerically as well. Reason is that any slow deformation 
(e.g.\ compression from below jamming) also leads 
to perturbations (like tapping leads to granular temperature): 
the stronger the system is perturbed, the better it will pack, 
so that usually $\phi_J > \volfracSJ$ is established.
Repeated perturbations, lead to a slow stretched exponential 
approach to an upper-limit jamming density
$\volfracJ \rightarrow \volfracJmax$ that itself increases slowly
with perturbation amplitude, see Fig.\ \ref{fullstretchedphiJlimits_prime}.
The observation of different $\volfracJ$ of a single material, was referred to 
as \textit{J-segment} \citep{ciamarra2010recent, ohern2002random},  
and requires an alternative interpretation of the classical 
``jamming diagram'' \citep{bi2011jamming, liu1998nonlinear, grob2014jamming},
giving up 
the misconception of a single, constant jamming ``density''. 
Note that the J-segment is not just due to fluctuations, but it is due
to the deformation history, and with fluctuations superposed.
The state-variable $\volfracJ$ varies due to deformation, but possibly 
has a unique limit value that we denote for now as $\volfracSJ$.
Jammed states below $\volfracSJ$ might be possible too, but require 
different protocols \citep{hopkins2013disordered}, or different materials, 
and are thus not addressed here. 
The concept of shear jammed states \citep{bi2011jamming} 
below $\volfracJ$, is discussed next.

\begin{figure*} 
\centering
\subfigure[]{\includegraphics[width=0.35\textwidth, angle=270]{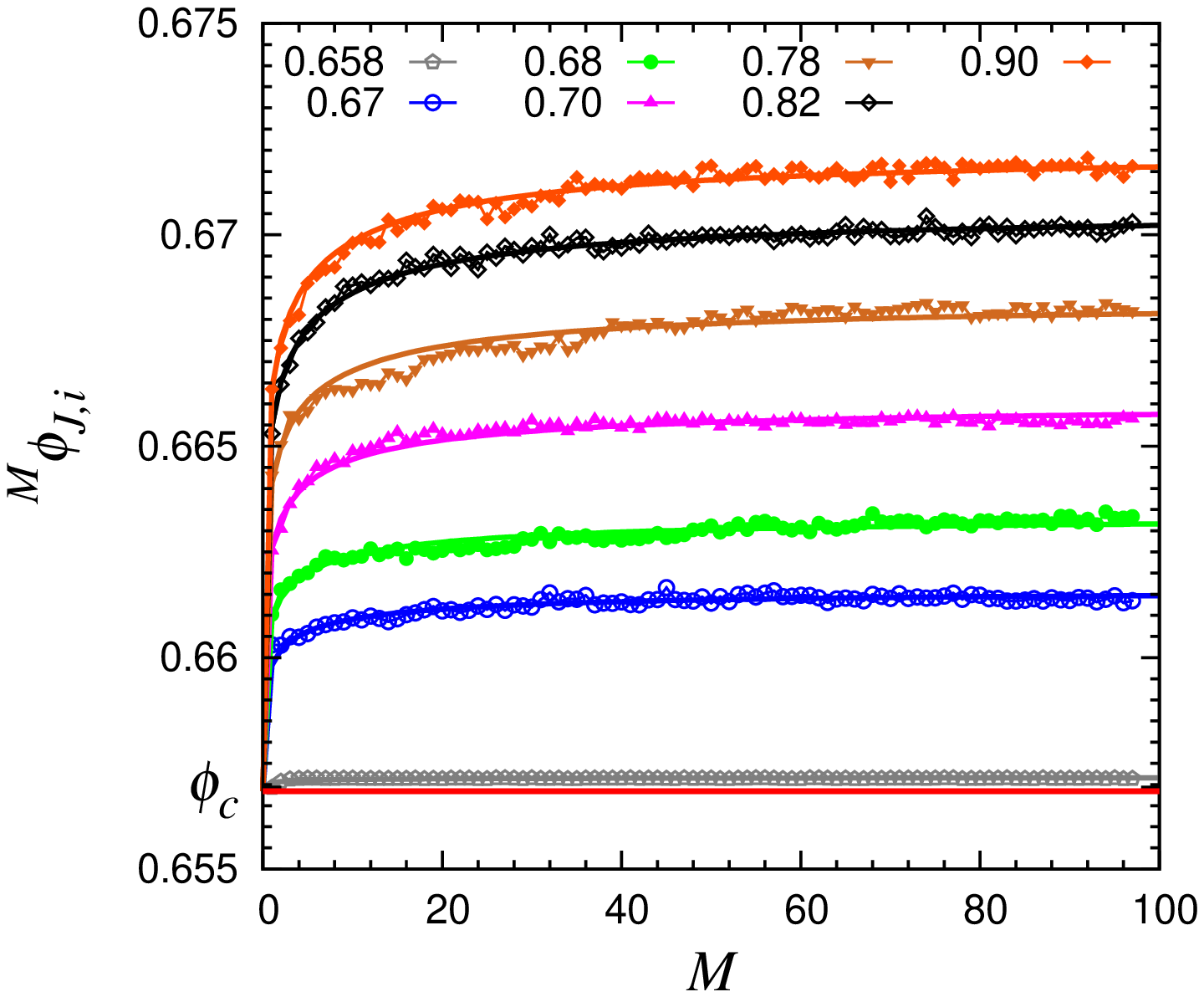}\label{diffmaxvol}}
\subfigure[]{\includegraphics[width=0.35\textwidth, angle=270]{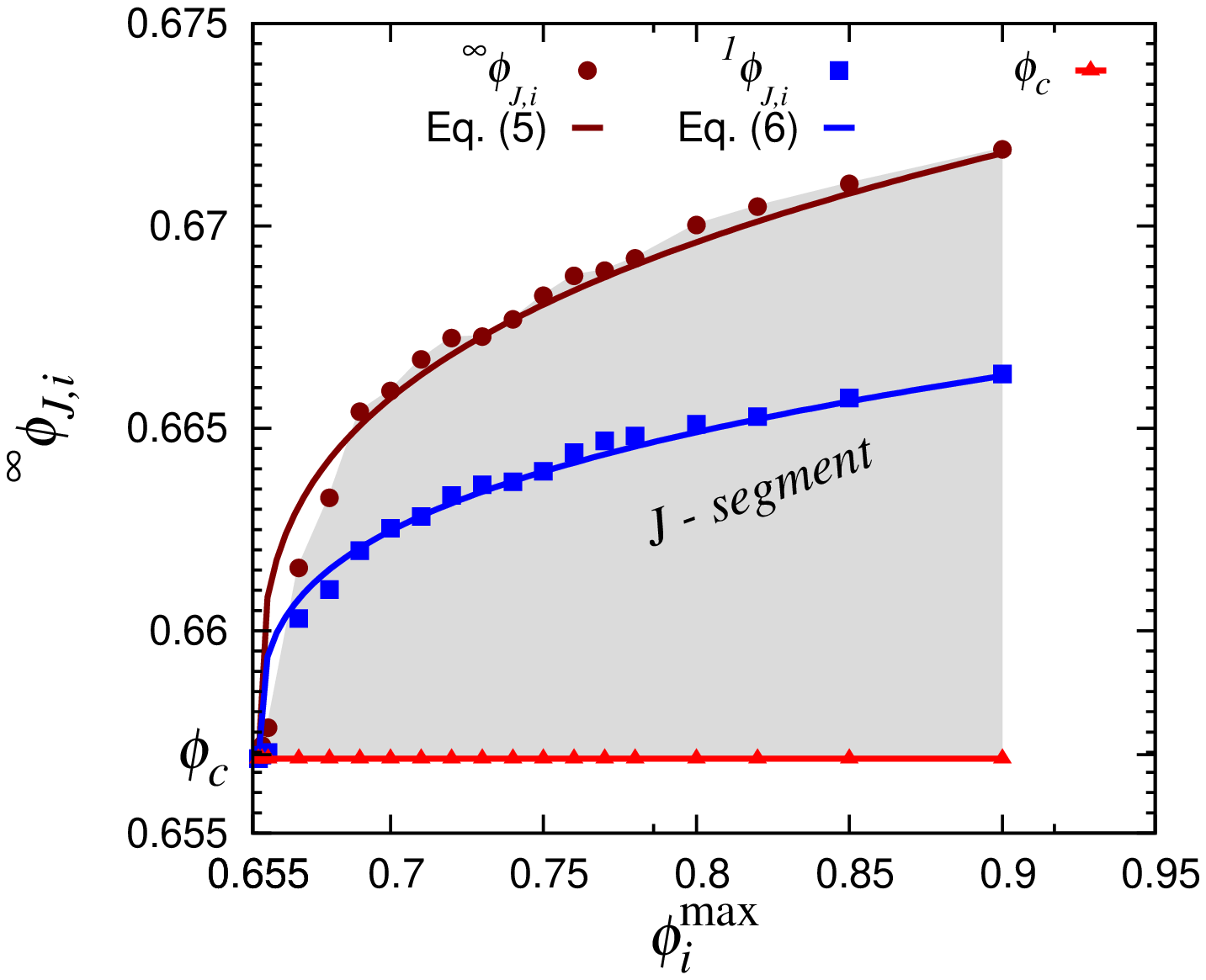}\label{fullstretchedphiJlimits_prime}}
\caption{(a) Evolution of isotropic jamming densities $\MvolfracJi$ after performing 
$M$ isotropic compression-decompression cycles up to 
different maximum volume fractions $\volfracmax_i$, as given in the inset. 
With increasing $\volfracmax_i$, the range of the established jamming densities 
$\MvolfracJi=\volfracJ(M,\volfracmax_i)$ increases. 
The minimum (lower bound) of all $\MvolfracJi$ is defined as the critical jamming 
limit point, $\volfracSJ= 0.6567$. %
The solid lines through the data are universal fits to a 
stretched exponential \citep{richard2005slow,knight1995density, rosato2010microstructure, andreotti2013granular} 
with only one single variable parameter $\volfracJmax$, i.e.,\ the upper
limit jamming density for $M \rightarrow \infty$, which depends 
on $\volfracmax_i$. 
(b) The first jamming density $\ovolfracJi$ (blue `$\blacksquare$') and after many over-compression ${}^{\infty}\volfracJi$ (brown `$\bullet$') are plotted against over-compression amplitude $\volfracmax_i$.
Solid lines represent Eqs.\ (\ref{eq:asymptoticrelation}) for ${}^{\infty}\volfracJi$ and (\ref{eq:asymptoticrelation_2}) for $\ovolfracJi$. 
The shaded region is the explorable range of jamming densities
$\MvolfracJi$, denoted as \textit{J-segment}. 
The red base line indicates the critical jamming density $\volfracSJ$. }
\label{diffmaxvol1}
\end{figure*}

\subsection{Shear deformation} 
\label{subsec:shear1}

To study shear jamming, we choose several unjammed states 
with volume fractions $\volfrac$ below their jamming densities $\ovolfracJi$, 
which were established after the first compression-decompression cycle, 
for different history, i.e.,\ various previously applied 
over-compression to $\volfracmax_i$. 
Each configuration is first relaxed and then subjected to four isochoric (volume conserving) pure shear cycles ((see section\ \ref{sec:prep})).  

\subsubsection{Shear jamming below $\bm\volfracJ(H)$} 
\label{subsubsec:shear2}

\begin{figure*}
\centering
\subfigure[]{\includegraphics[width=0.35\textwidth ,angle=270]{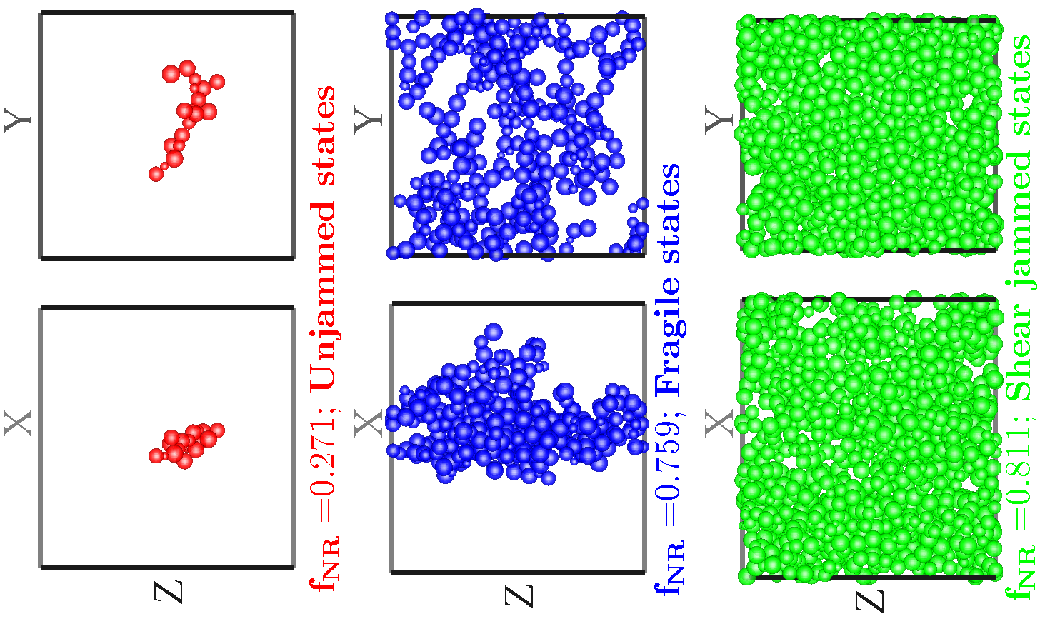}\label{snap_allpercoyz}}
\subfigure[]{\includegraphics[width=0.35\textwidth, angle=270]{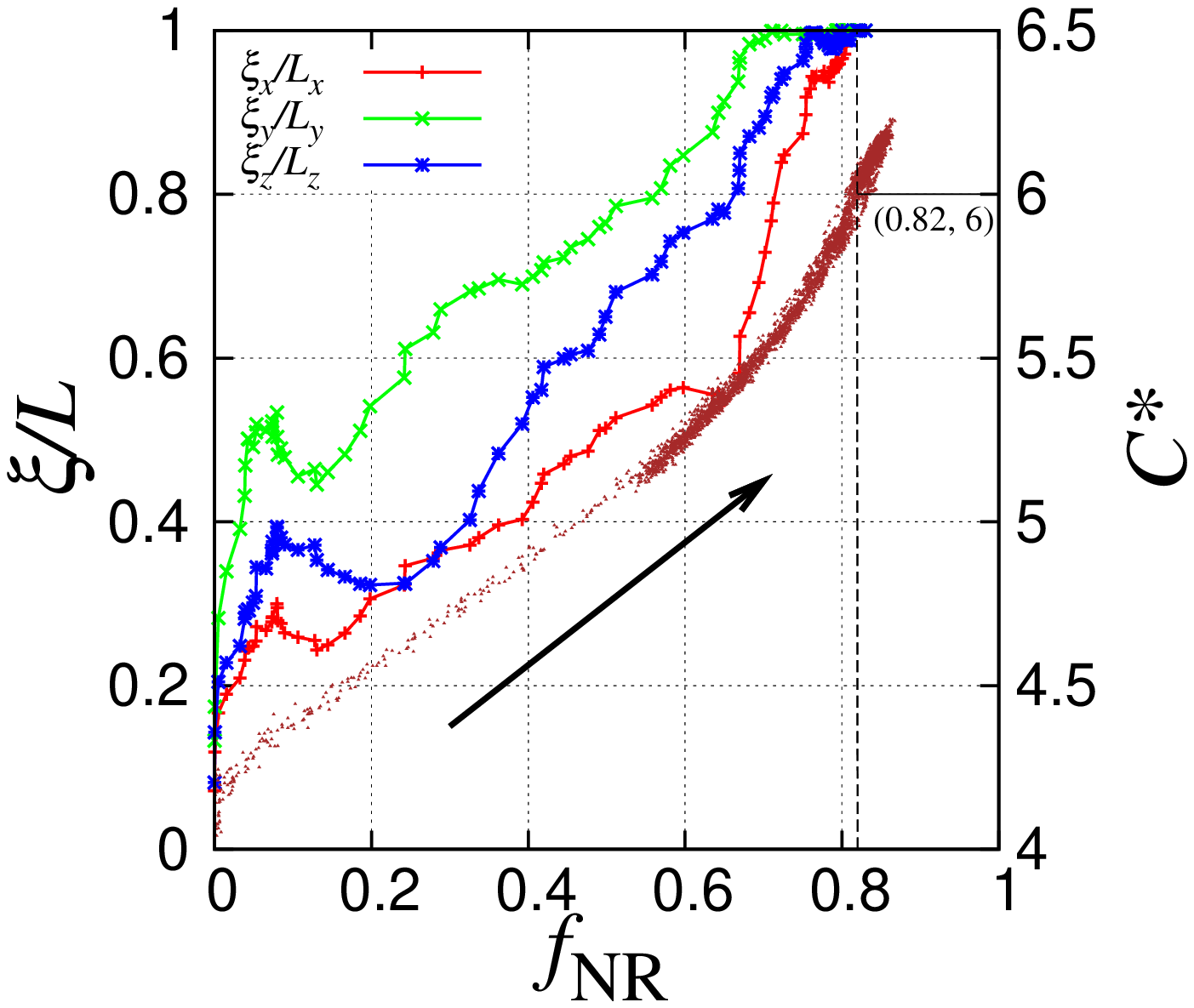}\label{percoandfNRCstar}}
\caption{(a) Snapshots of unjammed, fragile and shear jammed states, 
when the force networks are percolated in none, one or two, and 
all the three directions, respectively. Only the largest 
force network, connecting strong forces,  $f \ge k \langle f \rangle$, with $k = 2.2$ are shown for the three states for clarity, and hence the white spaces in the background. 
(b) Plot of $\Cstar$ and cluster sizes $\xi/L$ in the three 
directions for extension in $x-$ and compression in $y-$ directions 
against the non-rattler fraction $\fNR$, along the 
loading path for an isotropic unjammed initial state with volume fraction $\volfrac =0.6584$ and $\volfracJ\left(\volfracmax_i = 0.82, M=1\right) =: \ovolfracJi = 0.6652 $. 
The upward arrow indicates the direction of loading shear strain.  
}
\label{snapSJandFJ}
\end{figure*}

We confirm shear jamming, e.g.,\ by a transition in the coordination number $\Cstar$, from below to above 
its isostatic limit, $\Cstariso = 6$, for frictionless grains 
\citep{wang2012edwards, imole2013hydrostatic, peyneau2008frictionless, snoeijer2006sheared}.
This was consistently (independently) reconfirmed by using percolation analysis \citep{radjai1998bimodal, bi2011jamming}, 
allowing us to distinguish the three different regimes namely, 
unjammed, fragile and shear jammed states during 
(and after) shear \citep{grob2014jamming}, 
as shown in Fig.\ \ref{snap_allpercoyz}. 
For this, first we study the percolation analysis, that 
allows to distinguish the three regimes namely, 
unjammed, fragile and shear jammed states during 
(and after) shear, 
as shown in Fig.\ \ref{snap_allpercoyz}. 
We study how the $k-$cluster, defined as the largest 
force network, connecting strong forces, $f \ge k \favg$ \citep{hidalgo2002evolution, smith2011isostacity}, with $k=2.2$, different from $k=1$ for 2D frictional systems \citep{bi2011jamming},
percolates when the initially unjammed isotropic system is sheared. 
More quantitatively, for an exemplary volume fraction $\volfrac\left(\volfracmax_i = 0.82, M=1\right) = 0.6584$, very close to $\volfracSJ$, Fig.\ \ref{percoandfNRCstar} shows that $\fNR$ 
increases from initially zero to large values well below unity due to 
the always existing rattlers. The compressive direction percolating network 
$\xi_y/L_y $ grows faster than the extension direction network 
$\xi_x/L_x $, while the network in the non-mobile direction, 
$\xi_z/L_z $, lies in between them. 
For $\fNR>0.82\pm0.01$, we observe that the growing force 
network is percolated in all three directions (Fig.\ \ref{snap_allpercoyz}), which is astonishingly similar to the value reported for the 2D systems \citep{bi2011jamming}. 
The jamming by shear of the material corresponds (independently) to the 
crossing of $\Cstar$ from the isostatic limit of $\Cstariso =6$, 
as presented in Fig.\ \ref{percoandfNRCstar}.

From this perspective, when an unjammed material is sheared at constant 
volume, and it jams after application of sufficient shear strain, 
clearly showing that the jamming density has moved to a lower value. 
Shearing the system also perturbs it, just like over-compression; however, in addition,
finite shear strains enforce shape- and structure-changes and
thus allow the system to explore new configurations; typically,
the elevated jamming density $\volfracJ$ of a previously compacted 
system will rapidly decrease and exponentially
approach its lower-limit, the {\em critical} jamming density $\volfracSJ$,
below which no shear jamming exists. 
Note that we do {\em not} exclude
the possibility that jammed states below $\volfracSJ$ could be 
achieved by other, special, careful preparation procedures \citep{atkinson2014existence}.

\begin{figure*}
\centering
\includegraphics[width=0.6\textwidth, angle=270]{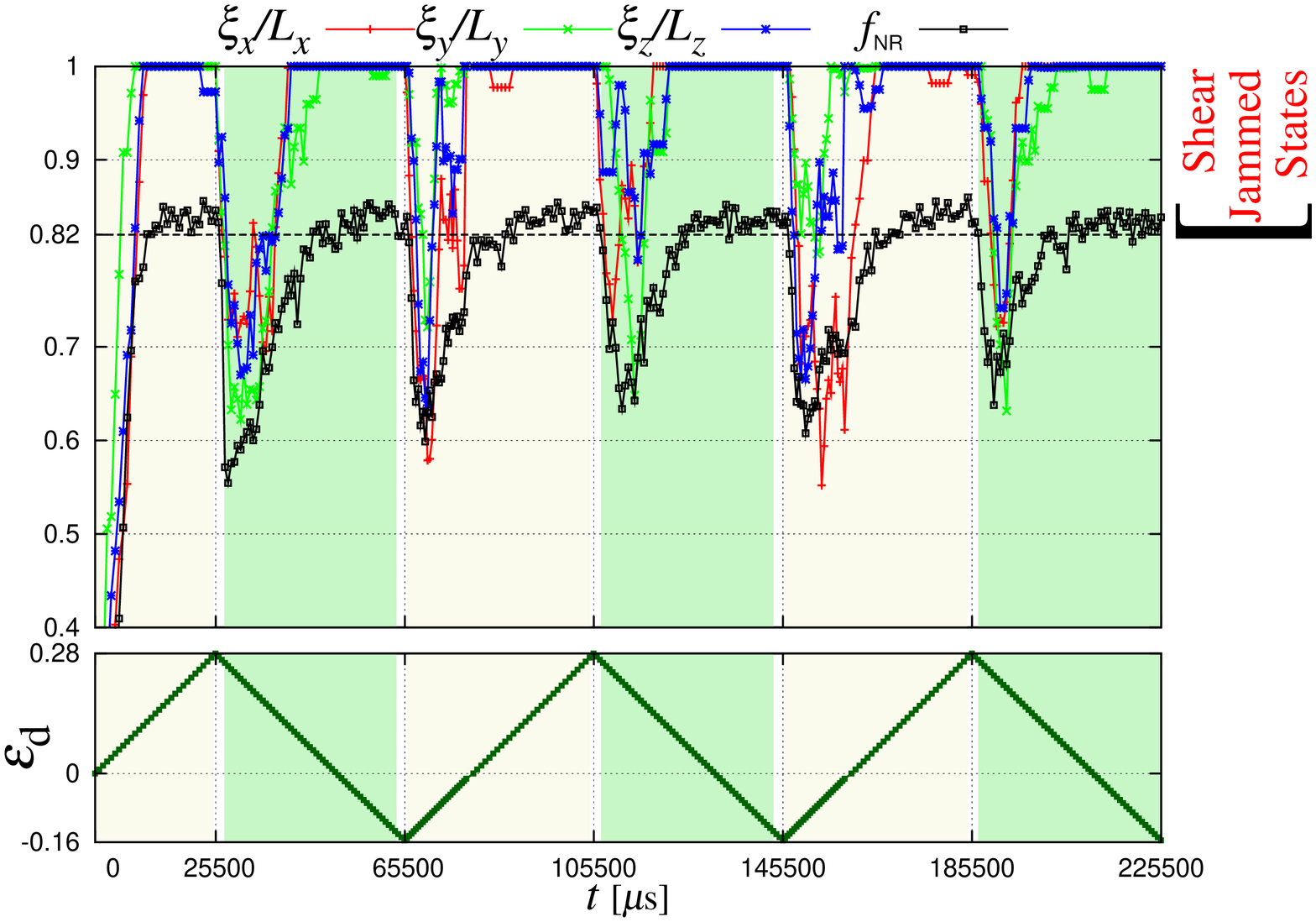}
\caption{Cluster sizes, $\fNR$ (top panel), over three strain cycles 
bottom for $\volfrac=0.6584$ and jamming density $\volfracJ\left(\volfracmax_i = 0.82, M=1\right) =: \ovolfracJi = 0.6652$. 
Dashed horizontal black line represents transition from unjammed to shear jammed states. 
The cluster sizes are smoothed averages over two past and future snapshots. }
\label{cycle_perco}
\end{figure*}

Next, we present the evolution of the strong force networks in each direction during cyclic shear, as shown in Fig.\ \ref{cycle_perco}, for the same initial system. 
After the first loading, at reversal $\fNR$ drops below the 0.82 threshold, which indicates the breakage/disappearance of strong clusters, i.e. the system unjams. 
The new extension direction $\xi_y/L_y $ drops first with the network in the non-mobile directions, 
$\xi_z/L_z $, lying again in between the two mobile direction. With further applied strains, $\fNR$ increases and again, the cluster 
associated with the compression direction grows faster than in the extension direction. 
For $\fNR$ above the threshold, the cluster percolates the full system, leading to shear jammed states again. 
At each reversal, 
the strong force network breaks/fails in all directions, and the system gets ``soft'' or even unjams temporarily. 
However, the network is rapidly re-established in the perpendicular 
direction, i.e.,\ the system jams and the strong, anisotropic force network again sustains the load. 
Note that some systems with volume fraction higher and away from $\volfracSJ$ can resist shear strain reversal as described and modeled in section \ref{subsubsec:model4}.

\subsubsection{Relaxation effects on shear jammed states}
\label{subsubsec:shear3}

\begin{figure*}
\centering
{\subfigure[]{\includegraphics[width=0.3\textwidth, angle=270]{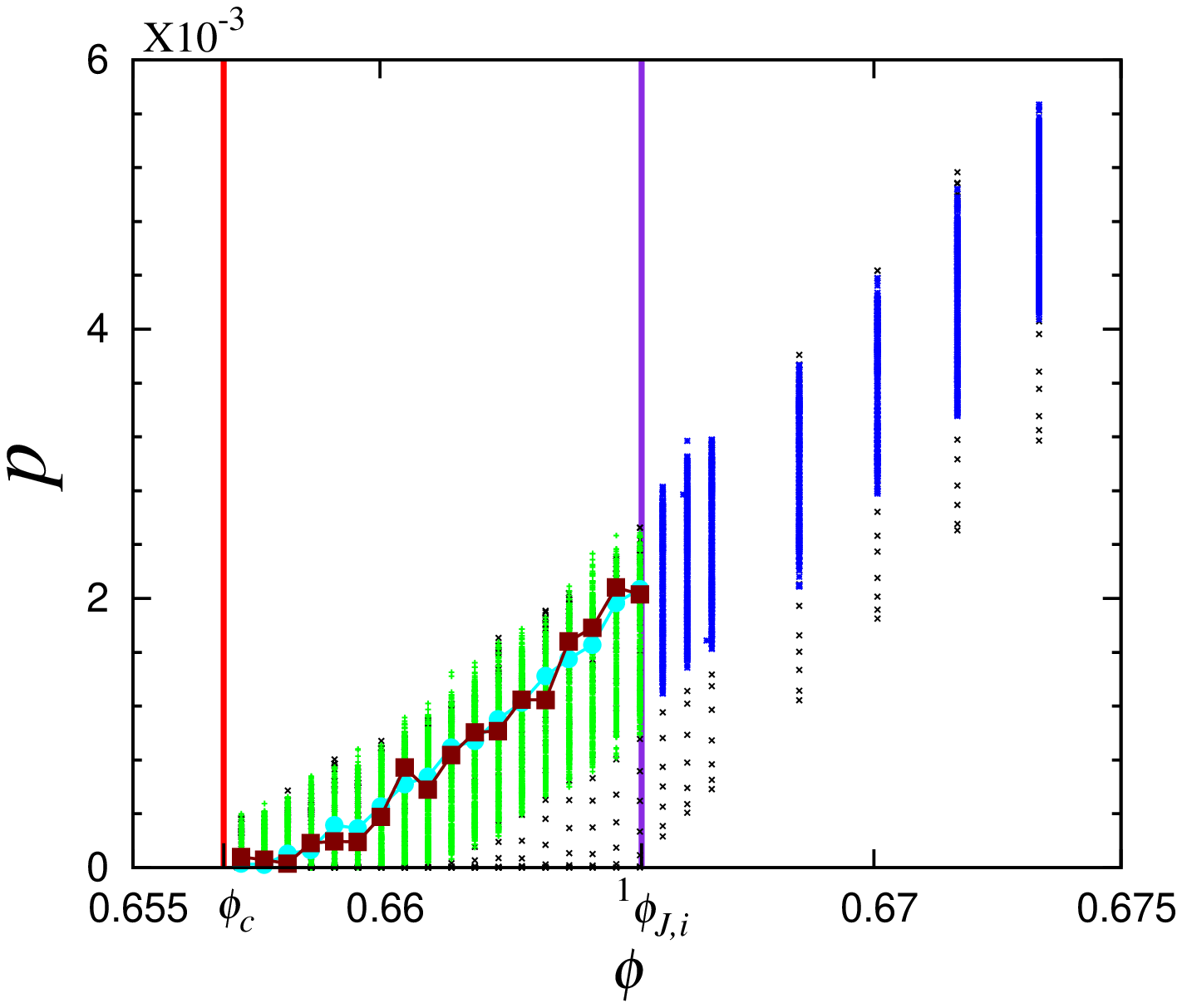}\label{p_states}}}
{\subfigure[]{\includegraphics[width=0.3\textwidth, angle=270]{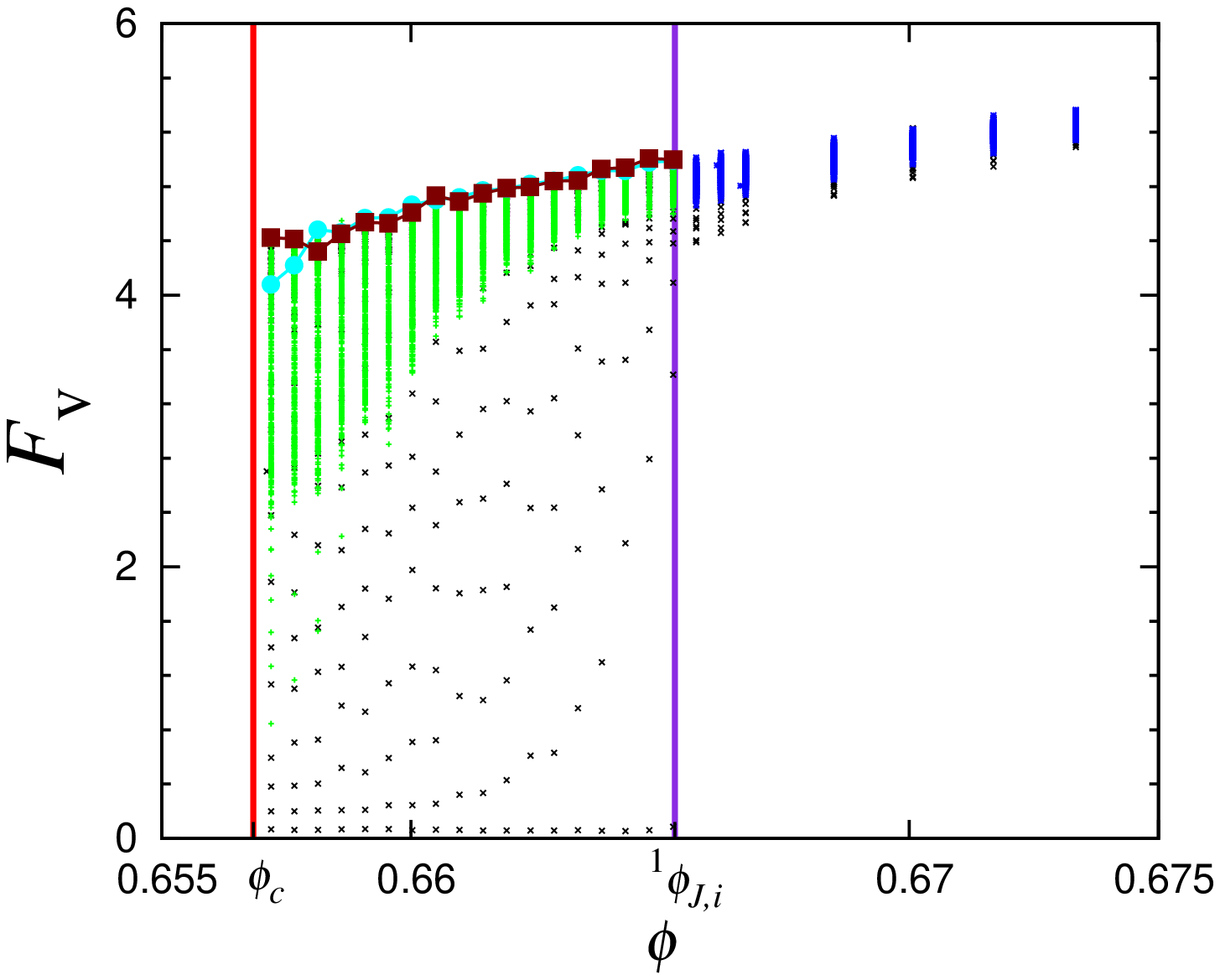}\label{fv_states}}}\\
{\subfigure[]{\includegraphics[width=0.25\textwidth, angle=270]{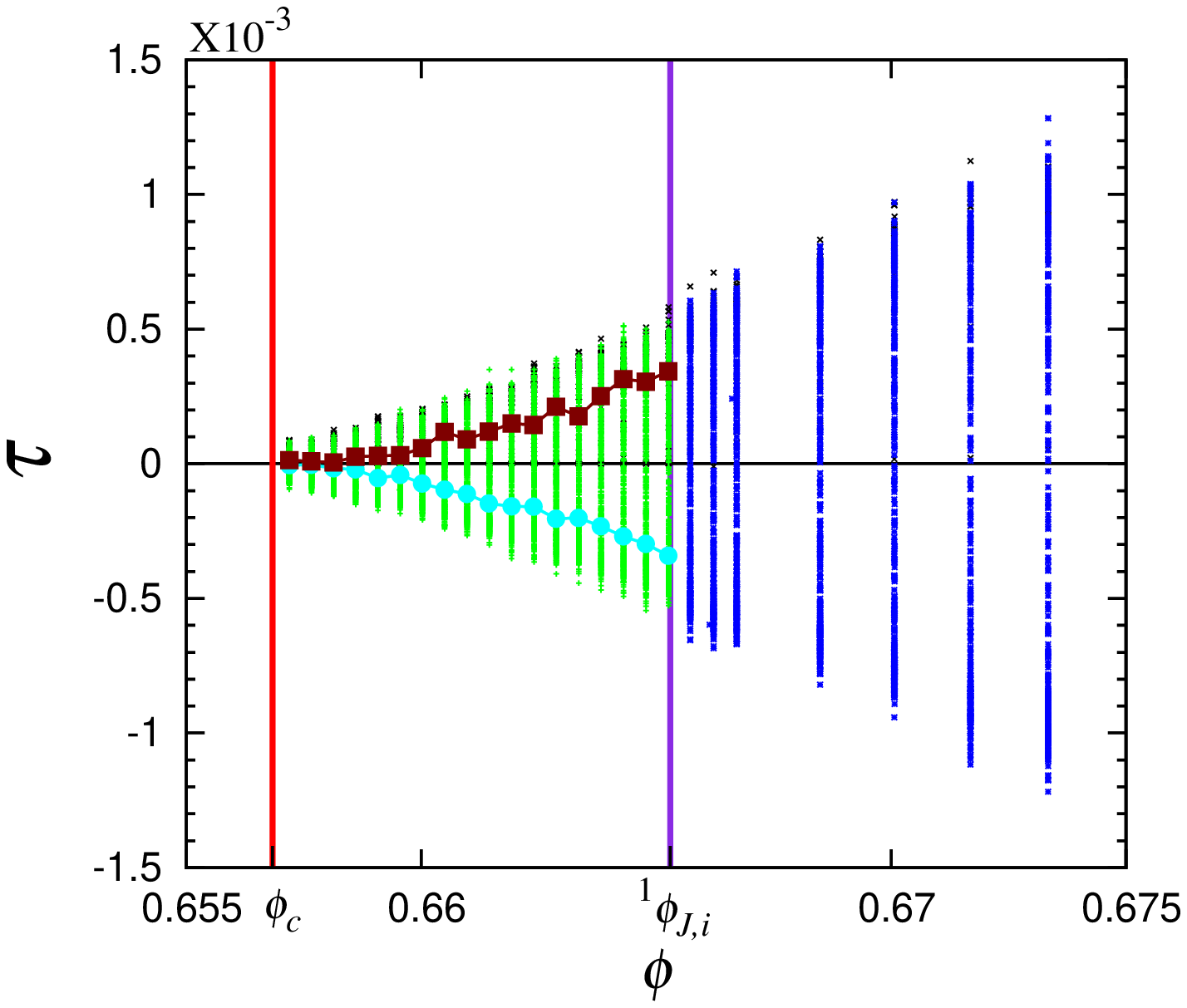}\label{sd_sign_states}} }
{\subfigure[]{\includegraphics[width=0.25\textwidth, angle=270]{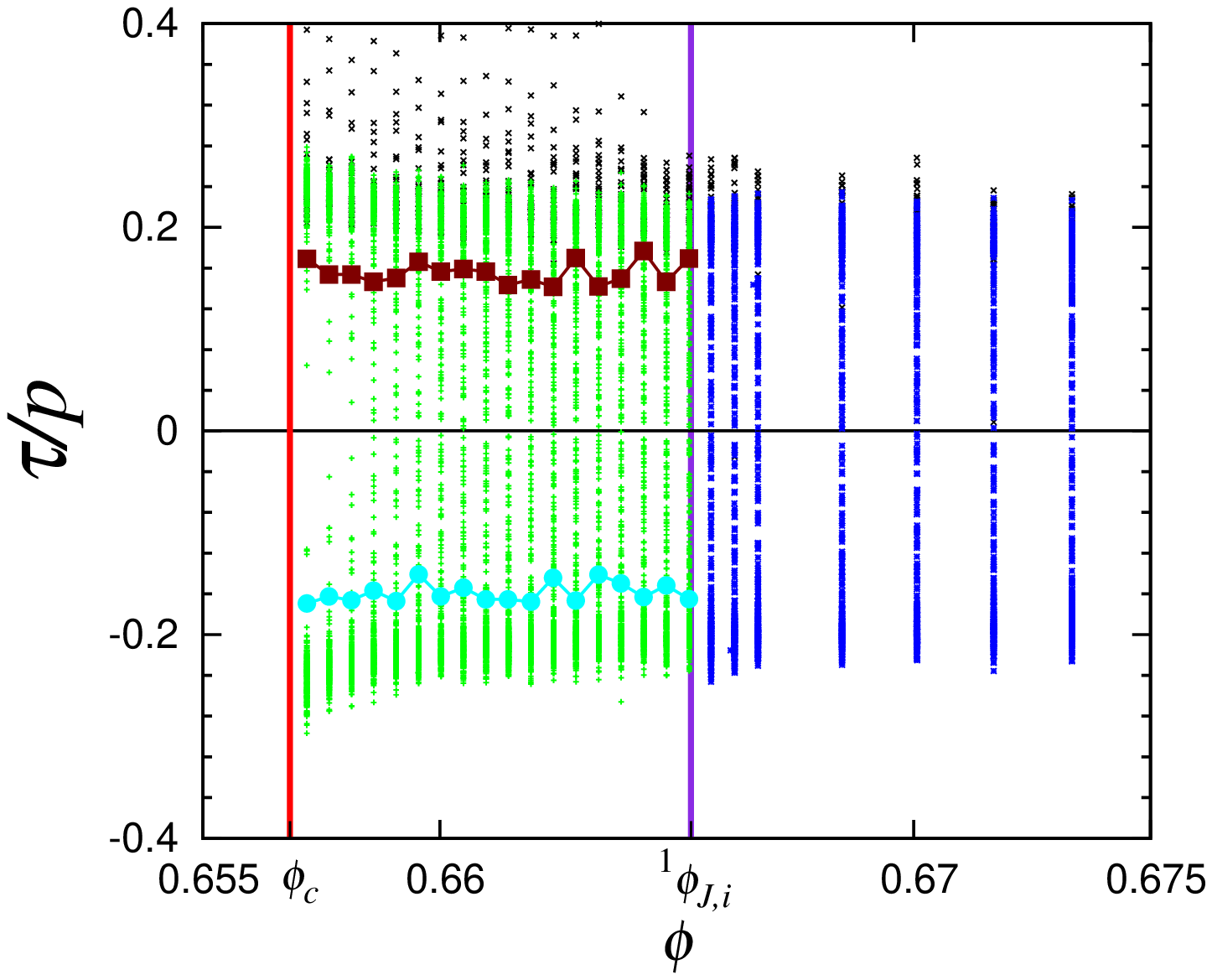}\label{sdp_sign_states}} }
{\subfigure[]{\includegraphics[width=0.25\textwidth, angle=270]{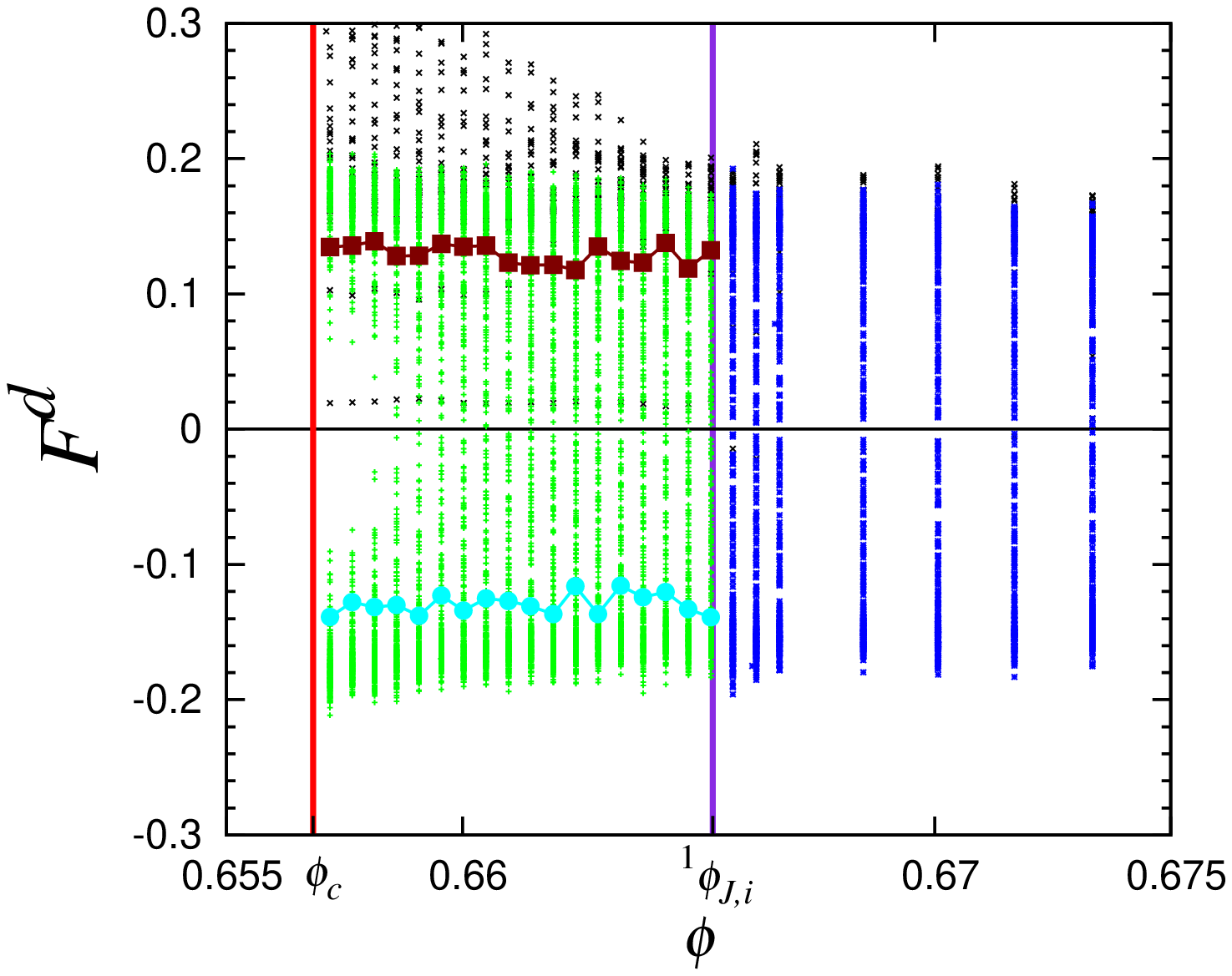}\label{fd_sign_states}} }
\caption{Scatter plots of isotropic quantities 
(a) pressure $\norPres$, (b) isotropic fabric $\Fv$ and deviatoric quantities 
(c) shear stress $\nstress$, (d) shear stress ratio $\stressrat$, and
(e) deviatoric fabric $\anisoF$ for various 
$\volfrac$ and jamming density $\volfracJ\left(\volfracmax_i = 0.82, M=1\right) =: \ovolfracJi = 0.6652$. 
Black `$\mathrm{x}$' symbols represent the initial loading cycle, while the green `$+$' and blue `$*$' represent states attained for 
$\volfrac<\volfracJ$ and $\volfrac>\volfracJ$, respectively for the subsequent shear. 
Cyan `$\bullet$' and the brown `$\blacksquare$' are states chosen after large strain during loading and unloading shear respectively, and are relaxed. 
The red and purple lines indicate the critical jamming density $\volfracSJ = 0.6567$ and the jamming density $\ovolfracJi$ respectively. 
}
\label{phasecompar}
\end{figure*}

Here, we will discuss the system stability by looking at the macroscopic quantities in the saturation state (after large shear strain), 
by relaxing them sufficiently long to have non-fluctuating values in the microscopic and macroscopic quantities. 
Every shear cycle after defining e.g. the $y-$direction as the initial active loading direction, 
has two saturation states, one during loading and, after reversal, the other during unloading. 
In Fig.\ \ref{phasecompar}, we show values attained by the isotropic quantities pressure $\norPres$, isotropic fabric $\Fv$ and the deviatoric quantities 
shear stress $\nstress$, shear stress ratio $\stressrat$, and 
deviatoric fabric $\anisoF$ for various $\volfrac$ given the same initial jamming density $\volfracJ\left(\volfracmax_i = 0.82, M=1\right) =: \ovolfracJi = 0.6652$. 
Data are shown during cyclic shear as well as at the two relaxed saturation states (averaged over four cycles), leading to following observations:  \\
(i) With increasing volume fraction, $\norPres$, $\Fv$ and $\nstress$ increase, while a weak decreasing trend in stress ratio $\stressrat$ and deviatoric fabric $\anisoF$ is observed. \\
(ii) There is almost no difference in the relaxed states in isotropic quantities, 
$\norPres$ and $\Fv$ for the two directions, whereas it is symmetric about zero for deviatoric quantities, $\nstress$, $\stressrat$, and $\anisoF$. 
The decrease in pressure during relaxation is associated with 
dissipation of kinetic energy and partial opening of the contacts to ``dissipate'' the related part of the contact potential energy.
However, $\Fv$ remains at its peak value during relaxation. 
It is shown in section\ \ref{subsec:tensorial} that  
$\Fv = g_3 \volfrac C$, as taken from \citet{imole2013hydrostatic}, with $g_3\cong1.22$ for the polydispersity used in the present work. 
Thus we conclude that the contact structure is almost unchanged and the network remains stable during relaxation, since during relaxation $\phi$ does not change.\\
(iii) For small volume fractions, close to $\volfracSJ$, the system becomes strongly anisotropic in stress ratio $\stressrat$, and fabric $\anisoF$ rather quickly, during (slow) shear 
(envelope for low volume fractions in Figs.\ \ref{sdp_sign_states} and \ref{fd_sign_states}), before it reaches the steady state \citep{walker2014uncovering}. \\ 
(iv) It is easy to obtain the critical (shear) jamming density $\volfracSJ$ from the relaxed critical (steady) state pressure $\norPres$, and 
shear stress $\nstress$, by extrapolation to zero, as the envelope of relaxed data in Figs.\ \ref{p_states} and \ref{sd_sign_states}.

We use the same methodology using Eq.\ (\ref{eq:pstar}), to extract the critical jamming density $\volfracSJ$.
When the relaxed $\norPres$ is 
normalized with the contact density $\volfrac C$, we obtain $\volfracSJ = 0.6567 \pm 0.0005$ by linear extrapolation. A similar value of $\volfracSJ$ is obtained 
from the extrapolation of the relaxed $\nstress$ data set, and is consistent with other methods using the coordination number $\Cstar$, or the energy \citep{goncu2009jamming}.
The quantification of history dependent jamming densities $\volfracJ(H)$, due to shear complementing the slow changes by cyclic isotropic (over)compression in Eq.\ (\ref{eq:strexp}), is discussed next.

\subsection{Jamming phase diagram with history $H$} 
\label{subsec:shear4}

We propose a jamming phase diagram with shear strain, and present a new, quantitative history dependent model that 
explains jamming and shear jamming, but also predicts that shear jamming
vanishes under some conditions, namely when the system is not tapped, tempered or over-compressed before shear is applied.
Using $\shstrain$ and $\volfrac$ as parameters, Fig.\ \ref{phaseplot} shows that for one initial the history dependent jamming state at $\ovolfracJi$, 
there exist sheared states within the 
range $\volfracSJ\le\volfrac\le\volfracJ(H)$, which are 
isotropically unjammed. 
After small shear strain they become
fragile, and for larger shear strain jam and remain jammed,
i.e.,\ eventually showing the critical state flow regime \citep{guo2013signature,zhao2013unique}, where pressure, shear stress
ratio and structural anisotropy have reached their saturation
levels and forgotten their initial state (data not shown). 
The transition to fragile states is accompanied by partial percolation 
of the strong force network, while percolation in all directions indicates the shear jamming 
transition. Above jamming, the large fraction of non-rattlers provides a persistent 
mechanical stability to the structure, even after shear is stopped.

\begin{figure*} 
\centering
\hspace{-15mm}
{\subfigure[]{\includegraphics[width=0.35\textwidth, angle=270]{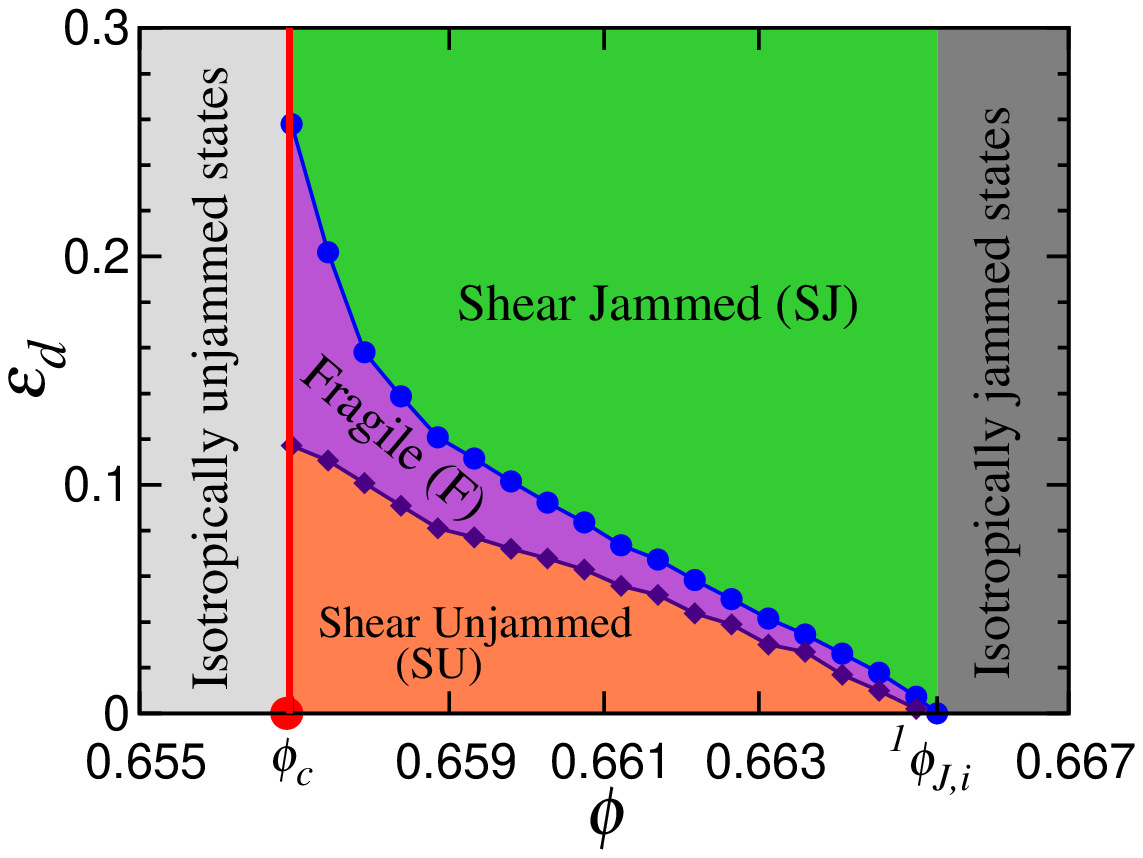}\label{phaseplot}}} 
{\subfigure[]{\includegraphics[width=0.35\textwidth, angle=270]{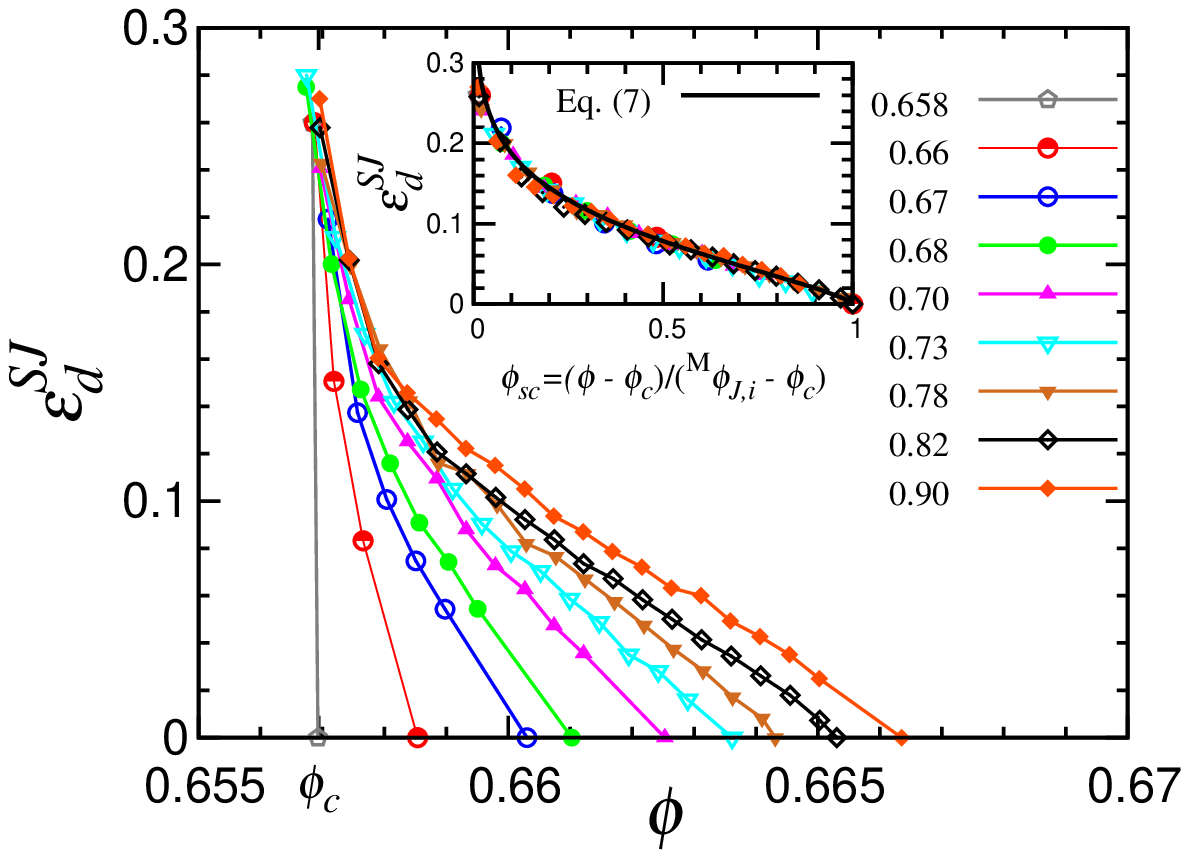}\label{phaseplot_maxvol}}}
\caption{{Phase diagram and scaling with $\bm{\volfracSJ}$ to replace the $\bm{\MvolfracJi}$'s. } 
(a) Phase diagram showing the different states: unjammed, isotropic jammed, shear unjammed, fragile and shear jammed, for one particular case of $\volfracJ\left(\volfracmax_i = 0.82, M=1\right) =: \ovolfracJi = 0.6652$.  
(b) Plot of minimum strain needed to jam states prepared from the first over-compression cycle 
with different $\volfracmax_i$, as given in the legend. 
The inset shows the collapse of the states using a scaled definition that includes distance from 
both isotropic jamming density $\MvolfracJi$ and critical jamming density $\volfracSJ$, using Eq.\ (\ref{eq:scalejamming}). 
We only show data for the states for $\volfrac < \ovolfracJi$ that after the first isotropic compression 
decompression cycle jam by applying shear. }
\label{phaseplot_overcompress}
\end{figure*}
For $\volfrac$ approaching $\volfracSJ$, the required shear strain to jam $\shstrain^{SJ}$ increases, i.e.,\ there exists a divergence ``point'' $\volfracSJ$, 
where `infinite' shear strain might jam the system, 
but below which no shear jamming was observed. 
The closer the (constant) volume fraction $\volfrac$ is to the initial 
$\ovolfracJi$, the smaller is $\shstrain^{SJ}$. 
States with $\volfrac\ge\ovolfracJi$ are isotropically jammed already before shear is applied. 
%

Based on the study of many systems, prepared via isotropic over-compression 
to a wide range of volume fractions $\volfracmax_i \ge \volfracSJ$, 
and subsequent shear deformation, Fig.\ \ref{phaseplot_maxvol} shows 
the strains required to jam these states by applying pure shear. 
A striking observation is that \textit{independent} of the isotropic jamming density 
$\ovolfracJi$, all curves approach a \textit{unique} 
critical jamming density at $\volfracSJ \sim 0.6567$ (see section\ \ref{subsubsec:shear3}). 
When all the curves are scaled with their original isotropic jamming density 
$\MvolfracJi$ as
$\volfracscaled = \left(\volfrac-\volfracSJ\right)/\left(\MvolfracJi-\volfracSJ\right)$ 
they collapse on a unique master curve
\begin{equation}
\label{eq:scalejamming}
\left(\shstrain^{SJ}/\shstrain^{0}\right)^{\alpha} ={-} \log{\volfracscaled} 
 = {-} \log{\left(\frac{\volfrac-\volfracSJ}{\MvolfracJi-\volfracSJ}\right) },
\end{equation}
shown in the inset of Fig.\ \ref{phaseplot_maxvol},
with power $\alpha=1.37 \pm 0.01$ and shear strain scale $\shstrain^{0}=0.102 \pm 0.001$ as the fit parameters. 
Hence, if the initial jamming density $\MvolfracJi$ or $\volfracJ(H)$ is known based on the 
past history of the sample, the shear jamming strain 
$\shstrain^{SJ}$ can be predicted. 

From the measured shear jamming strain, Eq.\ (\ref{eq:scalejamming}), knowing the initial and the limit
value of $\volfracJ$, we now postulate its evolution under isochoric pure shear strain:
\begin{equation}
\label{eq:scalejamming2}
\volfracJ(\shstrain) = \volfracSJ + \left(\volfrac - \volfracSJ\right)
   \exp \left[\left(\frac{       \left(\shstrain^{SJ}\right)^\alpha   -    \left(\shstrain\right)^\alpha }{    \left(\shstrain^{0}\right)^\alpha     }\right)   \right].
\end{equation}
Inserting, $\shstrain =0$, $\shstrain =\shstrain^{SJ}$ and $\shstrain =\infty$ leads to $\volfracJ = \MvolfracJi$, $\volfracJ = \volfrac$ and $\volfracJ = \volfracSJ$, respectively.
This means the jamming density evolution due to shear strain $\shstrain$ is faster than exponential 
(since $\alpha > 1$) decreasing to its lower limit $\volfracSJ$. This is qualitatively
different from the stretched exponential (slow) relaxation dynamics that leads to
the increase of $\volfracJ$ due to over-compression or tapping, see Fig.\ \ref{phievol} for both cases. 

\section{Meso-scale stochastic slow dynamics model} 
\label{sec:slowmodel}

The last challenge is to unify the observations in a qualitative model that accounts for the 
changes in the jamming densities for both isotropic and shear deformation modes. 
Over-compressing a soft granular assembly is analogous to small-amplitude
tapping \citep{zhang2010jamming, rosato2010microstructure, coulais2014ideal} 
of more rigid particles, in so far that both methods lead to more compact (efficient) 
packing structures, i.e., both representing more isotropic perturbations, rather than 
shear, which is deviatoric (anisotropic) in nature.
\begin{figure*}
\centering
{\subfigure[]{\includegraphics[width=0.3\textwidth, angle=270]{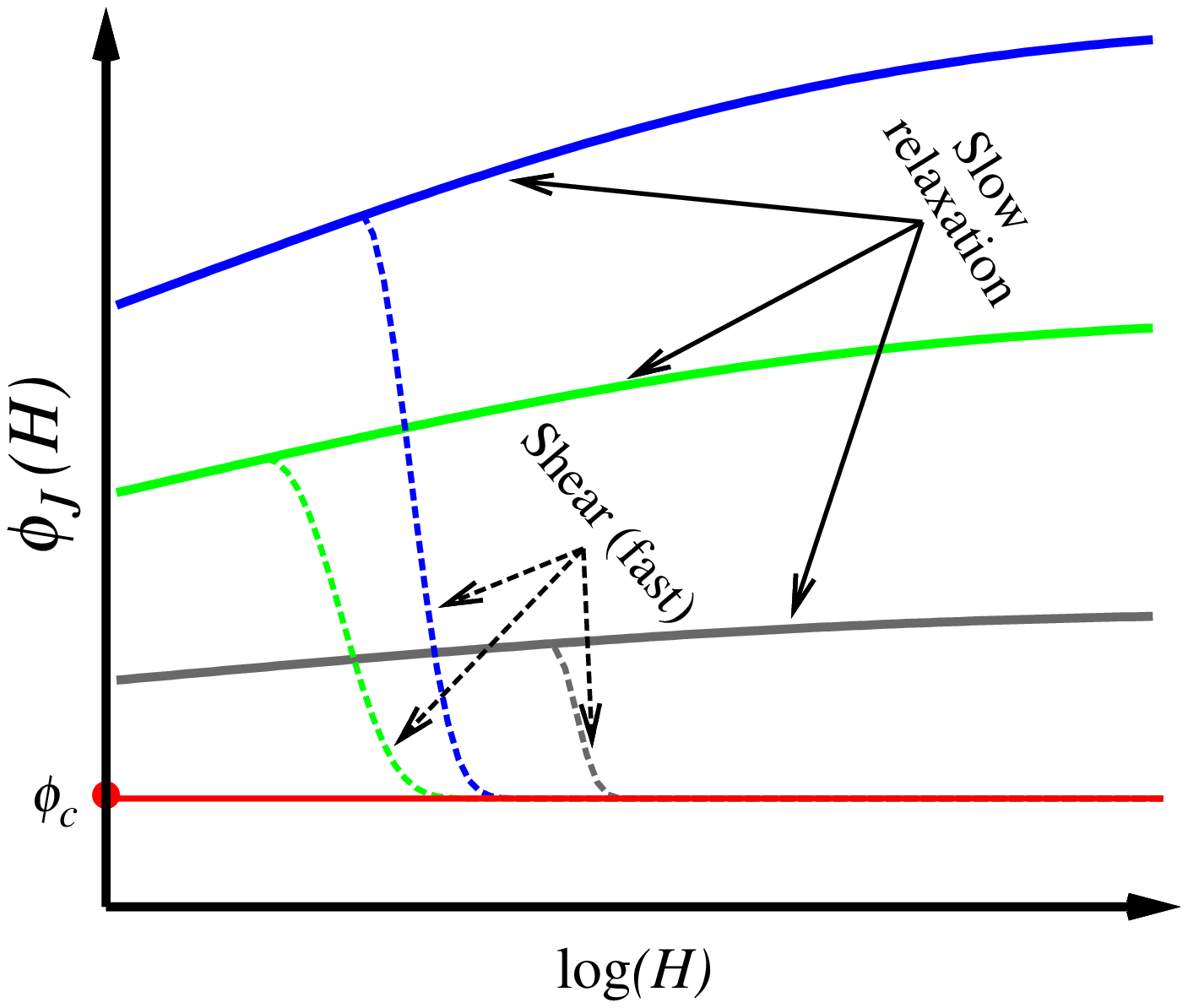}\label{phievol}}}
{\subfigure[]{\includegraphics[width=0.3\textwidth, angle=270]{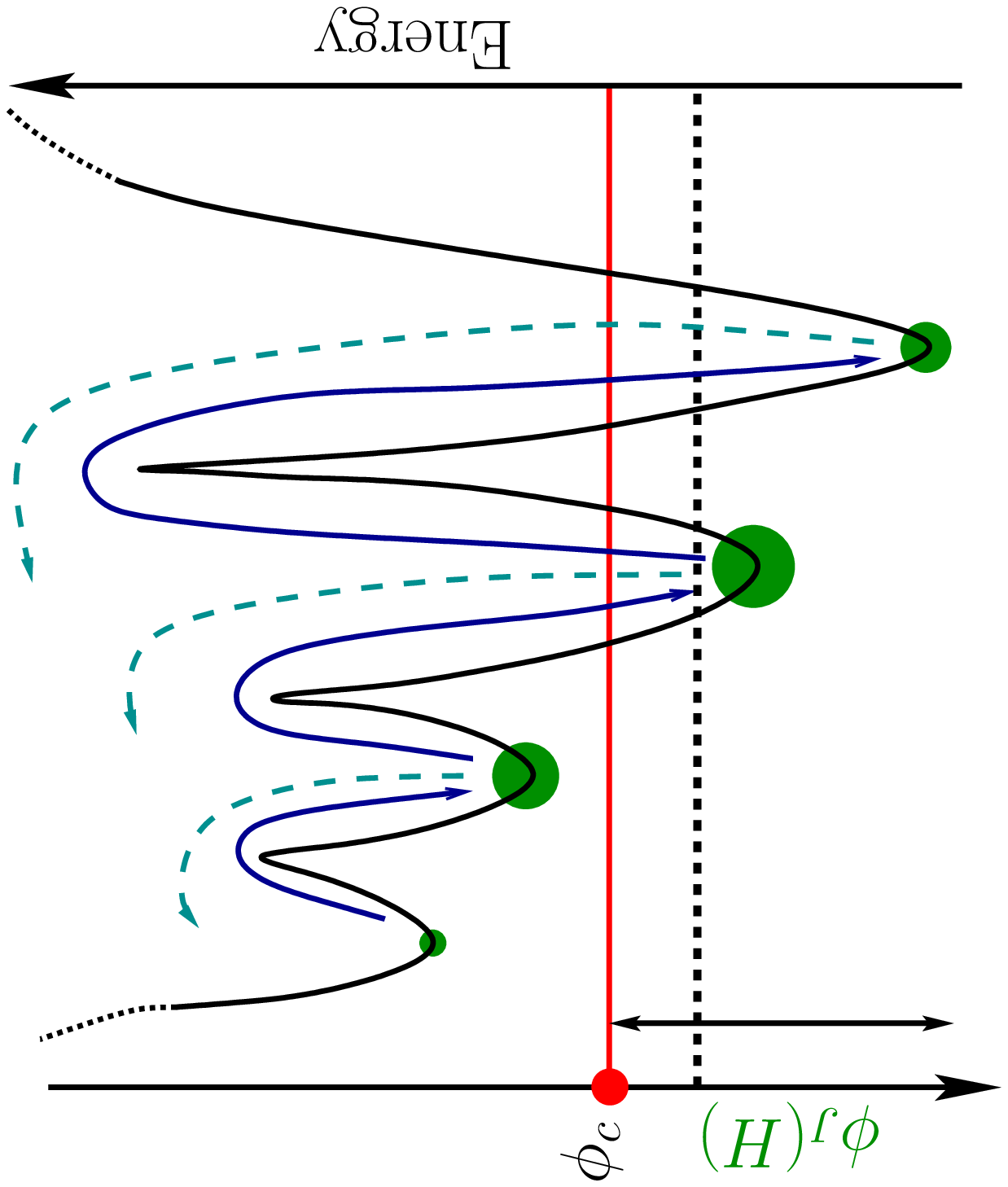}\label{landscape}}}
\caption{{Relaxation mechanisms and dynamics in an energy landscape due to memory effects.} 
(a) Schematic evolution of the jamming densities $\volfracJ(H)$ due to history $H$.  
Solid lines represent many isotropic compression decompression cycles for three different $\volfracmax_i$, 
leading to an increase in $\volfracJ(H)$ by slow stretched exponential relaxation, see Eq.\ (\ref{eq:strexp}). 
Dashed lines represent the much faster decrease in $\volfracJ(H)$ due to 
shear strain $\shstrain$, using Eq.\ (\ref{eq:scalejamming2}). 
(b) The sketch represents only a very small, exemplary part of the hierarchical, fractal-type energy landscape.
The red horizontal line represents the (quenched) average, while the dotted horizontal line indicates 
the momentary average $\volfracJ(H)$ (of the ensemble of states, where the population is represented by 
green circles). The blue solid arrows show (slow) relaxation due to perturbations, while the dashed arrows 
indicate (fast) re-arrangements (re-juvenation) due to finite shear strain. 
The green dots represent with their size the population after some relaxation, in contrast 
to a random, quenched population where all similar valleys would be equally populated \citep{xu2011direct}. } 
\label{prop}
\end{figure*}
These changes are shown in Fig.\ \ref{diffmaxvol}, where the originally reported
logarithmically slow dynamics for tapping 
\citep{knight1995density, andreotti2013granular, vzivkovic2011structural} 
is very similar to our results that are also very slow, with a stretched exponential behavior; 
such slow relaxation dynamics can be explained by a simple Sinai-Diffusion 
model of random walkers in a random, hierarchical, fractal, free energy 
landscape \citep{luding2000minimal, richard2005slow} in the (a-thermal) 
limit, where the landscape does not change -- for the sake of simplicity.

The granular packing is represented in this picture by an ensemble of random
walkers in (arbitrary) configuration space with (potential) energy according to the 
height of their position on the landscape. (Their average energy
corresponds to the jamming density and a decrease in energy 
corresponds to an increase in $\volfracJ(H)$, thus representing the 
``memory'' and history dependence with protocol $H$.) 
Each change of the ensemble represents a rearrangement of packing 
and units in ensemble represent sub-systems. 
Perturbations, such as tapping 
with some small-amplitude (corresponding to ``temperature'') allow the ensemble
to find denser configurations, i.e.,\ deeper valleys in the landscape, 
representing larger (jamming) densities \citep{mobius2014ir, reichhardt2014aspects}.
Similarly, over-compression is squeezing the ensemble ``down-hill'',
also leading to an increase of $\volfracJ$, as presented 
in Fig.\ \ref{landscape}. Larger amplitudes will allow the ensemble
to overcome larger barriers and thus find even deeper valleys. 
Repetitions have a smaller chance to do so -- since the easy reorganizations have been realized previously -- which explains the slow dynamics 
in the hierarchical multiscale structure of the energy landscape.

In contrast to the isotropic perturbations, where the random walkers follow
the ``down-hill'' trend, shear is anisotropic and thus pushing parts of the ensemble in ``up-hill' direction'.
For example, under planar simple shear, one (eigen) direction is extensive (up-hill) 
whereas an other is compressive (down-hill). If the ensemble is random, shear will
only re-shuffle the population. But if the material was previously forced or relaxed
towards the (local) land-scape minima, shear can only lead to 
a net up-hill drift of the ensemble, i.e.,\ to decreasing 
$\volfracJ$, referred to as dilatancy under constant stress boundary conditions.

For ongoing over-compression, both coordination number and pressure slowly 
increase, as sketched in Fig.\ \ref{schematic_iso1}, while the jamming density drifts
to larger values due to re-organization events that make the packing more effective, 
which moves the state-line to the right (also shown in Fig.\ \ref{phievol}). 
For decompression, we assume that there
are much less re-organization events happening, so that the pressure moves down 
on the state-line, until the system unjams.
For ongoing perturbations, at constant volume, as tapping or a finite temperature, $T_g$,
both coordination number and pressure slowly decrease (data not shown), whereas for fixed 
confining pressure the volume would decrease (compactancy, also not shown). 

For ongoing shear, the coordination number, the pressure and the shear stress
increase, since the jamming density decreases, as sketched in Fig.\ \ref{schematic_dev1} until a steady state is reached.
This process is driven by shear strain amplitude and is much faster than the relaxation dynamics.
For large enough strain the system will be sufficiently re-shuffled, randomized, or ``re-juvenated'' 
such that it approaches its quenched, random state close to $\volfracSJ$ (see Fig.\ \ref{phievol}).

If both mechanisms, relaxation by temperature, and continuous shear are occurring 
at the same time, one can reach another (non)-``equilibrium'' steady state, where the jamming
density remains constant, balancing the respective increasing and decreasing trends,
as sketched in Fig.\ \ref{dev1e}.

\begin{figure*}
\centering
{\subfigure[]{\includegraphics[width=0.19\textwidth, angle=0]{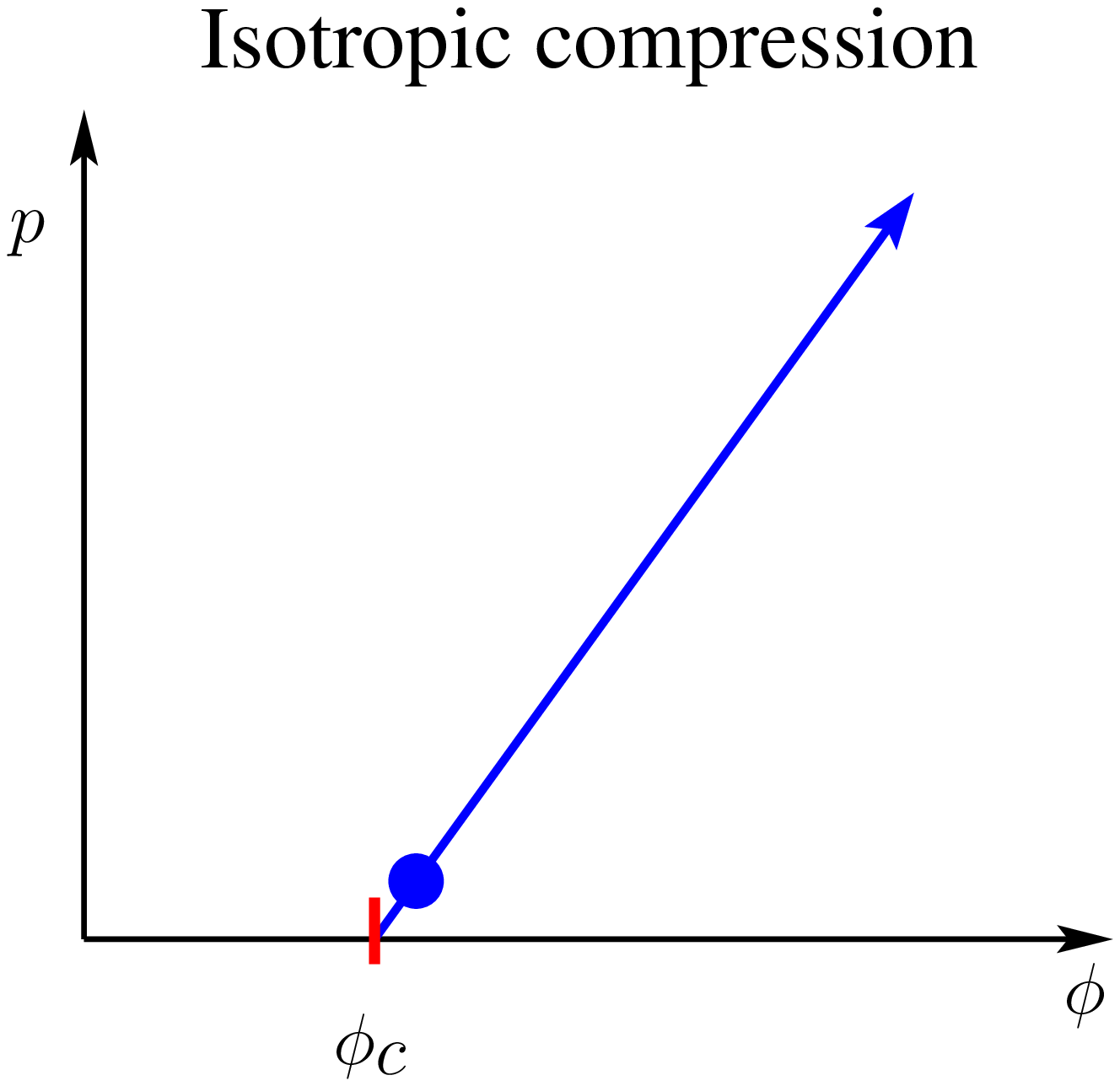}\label{iso1a}}}
{\subfigure[]{\includegraphics[width=0.19\textwidth, angle=0]{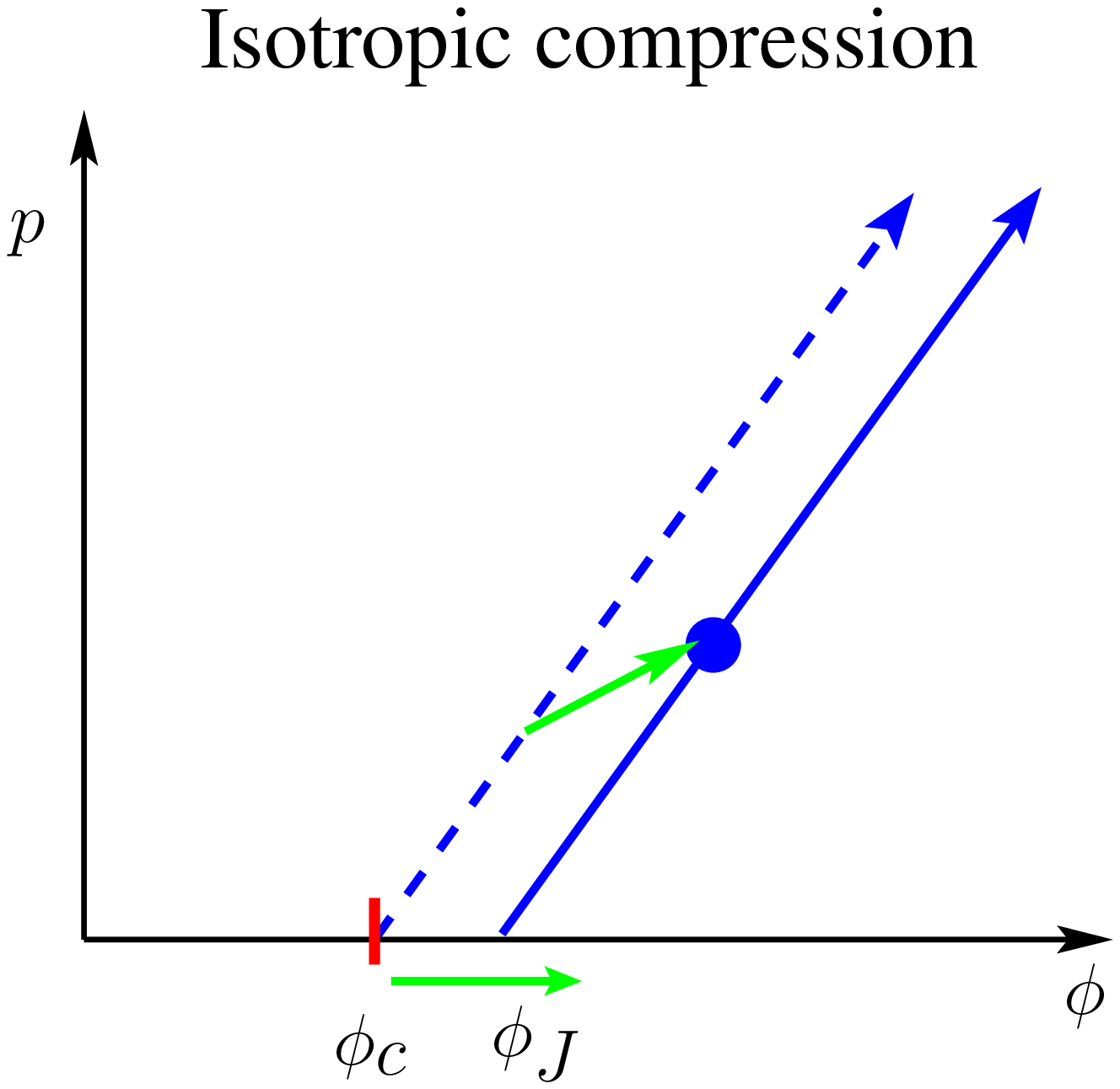}\label{iso1b}}}
{\subfigure[]{\includegraphics[width=0.19\textwidth, angle=0]{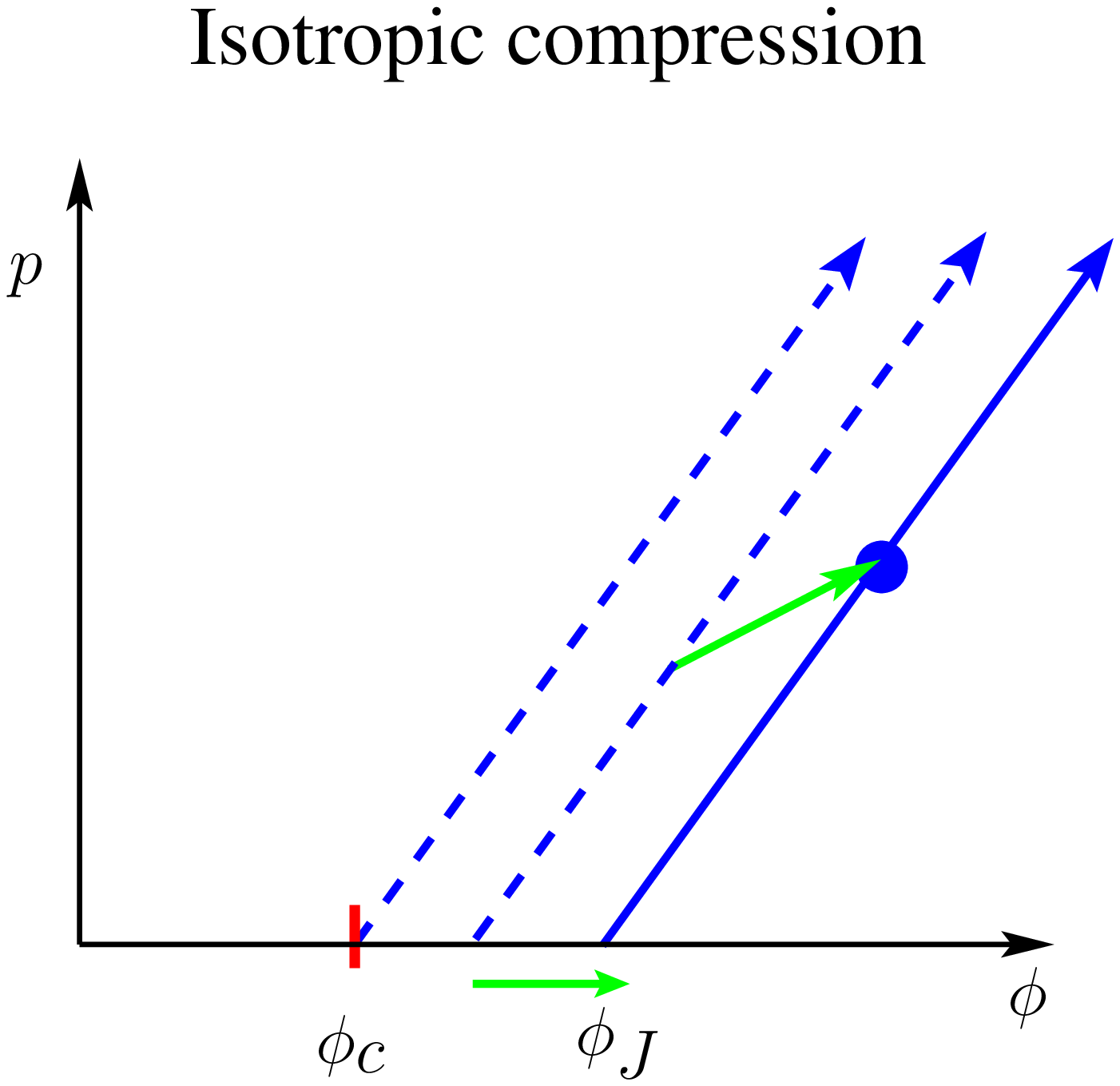}\label{iso1c}}}
{\subfigure[]{\includegraphics[width=0.19\textwidth, angle=0]{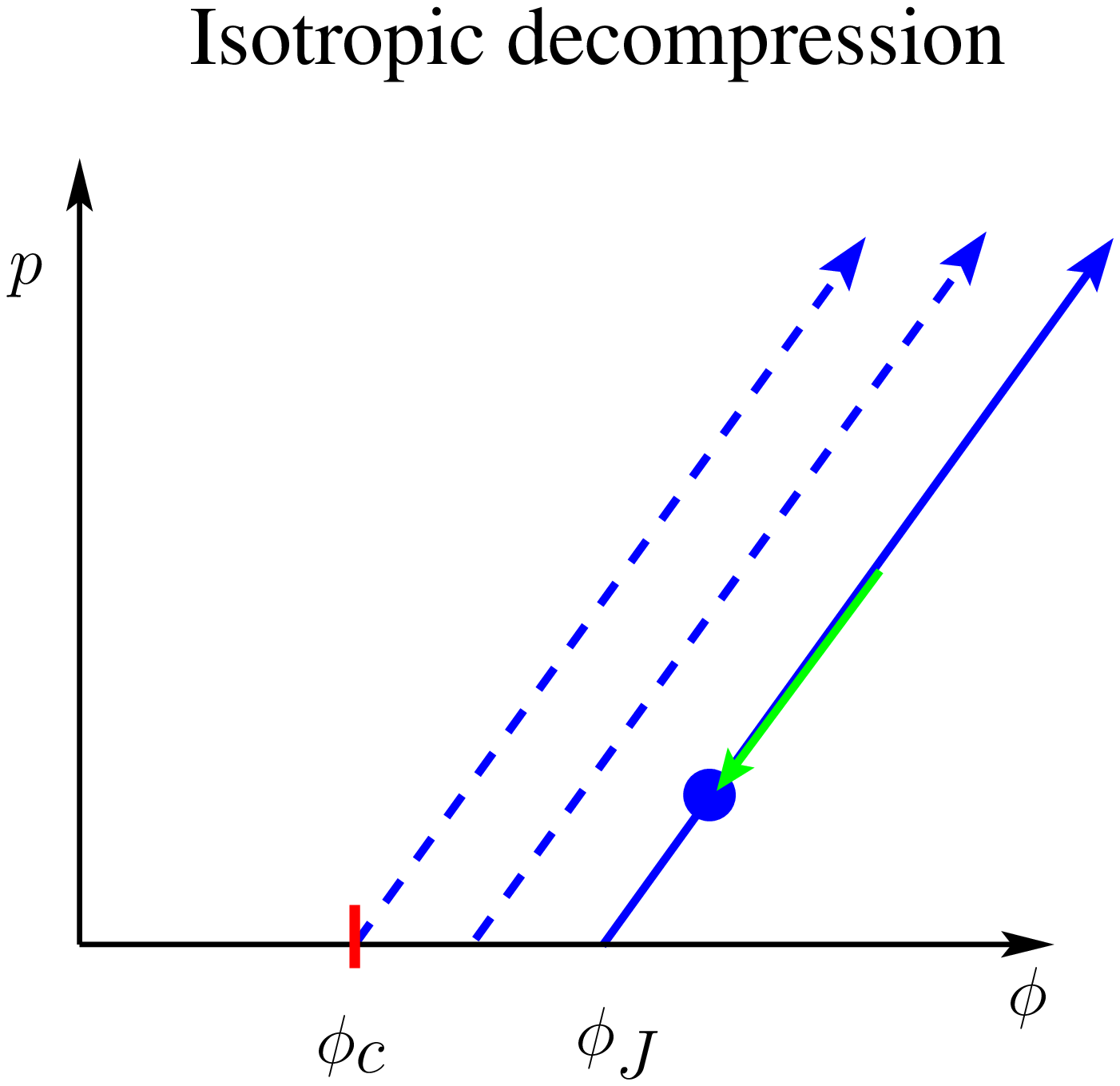}\label{iso1d}}}
{\subfigure[]{\includegraphics[width=0.19\textwidth, angle=0]{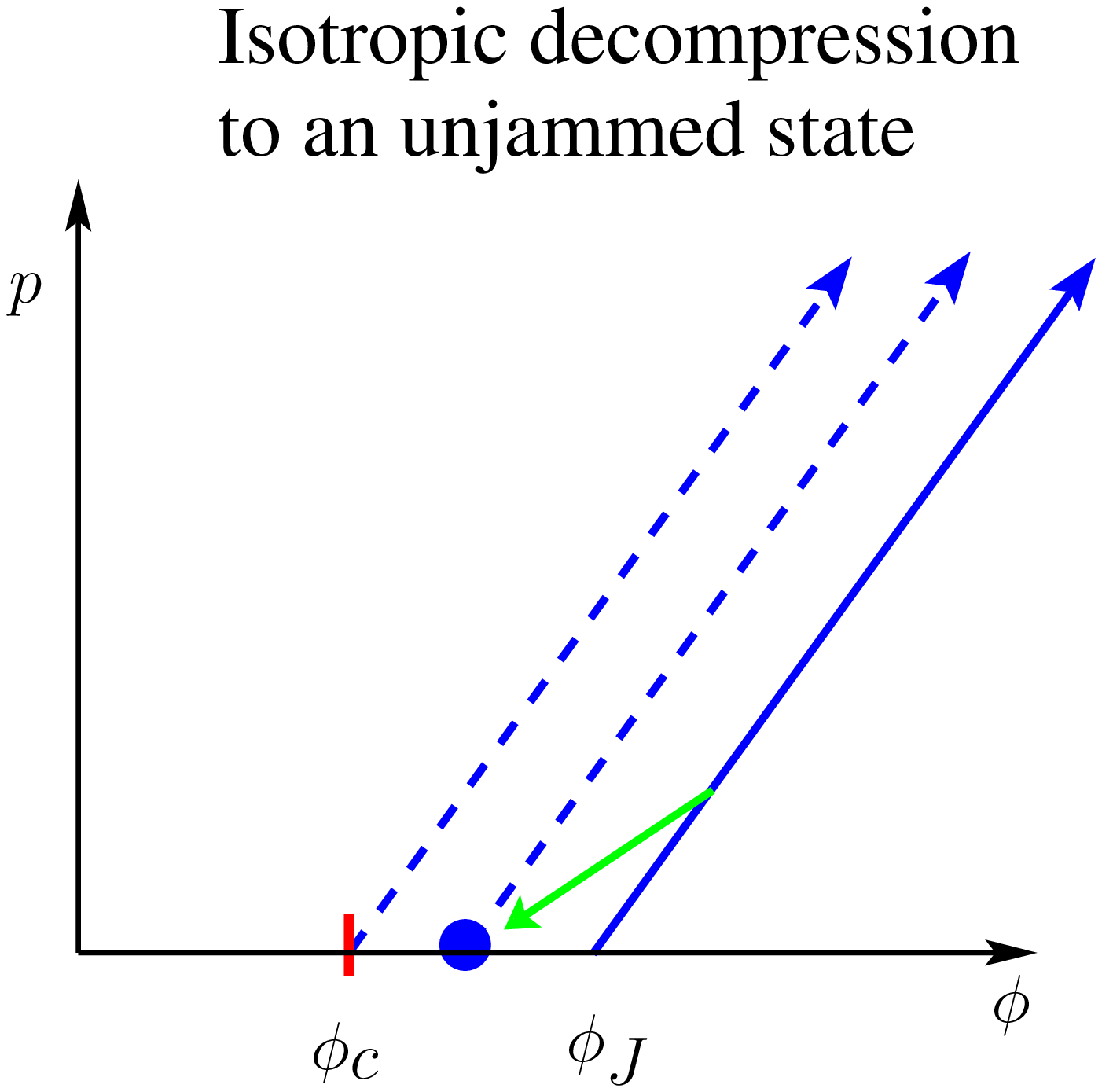}\label{iso1e}}}
\caption{
Schematic sketch of the evolution of the system in stress-density space, e.g., pressure,
(a) starting from a state (point) slightly above jamming, under 
(b) isotropic compression, and 
(c) further compression, the system reaches a higher stress level, while the 
jamming density moves to the right (larger densities). 
(d) For isotropic decompression 
(extension) the system reduces pressure and the jamming density remains (almost) 
constant, until for (e) ongoing decompression, the system unjams and reaches 
a density below the jamming density.
(For tapping (not shown), the density of the system would remain fixed, the jamming
density would increase for ongoing perturbations, so that the stress would reduce and
the system could even unjam if the density is low enough.)}
\label{schematic_iso1}
\end{figure*}
\begin{figure*}
\centering
{\subfigure[]{\includegraphics[width=0.19\textwidth, angle=0]{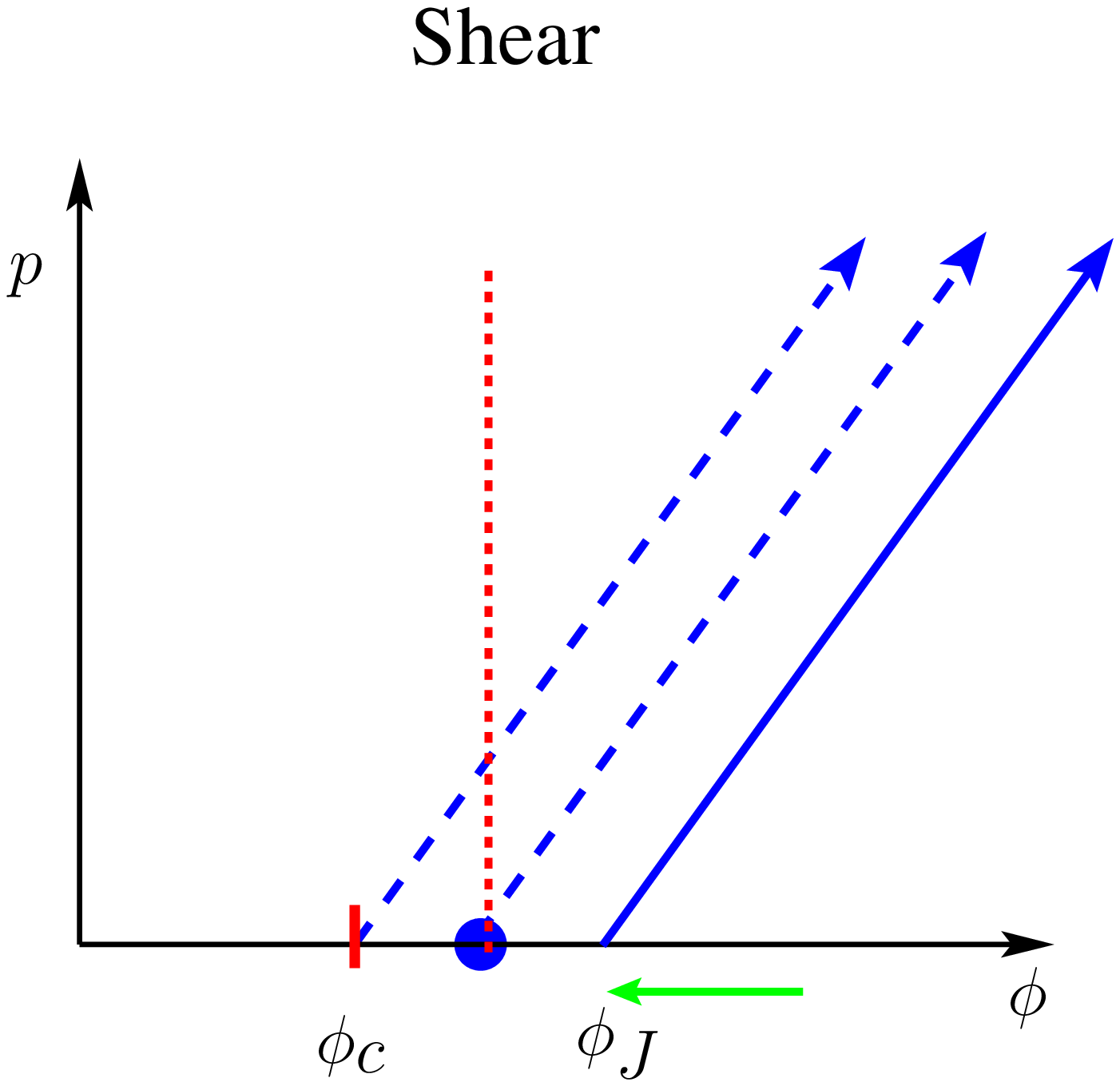}\label{dev1a}}}
{\subfigure[]{\includegraphics[width=0.19\textwidth, angle=0]{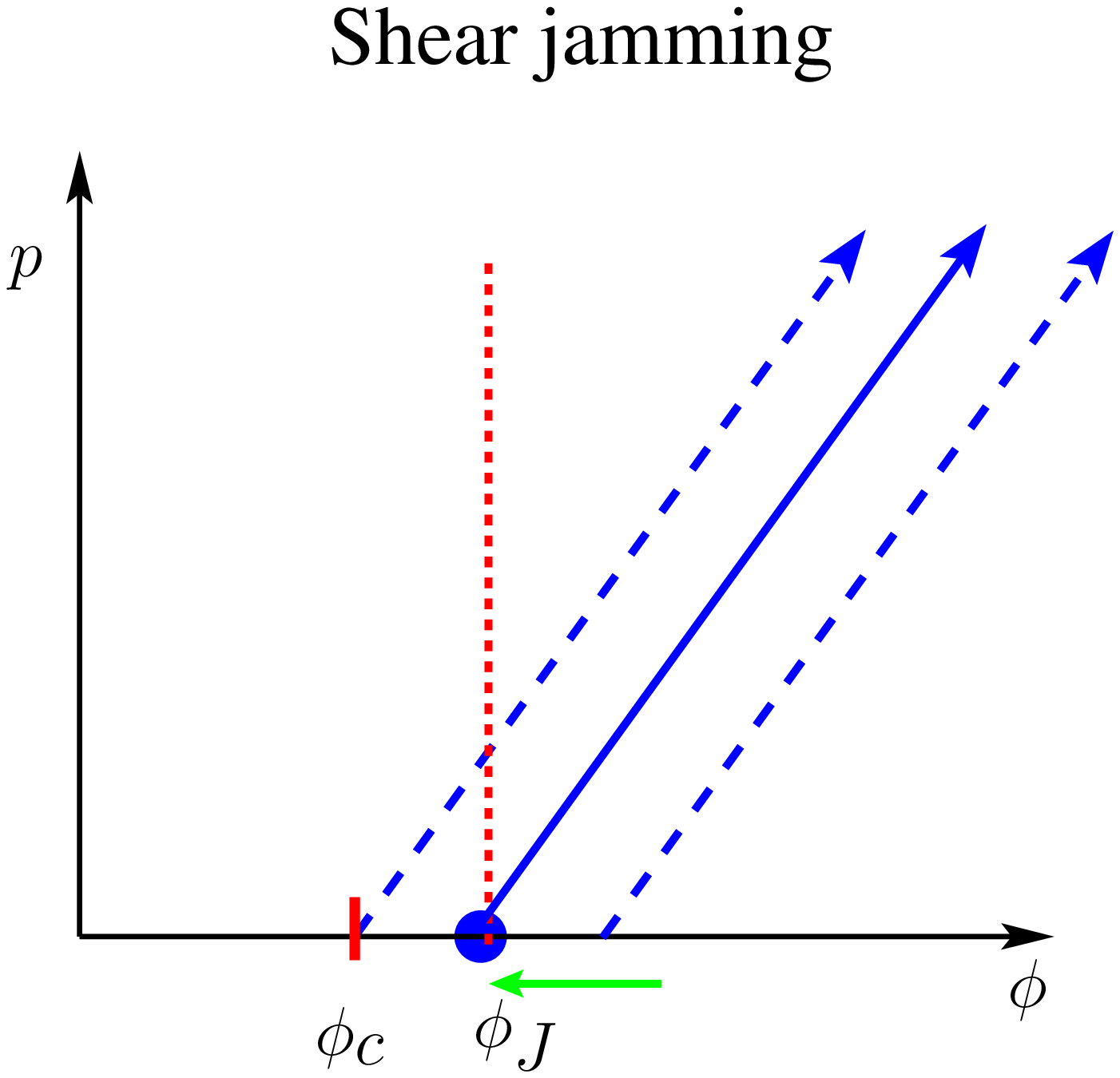}\label{dev1b}}}
{\subfigure[]{\includegraphics[width=0.19\textwidth, angle=0]{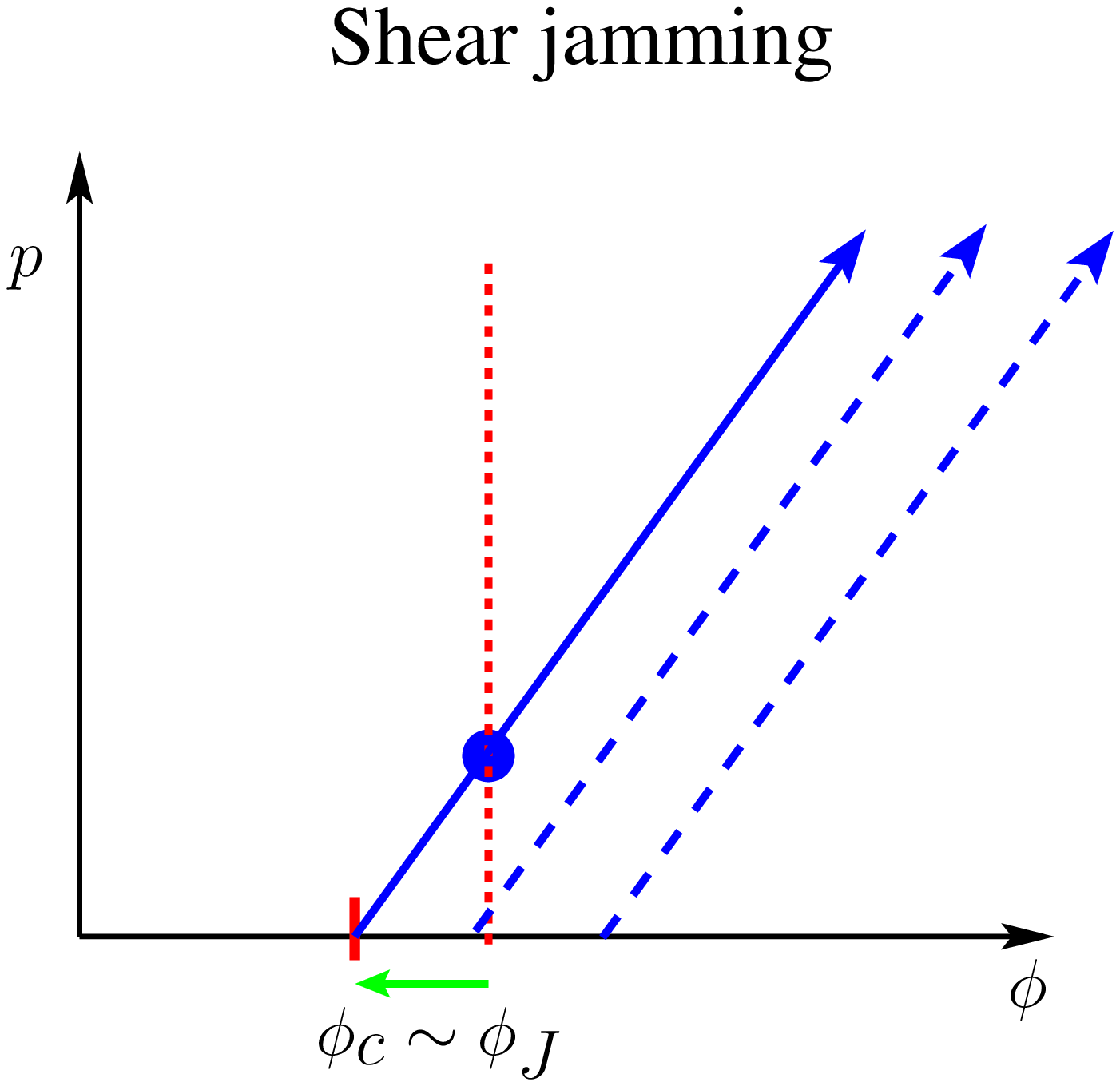}\label{dev1c}}}
{\subfigure[]{\includegraphics[width=0.19\textwidth, angle=0]{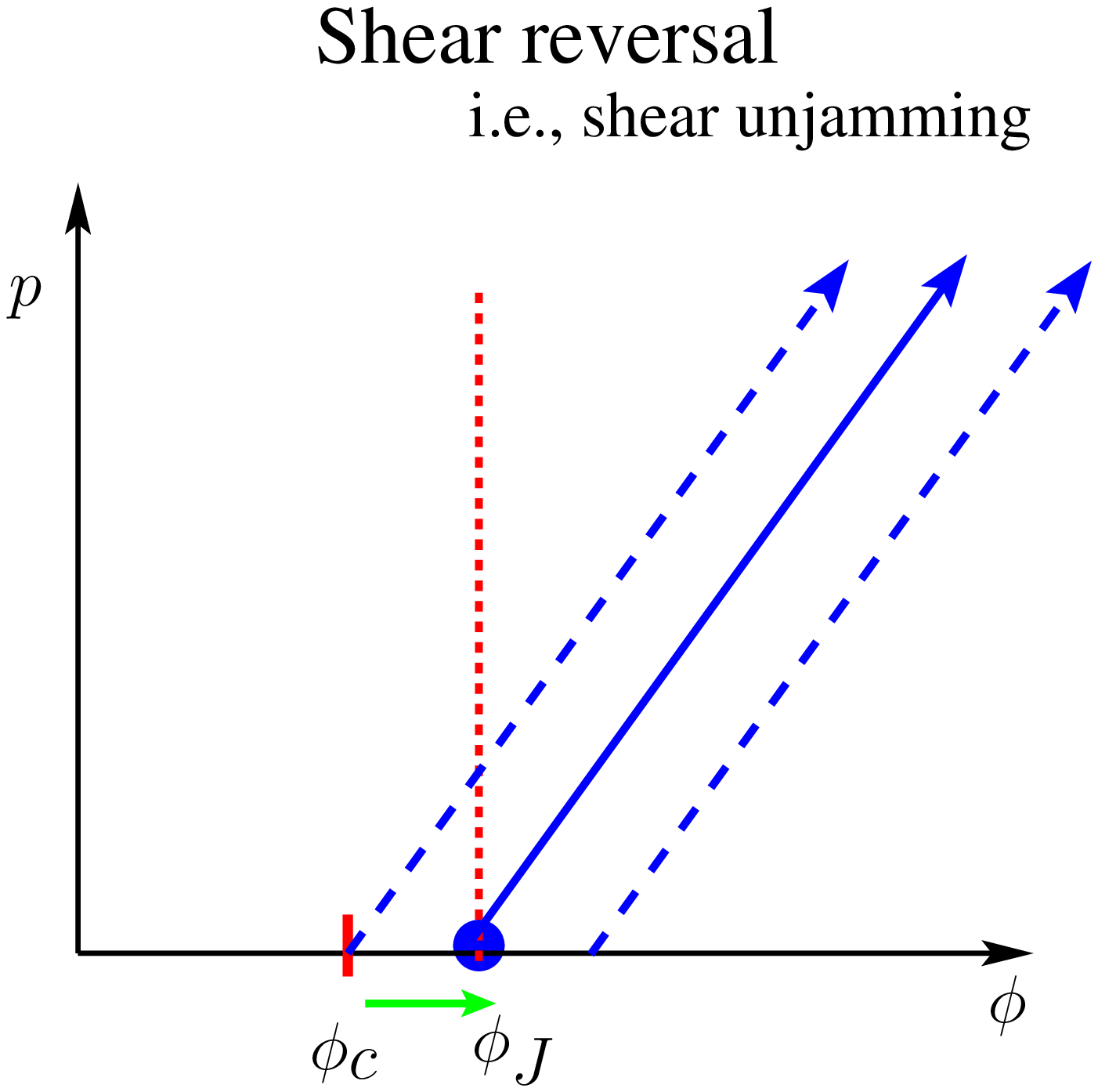}\label{dev1d}}}
{\subfigure[]{\includegraphics[width=0.19\textwidth, angle=0]{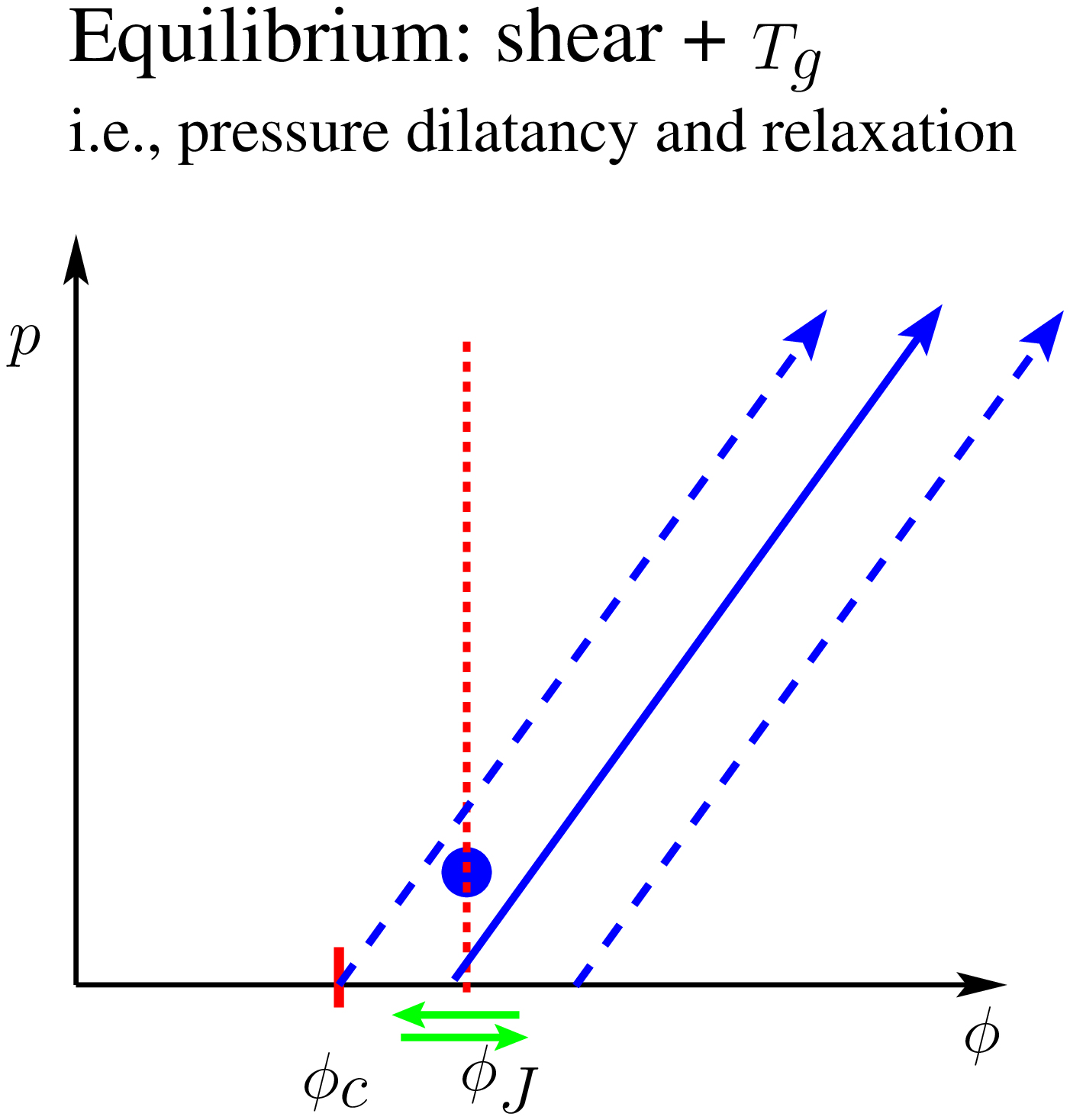}\label{dev1e}}}
\caption{
Schematic sketch of the evolution of the system under isochoric (volume conserving, represented by the dashed vertical red line) shear 
in stress-density space, think of shear stress, which is just proportional to pressure,
(a) starting from the state (point \ref{iso1e}) slightly below jamming, 
which was previously over-compressed.
Under shear 
(b) the jamming density shifts to the left until it reaches the actual density,
at which 
(c) shear jamming kicks in, i.e., stress increases above zero. From this state, for 
(d) shear reversal, the jamming density moves to the right again
and the system can unjam.
For ongoing shear, 
(e) at a higher density, at finite granular temperature $T_g$, 
the jamming density is increased by the perturbations due to $T_g$ while shear,
at the same time, decreases the jamming density, as indicated by the two arrows,
which resembles a steady state. A change of either shear rate or temperature 
will then lead to either transient shear-thickening or shear-thinning, before a new steady state path is reached.}
\label{schematic_dev1}
\end{figure*}

\section{Macroscopic constitutive model} 
\label{sec:model}

In this section, we present the simplest model equations, as used for the predictions, involving a history dependent $\volfracJ(H)$, 
as given by Eq.\ (\ref{eq:strexp}) for isotropic deformations and Eq.\ (\ref{eq:scalejamming2}) for shear deformations. 
The only difference to \citet{imole2013hydrostatic}, where these relations are taken from, based on purely isotropic unloading, is a variable $\volfracJ = \volfracJ(H)$. 

\subsection{Presentation and model calibration} 
\label{subsec:model1}

\subsubsection{During cyclic isotropic deformation} 
\label{subsec:model2}

During (cyclic) isotropic deformation, the evolution equation for the corrected coordination number $\Cstar$ is:
\begin{equation}
\label{eq:cstareqn}
\Cstar = C_0 + C_1\left(\frac{\volfrac}{\volfracJ(H)}-1 \right)^\theta,
\end{equation}
with $C_0=6$ for the frictionless case and parameters $C_1$ and $\theta$ are presented in Table~\ref{cpfnrparamter}.
The fraction of non-rattlers $\fNR$ is given as:
\begin{equation}
\label{eq:fNReqn}
\fNR = 1 - \varphi_c\mathrm{exp}\left[-\varphi_v    \left(\frac{\volfrac}{\volfracJ(H)}-1 \right) \right],
\end{equation}
with parameters $\varphi_c$ and $\varphi_v$ presented in Table~\ref{cpfnrparamter}.
We modify Eq.\ (\ref{eq:pstar}) for the evolution of $\norPres$ together with the history dependent $\volfracJ=\volfracJ(H)$ so that, 
\begin{equation}
p=\frac{\volfrac C}{\volfracJ(H)}p_0 (-\varepsilon_\mathrm{v})  \left [ 1-\gamma_p (-\varepsilon_\mathrm{v}) \right ],
\label{eq:pstareqn}
\end{equation}
with parameters $p_0$ and $\gamma_p$ presented in Table~\ref{cpfnrparamter}, and the true or logarithmic volume change 
of the system is $-\varepsilon_\mathrm{v} = \log(\volfrac/\volfracJ(H))$, relative to the momentary jamming density. 
The non-corrected coordination number
is $C = \Cstar\fNR$, as can be computed 
using Eqs.\ (\ref{eq:cstareqn}) and (\ref{eq:fNReqn}). 
Also the parameters $C_1$, $\theta$ for $\Cstar$, $\varphi_c$, $\varphi_v$ for $\fNR$, and $p_0$, $\gamma_p$ for pressure $p$ are similar to \citet{imole2013hydrostatic}, with 
the second order correction parameter $\gamma_p$ most sensitive to the details of previous deformations; 
however, not being very relevant since it is always a small correction due to the product $\gamma_p (-\varepsilon_\mathrm{v})$.

The above relations are used to predict the behavior of the 
isotropic quantities: 
dimensionless pressure $\norPres$ and coordination number $\Cstar$, as shown in Fig.\ \ref{predicted}(a-b) during isotropic compression, as well as for the 
fraction of non-rattlers in Fig.\ \ref{cycle_perco_new} for cyclic shear, with corresponding parameters presented in Table~\ref{cpfnrparamter}.
Note that during isotropic deformation, $\volfracJ(H)$ was changed only during 
the compression branch, using Eq.\ (\ref{eq:strexp}) for fixed $M=1$ using $\volfracmax_i$ as variable, but is kept constant 
during unloading/expansion.

The above relations are used to predict the behavior of the isotropic quantities: 
dimensionless pressure $\norPres$ and coordination number $\Cstar$, by only 
adding the history dependent jamming density $\volfracJ(H)$ to the constitutive model, 
as tested below in section\ \ref{subsec:model5}.

\subsubsection{Cyclic (pure) shear deformation} 
\label{subsubsec:model3}

During cyclic (pure) shear deformation, a simplified equation for the shear stress ratio $\stressrat$ 
is taken from \citet{imole2013hydrostatic}, where the full model was introduced as rate-type evolution
equations, and further calibrated and tested by Kumar et al.\ \cite{kumar2014macroscopic}: 
\begin{equation}
\label{eq:stressratioeqn}
\stressrat = {\left(\stressrat\right)}^{\mathrm{max}}  - \left[  {\left(\stressrat\right)}^{\mathrm{max}} - \left(\stressrat\right)^{\mathrm{0}}  \right]  \exp \left[ -\beta_s \shstrain \right],
\end{equation}
with $\left(\stressrat\right)^{\mathrm{0}}$ and $\left(\stressrat\right)^{\mathrm{max}}$ 
the initial and maximum (saturation) shear stress ratio, respectively, and $\beta_s$ its growth rate
\footnote{Note that the model in the form used here is ignoring the presence of kinetic energy
fluctuations, referred to as granular temperature $T_g$, or fields like the so-called fluidity
\citep{KamrinKoval2012,henann2013predictive, darnige2011creep}, that introduce an additional relaxation 
time-scale, as is subject of ongoing studies.}.
Similarly, a simplified equation for the deviatoric fabric $\anisoF$ can be taken from 
Refs.\ \citep{imole2013hydrostatic, kumar2014macroscopic} as: 
\begin{equation}
\label{eq:fabriceqn}
\anisoF = {\anisoF}^{\mathrm{max}}  - \left[{\anisoF}^{\mathrm{max}} - {\anisoF}^{\mathrm{0}}  \right]  \exp \left[ -\beta_F \shstrain \right],
\end{equation}
with ${\anisoF}^{\mathrm{0}}$ and ${\anisoF}^{\mathrm{max}}$ 
the initial and maximum (saturation) values of the deviatoric fabric, respectively, and $\beta_F$ its growth rate. 
The four parameters $\left(\stressrat\right)^{\mathrm{max}}$, $\beta_s$ for $\stressrat$ and ${\anisoF}^{\mathrm{max}}$, $\beta_F$ for $\anisoF$ 
are dependent on the volume fraction $\volfrac$ and are well described by the general relation from \citet{imole2013hydrostatic} as:
\begin{equation}
\label{eq:QaQceqn}
Q = Q_a + Q_c \exp \left[-\Psi  \left(\frac{\volfrac}{\volfracJ(H)}-1
                   \right) \right],
\end{equation}
where $Q_a$, $Q_c$ and $\Psi$ are the fitting constants with values 
presented in Table~\ref{fittingparameter}.

For predictions during cyclic shear deformation, $\volfracJ(H)$ was changed with applied shear strain $\shstrain$ using Eq.\ (\ref{eq:scalejamming2}). 
Furthermore, the jamming density is set to a 
larger value just after strain-reversal, as discussed next.

\begin{table}
  \begin{center}
\def~{\hphantom{0}}
  \begin{tabular}{l@{\hskip .3in}c@{\hskip .3in}cc}
Evolution parameters & $Q_{a}$ & $Q_{c}$ & $\Psi$\\[4mm]
 $\left(\stressrat\right)^{\mathrm{max}}$ & 0.12 & 0.091 & 7.9 \\[4mm]
 $\beta_s$  & 30 & 40 & 16\\[4mm]
${\anisoF}^{\mathrm{max}}$ & 0 & 0.17 & 5.3\\[4mm]
$\beta_F$ &  0 & 40 & 5.3\\[4mm]
   \end{tabular}
 \caption{Parameters for Eqs.\ (\ref{eq:stressratioeqn}) and (\ref{eq:fabriceqn}) using Eq.\ (\ref{eq:QaQceqn}), with slightly different values than from \citet{imole2013hydrostatic}, 
 that are extracted using the similar procedure as in \citet{imole2013hydrostatic}, for states with volume fraction close to the jamming volume fraction.}
 \label{fittingparameter}
  \end{center}
\end{table}

\subsubsection{Behavior of the jamming density at strain reversal} 
\label{subsubsec:model4}

As mentioned in section\ \ref{subsec:shear1}, there are some states below $\volfracJ$, where application of shear strain jams the systems.
The densest of those can resist shear reversal, but below a certain $\volfraccr\approx 0.662<\volfracJ$, shear reversal unjams the system again \cite{ness2015two}.
With this information, we postulate the following: \\
(i) After the first phase, for large strain pure shear, 
the system should forget where it was isotropically compressed to before
i.e.,\ $\MvolfracJi$ is forgotten and $\volfracJ=\volfracSJ$ is realized. \\
(ii) There exists a volume fraction $\volfraccr$, above which the
systems can just resist shear reversal and remain always jammed in both forward and reverse shear.\\ 
(iii) Below this $\volfraccr$, reversal unjams the system. 
Therefore, more strain is needed to jam the system (when compared to the initial loading), 
first to forget its state before reversal, and then to re-jam it in opposite (perpendicular) shear direction. 
Hence, the strain necessary to jam in reversal direction should be higher than for the first shear cycle. \\
(iv) As we approach $\volfracSJ$, the reverse strain needed to jam the system increases.

We use these ideas and measure the reversal shear strain $\shstrain^{SJ,R}$, needed to re-jam the states below $\volfraccr$, as shown in Fig.\ \ref{strain_reversal}.
When they are scaled with $\volfraccr$ as
$\volfracscaled = \left(\volfrac-\volfracSJ\right)/\left(\volfraccr-\volfracSJ\right)$, they collapse on a unique master curve, very similar to Eq.\ (\ref{eq:scalejamming}):
\begin{equation}
\label{eq:scalejamming_cr}
\left(\shstrain^{SJ,R}/\shstrain^{0,R}\right)^{\alpha} ={-} \log{\volfracscaled} 
 = {-} \log{\left(\frac{\volfrac-\volfracSJ}{\volfraccr-\volfracSJ}\right) },
\end{equation}
shown in the inset of Fig.\ \ref{phaseplot_maxvol},
with the same power $\alpha=1.37 \pm 0.01$ as Eq.\ (\ref{eq:scalejamming}). 
Fit parameter strain scale $\shstrain^{0,R}=0.17 \pm 0.002 > \shstrain^{0} = 0.102$, is consistent with the above postulates (iii) and (iv).

\begin{figure} 
\centering
\hspace{-15mm}
\includegraphics[width=0.35\textwidth, angle=270]{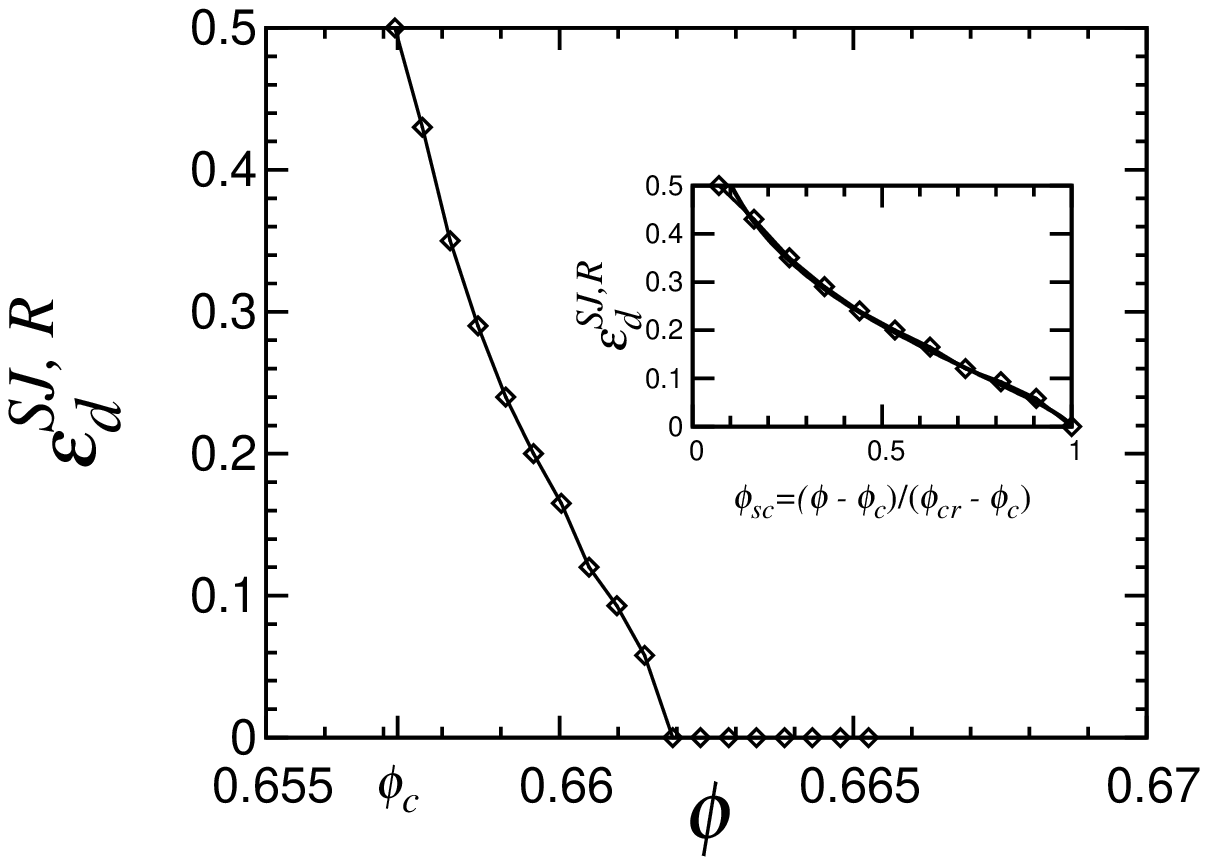}
\caption{Phase diagram showing the minimum reversal shear strain $\shstrain^{SJ,R}$ needed to jam the states below $\volfraccr$, 
for states prepared from the first over-compression cycle with different $\volfracmax_i$, as given in the legend. 
The inset shows a collapse of the states using a similar scaled definition as Eq.\ (\ref{eq:scalejamming})
that includes the distance from both $\volfraccr$ and critical jamming density $\volfracSJ$, 
using Eq.\ (\ref{eq:scalejamming_cr}).}
\label{strain_reversal}
\end{figure}

The above relations are used to predict the isotropic and the deviatoric quantities, 
during cyclic shear deformation, as described next, with the additional rule that all the quantities attain value zero for $\volfrac \le \volfracJ(H)$. 
Moreover, for any state with $\volfrac \le \volfraccr$, shear strain reversal moves the jamming density 
to $\volfraccr$, and the evolution of the jamming density follows Eq.\ (\ref{eq:scalejamming_cr}).

Any other deformation mode, can be written as a \textit{unique} superposition of pure isotropic 
and pure and axial shear deformation modes \cite{hartkamp2013constitutive}. 
Hence the combination of the above can be easily used 
to describe any general deformation, e.g.\ uniaxial cyclic compression (data not presented) where the axial strain can be decomposed in two plane strain modes.

\subsection{Prediction: minimal model} 
\label{subsec:model5}

\begin{figure*}
\centering
{\subfigure[]{\includegraphics[width=0.3\textwidth, angle=270]{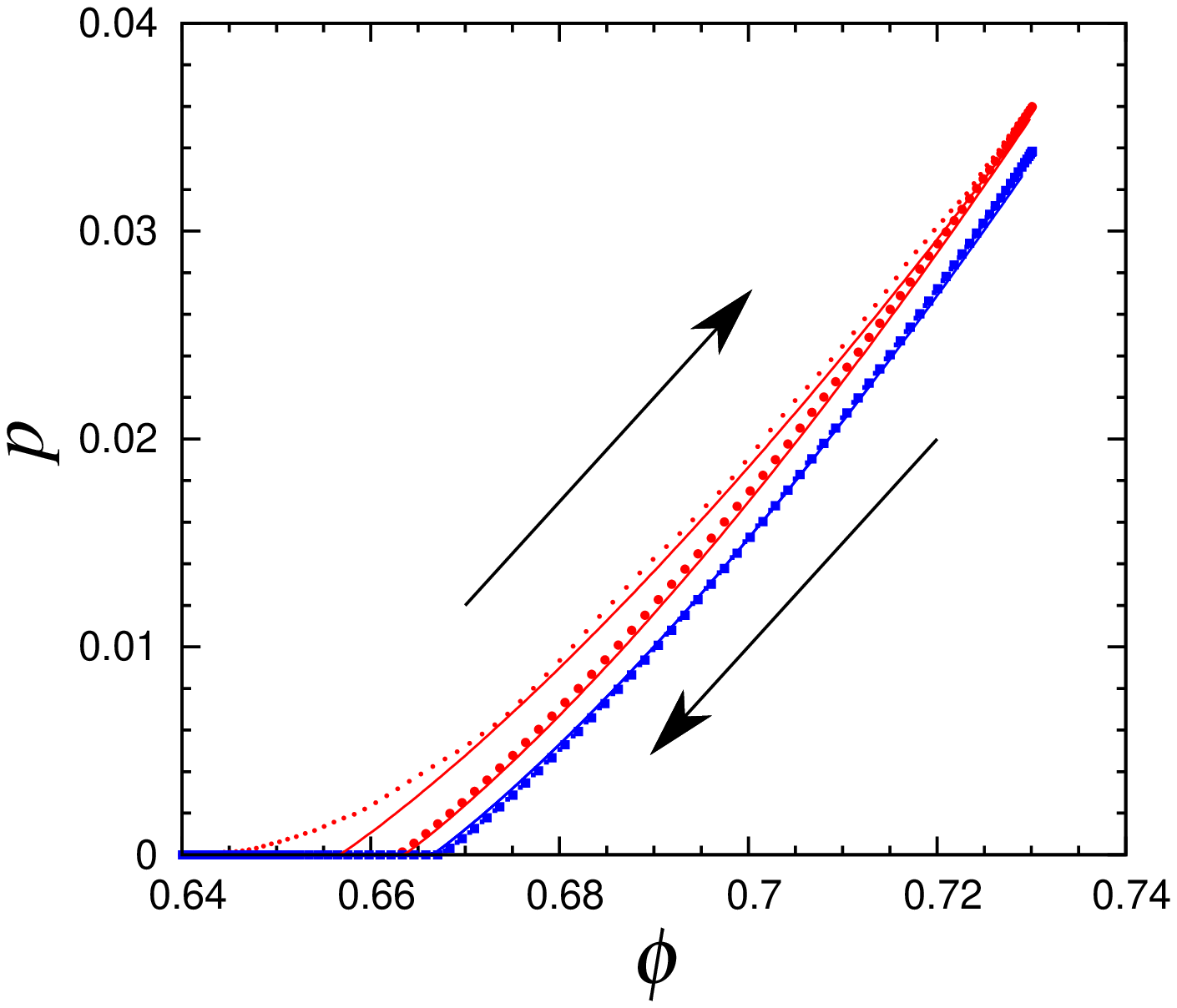}\label{pvsnu_predict}}}
{\subfigure[]{\includegraphics[width=0.3\textwidth, angle=270]{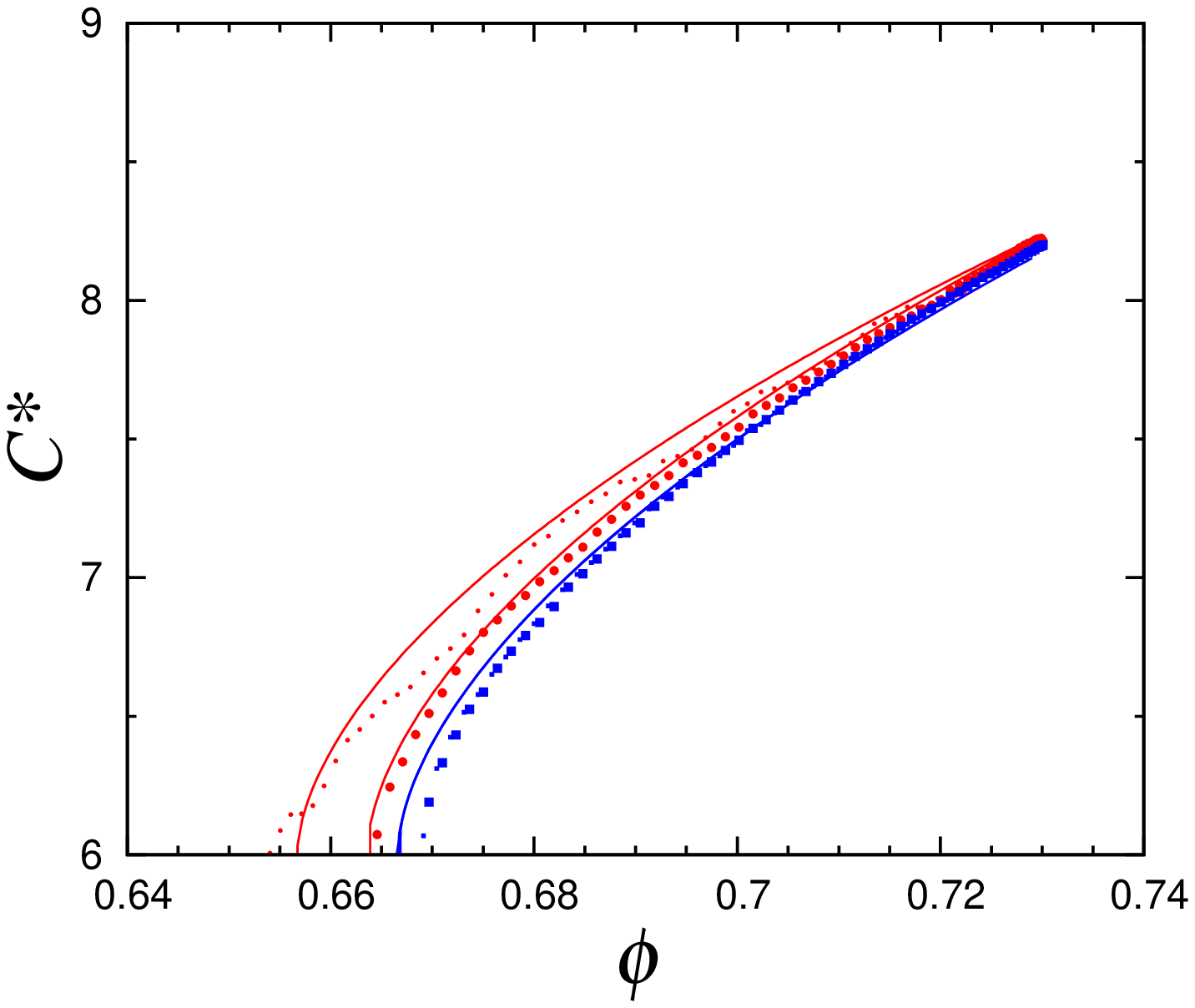}\label{cstarnu_predict}}}\\
{\subfigure[]{\includegraphics[width=0.50\textwidth, angle=270]{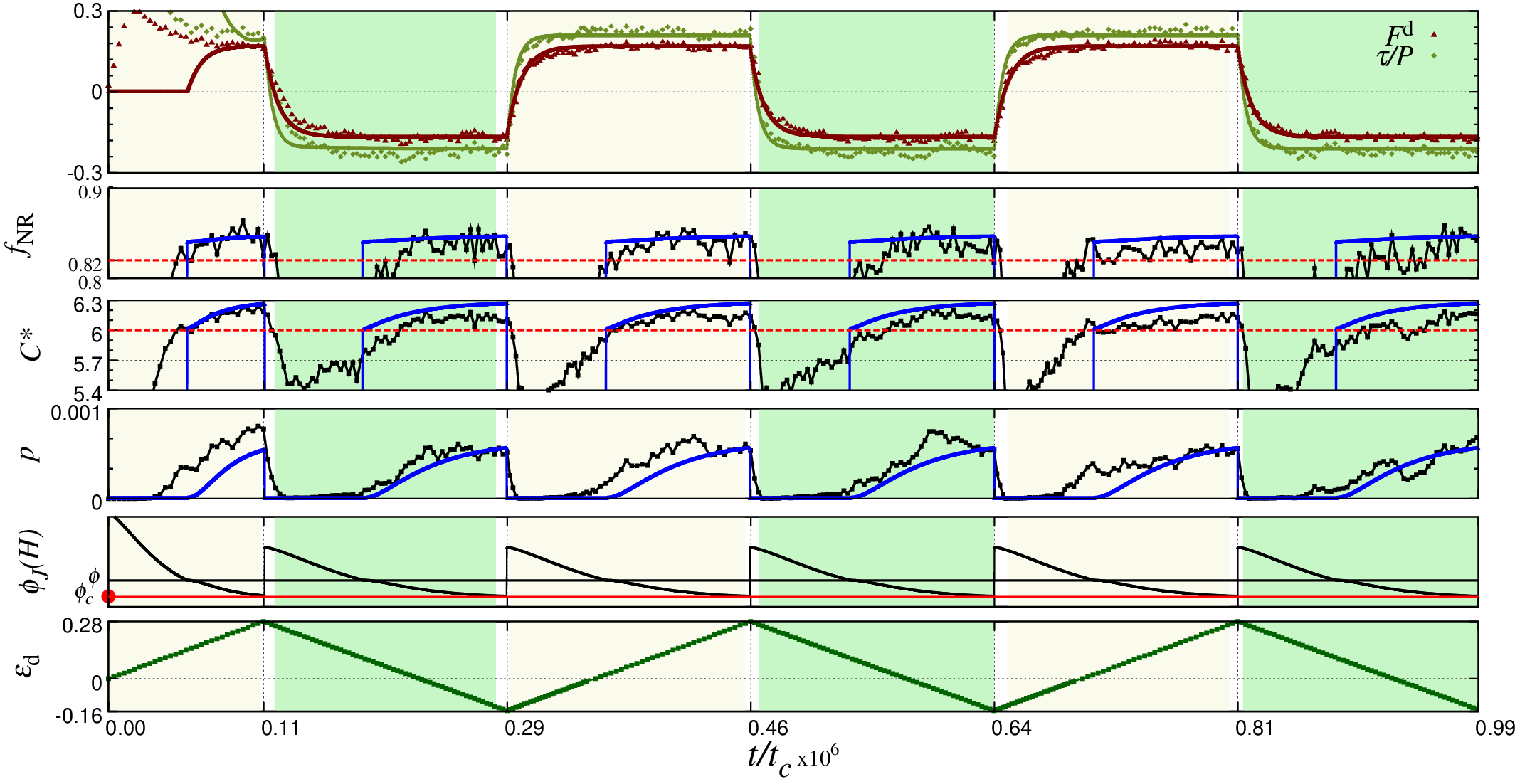}\label{cycle_perco_new}}}
\caption{
{Model prediction of cyclic loading:}
(a) Dimensionless pressure $\norPres$ and (b) coordination number $\Cstar$ 
plotted against volume fraction $\volfrac$ for an isotropic compression starting 
from $\volfract = 0.64$ to $\volfracmax_i = 0.73$ (small symbols)
and decompression (big symbols) back to $\volfract$, with ${}^{\infty}\volfracJi=0.667$, for $M=1$ (red `$\bullet$') and for $M=300$ (blue `$\blacksquare$'). 
(c) Deviatoric stress ratio $\stressrat$ and deviatoric fabric $\anisoF$, 
fraction of non-rattlers $\fNR$, coordination number $\Cstar$, pressure $p$ and history dependent jamming density $\volfracJ(H)$ 
over three pure shear strain cycles (bottom panel) for $\volfrac=0.6584$ and initial jamming density $\volfracJ\left(\volfracmax_i = 0.82, M=1\right) =: \ovolfracJi = 0.6652$. 
Solid lines through the data are the model prediction, involving the history dependent jamming density $\volfracJ(H)$, 
using Eq.\ (\ref{eq:strexp}) for isotropic deformation and Eq.\ (\ref{eq:scalejamming2}) for shear deformation, and others.
Dashed red lines in $\fNR$ and $\Cstar$ represent transition from unjammed to shear jammed states, 
whereas in $\volfracJ(H)$ the red line indicates the critical jamming density $\volfracSJ$.}
\label{predicted}
\end{figure*}

Finally, we test the proposed history dependent jamming density $\volfracJ(H)$ model, by predicting $\norPres$ and $\Cstar$, 
when a granular assembly is subjected to cyclic isotropic compression to $\volfracmax_i = 0.73$ for $M=1$ and for $M=300$ 
cycles, with ${}^{\infty}\volfracJi=0.667$, as shown in Fig.\ \ref{predicted}(a-b). 
It is observed that using the history dependence of $\volfracJ(H)$, the hysteretic behavior 
of the isotropic quantities, $\norPres$ and $\Cstar$, is very well predicted, qualitatively
similar to isotropic compression and decompression of real 2D frictional granular assemblies, 
as shown in by \citet{bandi2013fragility} and \citet{reichhardt2014aspects}. 
%

In Fig.\ \ref{cycle_perco_new}, we show the evolution of the deviatoric quantities 
shear stress ratio $\stressrat$ and deviatoric fabric $\anisoF$, when a system 
with $\volfrac=0.6584$, close to $\volfracSJ$, and initial jamming density 
$\volfracJ(0)= 0.6652$, is subjected to three shear cycles (lowest panel). 
The shear stress ratio $\stressrat$ is initially undefined, but soon 
establishes a maximum (not shown) and {\em decays} to its saturation level 
at large strain. After strain reversal, $\stressrat$ drops suddenly and attains the 
same saturation value, for each half-cycle, only with alternating sign. 
The behavior of the anisotropic fabric $\anisoF$ is similar to 
that of $\stressrat$. During the first loading cycle, the system is unjammed for some strain, 
and hence $\anisoF$ is zero in the model (observations in simulations can be non-zero, when the data correspond to only few contacts, mostly coming from rattlers).
However, the growth/decay rate and the saturation values attained 
are different from those of $\stressrat$,
implying a different, independent stress- and structure-evolution 
with strain -- which is at the basis of recently 
proposed anisotropic constitutive models for quasi-static
granular flow under various deformation modes \citep{imole2013hydrostatic}. 
The simple model with $\volfracJ(H)$, is able to predict quantitatively the behavior the $\stressrat$ and $\anisoF$ after the first loading path, and is qualitatively close to the
cyclic shear behavior of real 2D frictional granular assemblies, 
as shown in Supplementary Fig.\ 7 by \citet{bi2011jamming}. 

At the same time, also the isotropic quantities are very well predicted by
the model, using the simple equations from section\ \ref{subsec:model1},
where only the jamming density is varying with shear strain, while all material 
parameters are kept constant. 
Some arbitrariness involves the sudden changes of 
$\volfracJ$ at reversal, as discussed in section\ \ref{subsec:model1}.
Therefore, using a history dependent $\volfracJ(H)$ gives hope to understand the 
hysteretic observations from realistic granular assemblies, and also provides a 
simple explanation of shear jamming. Modifications of continuum models like 
anisotropic models \citep{imole2013hydrostatic, kumar2014macroscopic}, or 
GSH type models \citep{jiang2008incremental,jiang2015applying}, by including 
a variable $\volfracJ$, can this way quantitatively explain various mechanisms around jamming.

\section{Towards experimental validation}
\label{appA}

The purpose of this section is two-fold: First, we propose ways to (indirectly) measure the jamming
density, since it is a virtual quantity that is hard to measure directly, just as the ``virtual, stress-free
reference state'' in continuum mechanics which it resembles.
Second, this way, we will introduce alternative state-variables, since by no means is the jamming
density the only possibility.

\begin{figure}
\centering
{\subfigure[]{\includegraphics[width=0.3\textwidth, angle=270]{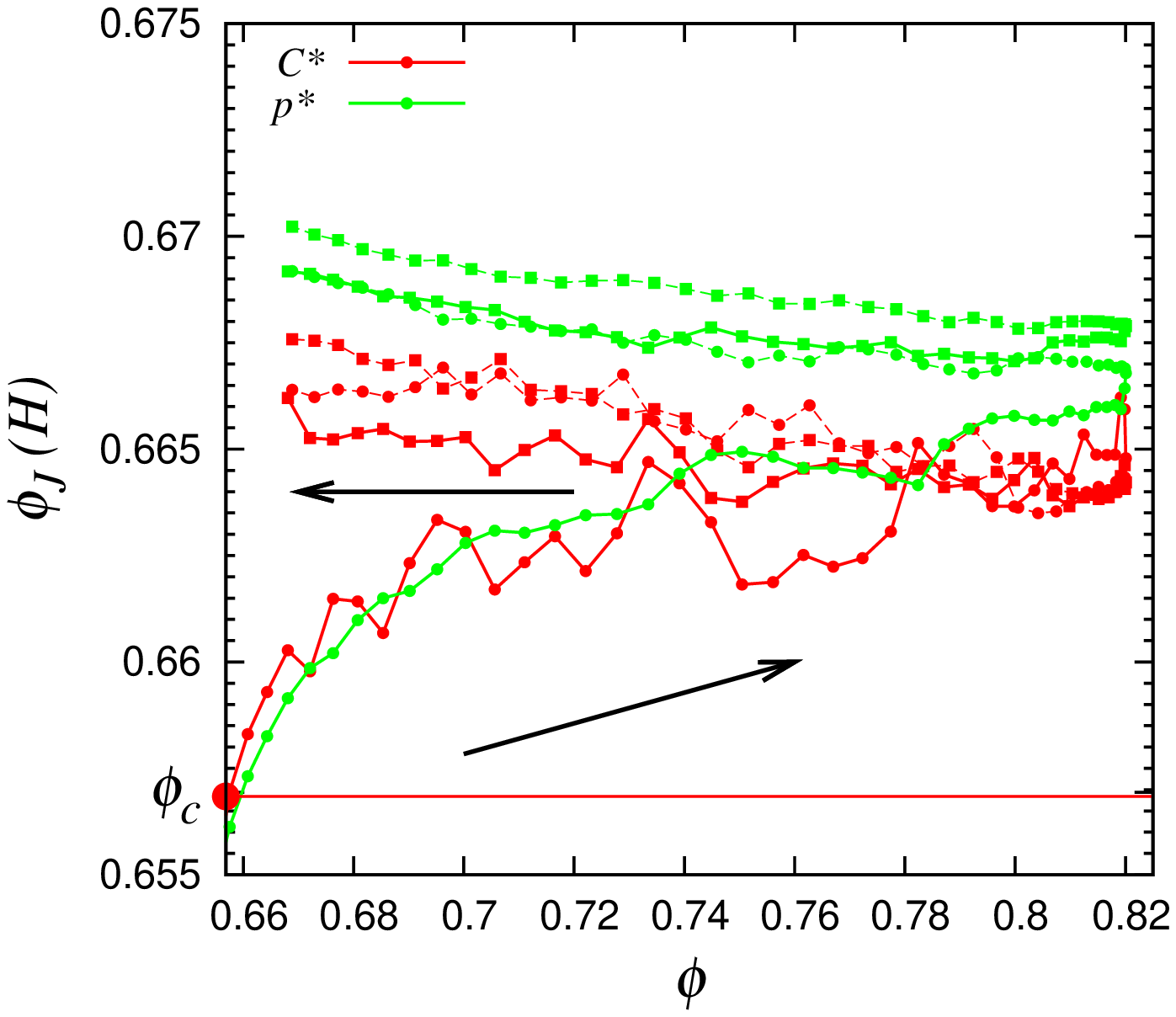}\label{iso2}}}\\
{\subfigure[]{\includegraphics[width=0.3\textwidth, angle=270]{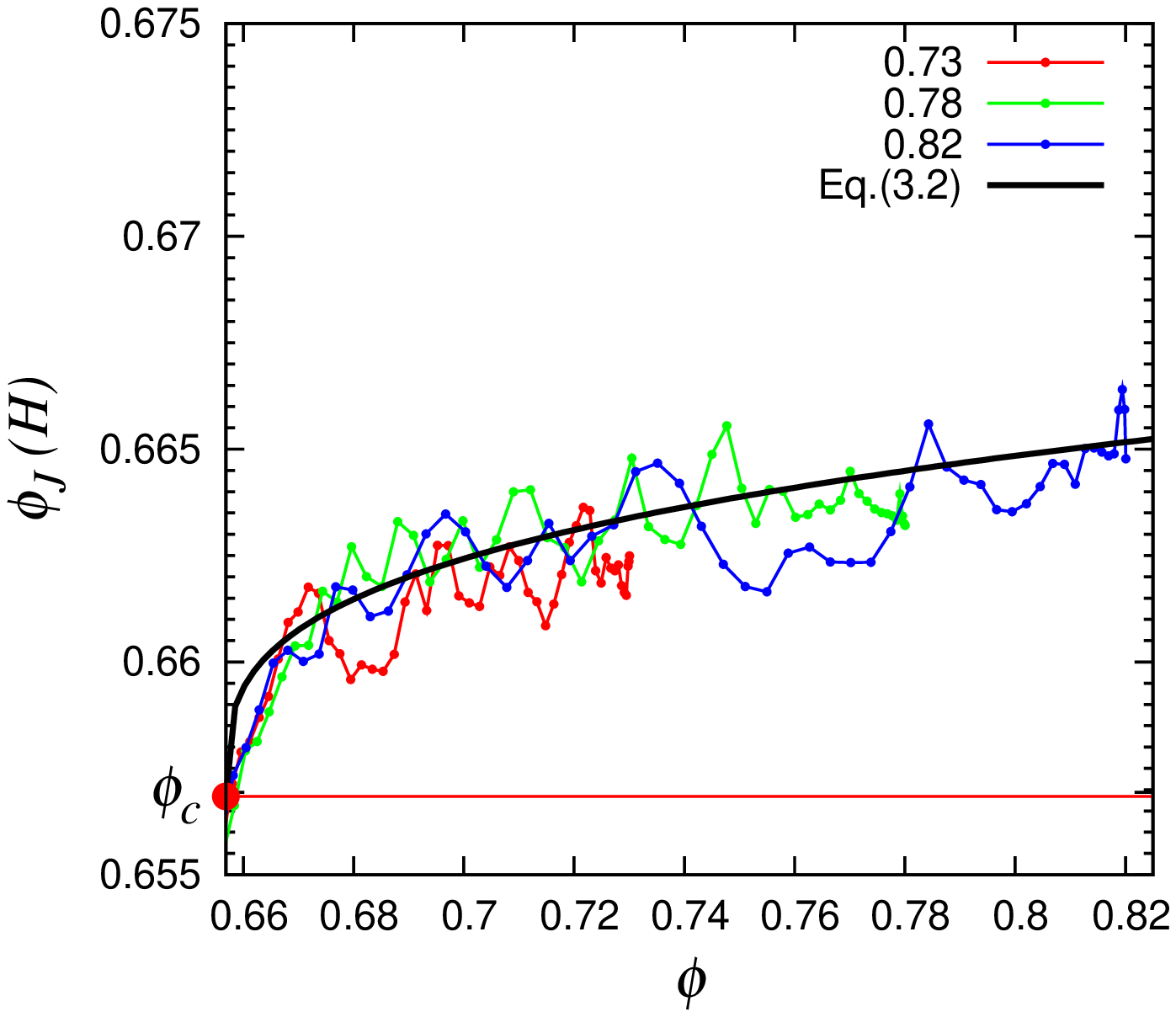}\label{iso_phiJ_moveC}}}
{\subfigure[]{\includegraphics[width=0.3\textwidth, angle=270]{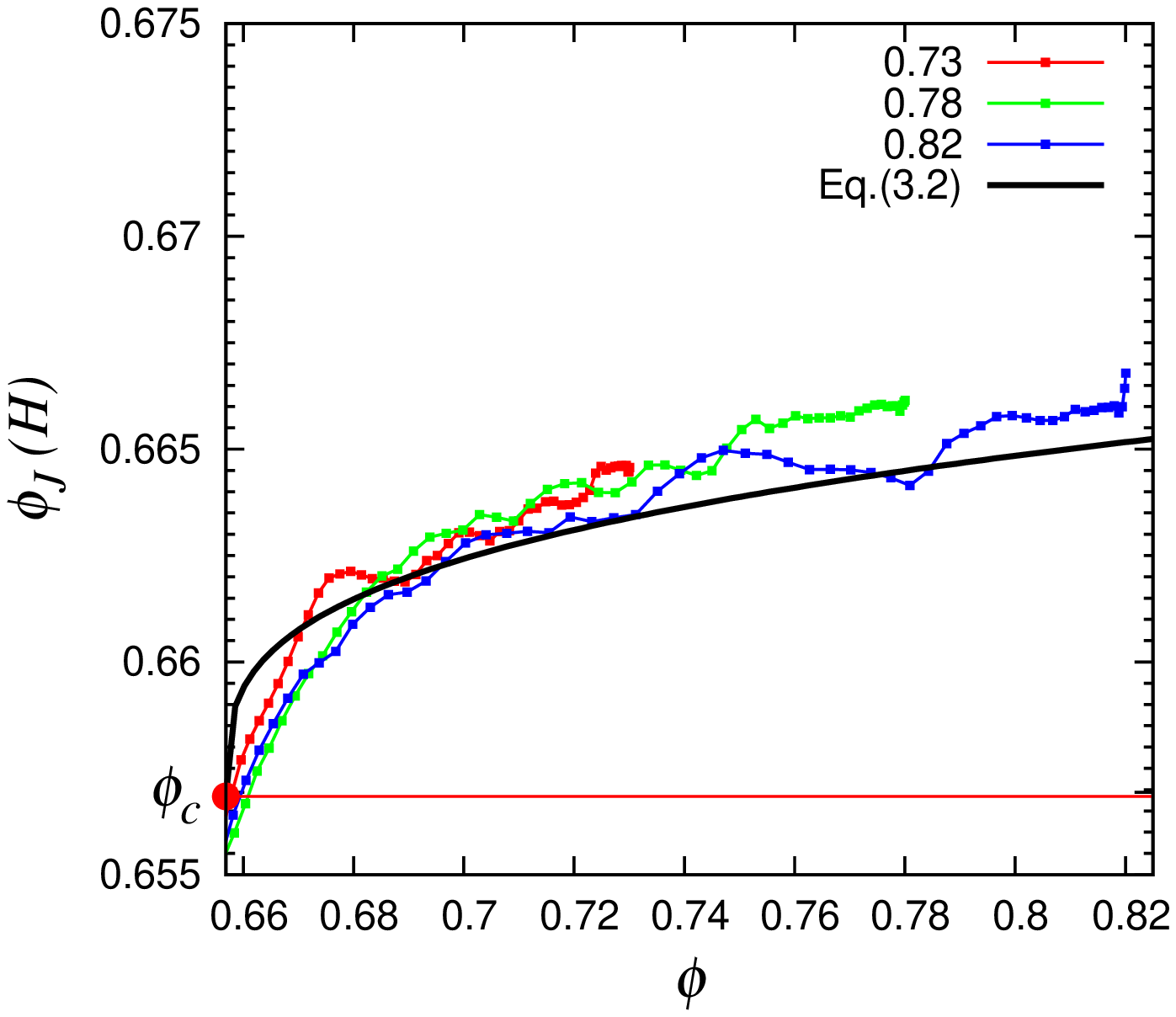}\label{iso_phiJ_movep}}}
\caption{(a) Evolution of the history dependent jamming density $\volfracJ(H)$ during isotropic over-compression to $\volfracmax_i=0.82$ for two cycles, 
calculated back from the measured quantities: coordination number $\Cstar$ (green) and pressure $p$ (red), 
using Eqs.\ (\ref{eq:cstareqn}) and (\ref{eq:pstareqn}) respectively. 
The `$\bullet$' and `$\blacksquare$' represent the first and second cycle respectively.
Solid lines are the loading path while the dashed lines represent the unloading path for the corresponding cycle. 
Evolution of history dependent jamming density $\volfracJ(H)$ using (b) coordination number $\Cstar$ and (c) pressure $p$ for three levels of over-compression $\volfracmax_i$, as shown in the inset. 
Solid black line represents Eq.\ (\ref{eq:strexp}) with $M=1$, and ${}^{\infty}\volfracJi$ calculated using Eq.\ (\ref{eq:asymptoticrelation}).}
\label{extract_phiJ_iso}
\end{figure}

\noindent \textbf{Measuring $\volfracJ$ from experiments}\\
Here we show the procedure to extract the history dependent jamming density $\volfracJ(H)$ from 
measurable quantities, indirectly obtained via Eqs.\ (\ref{eq:cstareqn}), 
(\ref{eq:fNReqn}), (\ref{eq:pstareqn}), and directly from Eq.\ (\ref{eq:scalejamming2}).
There are two reasons to do so: ($i$) the jamming density $\volfracJ(H)$ is only 
accessible in the unloading limit $p\to0$, which requires an experiment or a simulation
to ``measure'' it (however, during this measurement, it might change again); 
($ii$) deducing the jamming density from other quantities that are defined for 
an instantaneous snapshot/configuration for $p > 0$ allows to indirectly obtain 
it -- if, as shown next, these indirect ``measurements'' are compatible/consistent: 
Showing the equivalence of all the different $\volfracJ(H)$, proofs the consistency and completeness of the model and, even more important, provides a way to obtain 
$\volfracJ(H)$ indirectly from experimentally accessible quantities.

\noindent \textbf{For isotropic compression}\\
Fig.\ \ref{extract_phiJ_iso} shows the evolution of $\volfracJ(H)$, measured from the two experimentally accessible quantities: 
coordination number $\Cstar$ and pressure $p$, using Eqs.\ (\ref{eq:cstareqn}) and (\ref{eq:pstareqn}) respectively
for isotropic over-compression to $\volfracmax_i=0.82$ over two cycles. Following observations can be made: 
($i$) $\volfracJ$ for isotropic loading and unloading can be extracted from $\Cstar$ and $p$,
($ii$) it rapidly increases and then saturates during loading,
($iii$) it mimics the fractal energy landscape model in Fig. 4 from \citet{luding2000minimal} very well,
($iv$) while is was assumed not to change for unloading, it even increases, which we attribute to the perturbations and fluctuations (granular temperature) induced during the quasi-static deformations,
($v$) the indirect $\volfracJ$ are reproducible and follow the same master-curve
for first over-compression as seen in Figs.\ \ref{extract_phiJ_iso}, independent of the maximum -- all following deformation is dependent on the previous maximum density.

\noindent \textbf{For shear deformation}\\
\begin{figure}
\centering
\includegraphics[width=0.4\textwidth, angle=270]{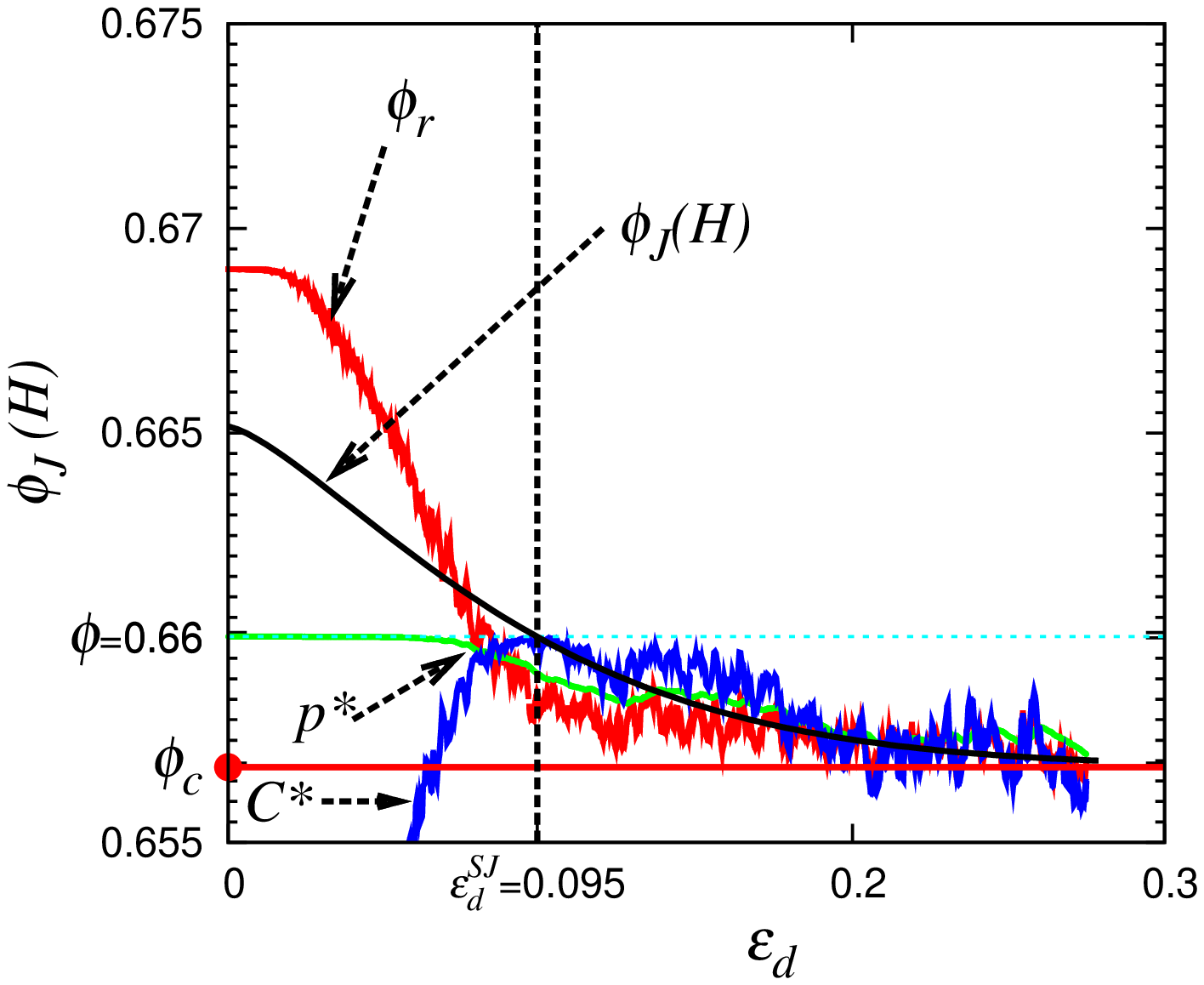}
\caption{Evolution of the history dependent jamming density $\volfracJ(H)$ during pure
shear, calculated back from the measured quantities: coordination number $\Cstar$, fraction of non-rattlers $\fNR$ and pressure $p$,
using Eqs.\ (\ref{eq:cstareqn}), (\ref{eq:fNReqn}), (\ref{eq:pstareqn}) respectively, as marked with arrows. 
The volume fraction is constant, $\volfrac=0.66$, and the initial jamming density 
$\volfracJ\left(\volfracmax_i = 0.82, M=1\right) =: \ovolfracJi = 0.6652$ is greater than $\volfrac$ (represented by horizontal cyan line). 
The solid black line represents Eq.\ (\ref{eq:scalejamming2}), and 
the dashed vertical line indicates the shear strain needed to jam 
the system, $\shstrain^{SJ}$, from which on -- for larger shear strain -- 
the system is jammed.  }
\label{extract_phiJ}
\end{figure}
Fig.\ \ref{extract_phiJ} shows the evolution of $\volfracJ(H)$,  measured from the two experimentally accessible quantities: 
coordination number $\Cstar$ and pressure $p$, using Eqs.\ (\ref{eq:cstareqn}) and (\ref{eq:pstareqn}) respectively during volume conserving shear with 
$\volfrac=0.66$, and the initial jamming density 
$\volfracJ\left(\volfracmax_i = 0.82, M=1\right) =: \ovolfracJi = 0.6652 > \volfrac$ and shows 
good agreement with the theoretical predictions using Eq.\ (\ref{eq:scalejamming2}) after shear jamming.
Thus the indirect measurements of 
$\volfracJ(H)$ can be applied if $\volfracJ(H)<\volfrac$; the result deduced from
pressure fits the best, i.e., it interpolates the two others and is smoother.

\section{Summary, Discussion and Outlook} 
\label{sec:interpretation}

In summary, this study presents a quantitative, predictive macroscopic constitutive model
that unifies a variety of phenomena at and around jamming, for quasi-static deformation modes. 
The most important ingredient is a scalar state-variable that characterizes the packing
``efficiency'' and responds very slowly to (isotropic, perturbative) deformation.
In contrast, it responds exponentially fast to finite shear deformation. This
different response to the two fundamentally different modes of deformation 
(isotropic or deviatoric, shear) is (qualitatively) explained by a stochastic 
(meso-scale) model with fractal (multiscale) character. 
All simulation results considered here
are quantitatively matched by the macroscopic 
model including both the isotropic and the anisotropic microstructure as state-variables.
Discussing the equivalence of alternative state-variables and ways to experimentally
measure the model parameters and apply it to other, more realistic materials, 
concludes the study.
The following subsections wrap up some major aspects of this study and also add some
partly speculative arguments about the wider consequences of our results for rheology as well as an outlook.

\subsection{Some questions answered}

The questions posed in the introduction can now be answered: 
$(i)$ The transition between the jammed and flowing (unjammed) regimes
is controlled by a single new, isotropic, history dependent state-variable, 
the jamming density $\volfracJ(H)$ (with history $H$ as shorthand place-holder
for any deformation path), which
$(ii)$ has a unique lower critical jamming density $\volfracSJ$ when $p \to 0$ without previous history $H$, 
or after very long shear without temperature $T_g$, so that
$(iii)$ the history (protocol dependence) of jamming is completely
encompassed by this new state-variable, and 
$(iv)$ jamming, unjamming and shear jamming can all occur in 3D without any friction, 
only by reorganizations of the micro-structure.

\subsection{Lower limit of jamming}

The multiscale model framework implies now
a minimum $\volfracSJ$ that represents the (critical)
steady state for a given sample in the limit of vanishing confining stress, 
i.e.,\ the \textit{lower limit of all jamming densities}. This is nothing 
but the lowest stable random density a sheared system ``locally'' 
can reach due to continuously ongoing shear, 
in the limit of vanishing confining stress. 

This lower limit is difficult to access in experiments and simulations,
since every shear also perturbs the system leading at the same time to (slow) 
relaxation and thus a competing increase in $\volfracJ(H)$. However,
it can be obtained from the (relaxed) steady state values of pressure, 
extrapolated to zero, i.e.,\ from the envelope of pressure in Fig.\ \ref{phasecompar}.
Note that special other deformation modes or careful preparation 
procedures e.g.\ energy minimization techniques or manual construction
\citep{ohern2003jamming, torquato2010jammed} 
may lead to jammed states at even lower density
than $\volfracSJ$, from which starting to
shear would lead to an increase of the jamming density 
(a mechanism which we could not clearly identify from our 
frictionless simulations due to very long relaxation times near jamming for soft particles). 
This suggests future studies in the presence of friction so that one has wider range of 
jamming densities and lower density states might be much more stable as compared 
to the frictionless systems. 
In this work, we focused on fixed particle size polydispersity with uniform size distribution. 
We expect the effects of polydispersity \citep{kumar2014effects} will have similar order of explorable jamming range as in this work, whereas friction etc. 
will cause larger explorable jamming range and bigger changes in the calibrated parameters. 

\subsection{Shear jamming as consequence of a varying $\volfracJ(H)$}

Given an extremely simple model picture, starting from
an isotropically unjammed system that was previously compressed
or tapped (tempered), shear jamming is not anymore
a new effect, but is just due to the shift of the state-variable
(jamming density) to lower values during shear. 
In other words, shear jamming occurs when the state-variable 
$\volfracJ(H)$ drops below the density $\volfrac$ of the system. 

Even though dilatancy is that what is typically expected under shear 
(of a consolidated packing), also compactancy is observed in some cases \citep{imole2013hydrostatic}
and can be readily explained by our model. Given a certain preparation protocol,
typically a jamming density $\volfracJ > \volfracSJ$ will be reached for a sample, 
since the critical limit $\volfracSJ$ is very difficult to reach. 
When next a shear deformation
is applied, it depends e.g.\ on the strain rate whether dilatancy or 
compaction will be observed: if the shear mode is ``slower'' than the 
preparation, or if $\volfracJ > \volfracSJ$, dilatancy is expected
as a consequence of the rapidly decreasing $\volfracJ$ of the sample. 
In contrast, 
for a relatively ``fast'', violent shear test (relative to the previous preparation 
and possibly relaxation procedure), compactancy also \textit{can be} the result, 
due to an increase of $\volfracJ$ during shear.

\subsection{Rheology}

The multiscale models presented in this study, based on data from frictionless particle simulations, 
implies that a superposition of the two fundamental deformation modes 
(isotropic and deviatoric, i.e.\ plane strain pure shear) is possible or, with other words, 
that the respective system responses are mostly decoupled as shown for the non-Newtonian rheology of simple
fluids in Ref.\ \cite{hartkamp2013constitutive}. 
Even though this decoupling is mostly consistent with our present data 
(the responses to isotropic and deviatoric deformations are mostly independent and 
can be measured independently),
this separation and superposition cannot be taken for granted for more realistic granular
and powder systems.

Nevertheless, the meso-scale model presented here, 
as based on a multi-scale energy landscape, 
explains compactancy and dilatancy, at constant confining stress,
as caused by an increasing jamming density, or a decreasing jamming density, 
respectively (not shown). Similarly, at constant volume, the pressure
either decreases or increases (pressure-dilatancy) due to an 
increasing or decreasing jamming density, respectively.

The model also allows to explain other rheological phenomena as
shear-thinning (e.g., due to an increasing jamming density, at constant volume) 
or shear-thickening (e.g., due to a decreasing jamming density, at constant volume).
As generalization of the present work, also the (granular) temperature (fluctuations of kinetic energy)
can be considered, setting an additional (relaxation) time-scale, which effects the interplay 
between (shear) strain-rate and the evolution of the jamming density, so that even
in a presumed ``quasi-static'' regime interesting new phenomena can be 
observed and explained.

\subsection{Towards experimental validation}

The history dependent jamming density $\volfracJ(H)$ is difficult to access directly,
but can consistently be extracted from other, experimentally measurable quantities, 
e.g.\ pressure $p$, coordination number $\Cstar$ or fraction of non-rattlers $\fNR$.
We explain the methodology to extract $\volfracJ(H)$ experimentally, and confirm 
by indirect measurement, as detailed in section \ref{appA},
that the jamming density is indeed increasing during isotropic deformation
and decreasing during shear, consistently also when deduced from these 
other quantities.

With other words, we do not claim that the jamming density is the only 
choice for the new state-variable that is needed. It can be replaced by any
other isotropic quantity as, e.g.\ the isotropic fabric, the fraction of non-rattlers,
the coordination number, or an empirical stress-free state that is extrapolated 
from pressure (which can be measured most easily), as long as this variable
characterizes the packing ``efficiency''.

Since an increased packing efficiency could be due to ordering (crystallization),
we tried to, but could not trace any considerable crystallization and definitely no phase-separation. 
We attribute this to the polydispersity of the sizes of the particles used being in the range 
to avoid ordering effects, as studied in detail in Ref.\ \citep{ogarko2013prediction}. 
Quantities like the coordination number, which can tremendously increase due to crystallization, 
did not display significant deviation from the random packing values and, actually, it even decreases
in the unloading phases, relative to the initial loading phase, see Fig.\ \ref{predicted}(b).
This is not a proof that there is no crystallization going on, it is just not strong enough to be
clearly seen. The reasons and micro-structural origin of the increased packing efficiency, 
as quantified by the new state-variable, are subject of ongoing research.

\subsection{Outlook}

Experiments should be performed to calibrate our model for suspended
soft spheres (e.g. gels, almost frictionless) and real, frictional materials 
\citep{brujic2007measuring, yu2014monitoring, brodu2015spanning}. 
Over-compression is possible for soft materials, but not expected 
to lead to considerable relaxation due to the small possible compressive strain 
for harder materials. 
However, tapping or small-amplitude shear can take the role of over-compression,
also leading to perturbations and increasing $\volfracJ$,  in contrast, 
large-amplitude shear leads to decreasing $\volfracJ$ and can be calibrated 
indirectly from different isotropic quantities. 
Note that the accessible range of $\volfracJ - \volfracSJ$ is expected to 
much increase for more realistic systems, e.g.,\ with friction, 
for non-spherical particle shapes, or for cohesive powders.

From the theoretical side, a measurement of the multiscale energy landscape, e.g.\ the valley width, 
depth/shapes and their probabilities \citep{xu2011direct} should be done to verify our model-picture, 
as it remains qualitative so far. Finally, applying our model to glassy dynamics, ageing and re-juvenation, 
and frequency dependent responses, encompassing also stretched exponential relaxation, see
e.g.\ \citet{lieou2012nonequilibrium}, is another open challenge for future research. All this involves
the temperature as a source of perturbations that affect the jamming density, and will thus also
allow to understand more dynamic granular systems where the granular temperature is finite
and not negligible as implied in most of this study for the sake of simplicity. A more complete 
theory for soft and granular matter, which involves also the (granular) temperature, is in preparation.

Last, but not least, while the macro/continuum model predicts a smooth evolution of the 
state variables, finite-size systems display (system-size dependent) fluctuations that 
only can be explained by a meso-scale stochastic model as proposed above, with 
particular statistics as predicted already by rather simple models in 
Refs.\ \citep{dahmen2009micromechanical,dahmen2011simple,saitoh2015master}.

\begin{acknowledgements}
We thank 
Robert Behringer, Karin Dahmen, Itai Einav, Ken Kamrin, Mario Liu, Vitaliy Ogarko, Corey O'Hern, 
and Matthias Sperl, for valuable scientific discussions;
critical comments and reviews from Vanessa Magnanimo and Olukayode Imole are gratefully acknowledged.  
This work was financially supported by the European Union funded Marie Curie Initial Training Network, FP7 (ITN-238577), PARDEM (\url{www.pardem.eu}) and the NWO STW-VICI project 10828. 
\end{acknowledgements}


%

\end{document}